\documentclass[aip, floatfix, amsmath, amssymb, reprint]{revtex4-1}

\usepackage{graphicx}
\usepackage{dcolumn}
\usepackage{bm}
\usepackage[utf8]{inputenc}
\usepackage[T1]{fontenc}
\usepackage{mathptmx}
\usepackage{etoolbox}

\usepackage[justification=raggedright]{caption}
\usepackage{subcaption}

\makeatletter
\def\@email#1#2{%
 \endgroup
 \patchcmd{\titleblock@produce}
  {\frontmatter@RRAPformat}
  {\frontmatter@RRAPformat{\produce@RRAP{*#1\href{mailto:#2}{#2}}}\frontmatter@RRAPformat}
  {}{}
}%
\makeatother
\begin{document}

\preprint{AIP/123-QED}

\title{Formation of nanoribbons by carbon atoms confined in single-walled carbon nanotube -- a molecular dynamics study}

\author{S. Eskandari*}
\affiliation{Department of Biological Physics, Eötvös University, Pázmány Péter sétány 1/A, 1117 Budapest, Hungary}

\author{J. Koltai}
\affiliation{Department of Biological Physics, Eötvös University, Pázmány Péter sétány 1/A, 1117 Budapest, Hungary}

\author{I. László}
\affiliation{Department of Theoretical Physics, Budapest University of Technology and Economics, Budafoki út 8, 1111 Budapest, Hungary}

\author{M. Vaezi}
\affiliation{Institute for Nanoscience and Nanotechnology (INST), Sharif University of Technology, Tehran, Iran}

\author{Jenő Kürti*}
\affiliation{Department of Biological Physics, Eötvös University, Pázmány Péter sétány 1/A, 1117 Budapest, Hungary}
\email{eskandari.somaye@gmail.com  and  jeno.kurti@ttk.elte.hu}

\date{\today}

\begin{abstract}
Carbon nanotubes can serve as one-dimensional nanoreactors for in-tube synthesis of various nanostructures. Experimental observations have shown that chains, inner tubes or nanoribbons can grow by thermal decomposition of organic/organometallic molecules encapsulated in carbon nanotubes. The result of the process depends on the temperature, the diameter of the nanotube, and the type and amount of material introduced inside the tube. Nanoribbons are particularly promising materials for nanoelectronics. Motivated by recent experimental results observing the formation of carbon nanoribbons inside carbon nanotubes, molecular dynamics calculations were performed with the open source LAMMPS code to investigate the reactions between carbon atoms confined within a single-walled carbon nanotube. 
Our results show that the interatomic potentials behave differently in quasi-one-dimensional simulations of nanotube-confined space than in three-dimensional simulations. In particular, the Tersoff potential performs better than the widely used ReaxFF potential, in describing the formation of carbon nanoribbons inside nanotubes. We also found a temperature window where the nanoribbons were formed with the fewest defects, i.e. with the largest flatness and the most hexagons, 
which is in agreement with the experimental temperature range.
\end{abstract}

{\maketitle}

\section{\label{sec:level1} Introduction }

In addition to their attractive electrical, optical and mechanical properties, carbon nanotubes are also exceptional materials because, when filled with various atoms or molecules, they can serve as one-dimensional nanocontainers or even nanoreactors for chemical reactions.

One-dimensional confinement in the nanotubes offers the possibility of new types of processes that do not occur in three dimensions, or occur in very different ways, and can thus provide new approaches to different chemical syntheses.

Many good reviews have been published on the encapsulation of various organic and inorganic materials in carbon nanotubes and their chemical reactions within nanotubes \cite{ref01,ref02,ref03,ref04,ref05}. Here we mention just a few of the rich array of relevant experimental studies. 

For the first time, lead was successfully introduced into multi-walled carbon nanotubes (MWCNTs) \cite{Inorg_1}. This has since been replicated with many other metals and other inorganic components \cite{Inorg_2, Inorg_3, Inorg_4, Inorg_5, Inorg_6, Inorg_7, Inorg_8}. The ability of single-walled carbon nanotubes (SWCNTs) to trap organic molecules in their interiors was first demonstrated experimentally on nanopeapods, where $C_{60}$ fullerenes were found to form a “string of pearls” inside SWCNTs of suitable diameter of about 1.4 nm \cite{Peapod_1}. It was shown later that by annealing peapods the fullerene molecules within the host nanotube coalesce and transform into an inner nanotube \cite{Peapod_2, Peapod_3}. The family of peapods that can be obtained by loading nanotubes with different fullerenes, endohedral fullerenes and molecules derived from fullerenes is very rich \cite{Peapod_4,Peapod_5,Peapod_6,Peapod_7,Peapod_8,Peapod_9,Peapod_10,Peapod_11,Peapod_12,Peapod_13,Peapod_14,Peapod_15,Peapod_16}. 
The filling of nanotubes with fullerenes even opens up the possibility of implementing quantum computing with SWCNTs filled with endohedral fullerenes \cite{benjamin2006towards}
Many other organic  \cite{Org_1,Org_2,Org_3,Org_4,Org_5,Org_6} 
and organometallic \cite{Mc_1,Mc_2,Mc_3,Mc_4,Mc_5,Mc_6} molecules can also be encapsulated in carbon nanotubes. 
Numerous experimental observations have shown that linear carbon chains \cite{Cchain_1,Cchain_2,Cchain_3,Cchain_4,Cchain_5,Cchain_6,Cchain_7}, internal carbon tubes \cite{DW_1,DW_2,DW_3,DW_4} or carbon nanoribbons \cite{GNR_1,GNR_2,GNR_3,GNR_4,GNR_5,GNR_7,GNR_8,GNR_9,GNR_10,GNR_11,GNR_12} can be grown by heating organic or organometallic molecules entrapped in SWCNTs.

Carbon nanoribbons are particularly promising materials for nanoelectronic applications because, unlike infinite graphene, they have a non-zero band gap due to their finite width \cite{GNR_13}.
In a recent paper \cite{GNR_12} the synthesis of narrow graphene nanoribbons (GNRs) by high-temperature (800-1200 K) vacuum annealing of ferrocene molecules inside single-walled carbon nanotubes (SWCNTs) was described. 
The formation of two specific armchair graphene nanoribbons, 6-AGNR and 7-AGNR, in (18,0) nanotube was experimentally demonstrated. (6 or 7 is the number of carbon atoms along the width of the ribbon.)
In another recent study, it was demonstrated by tip-enhanced Raman scattering and HRTEM that the encapsulation of 1,2,4-trichlorobenzene molecules in SWCNTs and subsequent annealing in the temperature range of 800-1100 K resulted in the formation of 6-AGNR graphene nanoribbons with lengths of tens of nanometers \cite{cadena2022molecular}.
It should be mentioned that the same kind of procedure but using boron nitride nanotubes instead of carbon nanotubes resulted also in the formation of graphene nanoribbons inside the tubes \cite{cadena2023encapsulation}. These experimental studies \cite{GNR_12,cadena2022molecular} were the main motivation for our molecular dynamics calculations.

In contrast to much experimental work, there is little theoretical work in the literature on the chemical reactions inside carbon nanotubes, and molecular dynamics simulations in particular are rare.

Kim and Tománek have theoretically investigated the microscopic mechanism of fullerene fusion in peapods by calculating total energy change along the optimum reaction path in phase space \cite{Tomanek2004_PRB}.
Their results highlighted the important role of confinement.

Nishio et al studied the formation of $Si$ nanowires inside $(13,0)$ and $(14,0)$ SWCNTs by molecular dynamics simulations \cite{nishio2008formation} where the $Si$ and $C$ atoms were modeled by Tersoff potential \cite{tersoff1988empirical,tersoff1989modeling}. 

In a previous work, we investigated the stability of bamboo defects on the inner tube of double-walled carbon nanotubes grown from peapods by molecular dynamics simulation using the molecular dynamics package DL\_POLY with Tersoff potential \cite{zolyomi2010characteristics}.

Diels-Alder reactions in confined space inside $(8,8)$ and $(9,9)$ SWCNT were investigated by molecular dynamics simulations using CHARMM27 force field in \cite{smith2014confined}.   

Molecular dynamics simulations were carried out by Calvaresi and Zerbetto in order to get insights into adsorption, packing, and fluxes of molecules inside carbon nanotubes \cite{Calvaresi2014_JMaterChemA}.

Marforio et al investigated the aromatic bromination of N-phenylacetamide inside carbon nanotubes by ONIOM calculations \cite{Marforio2017_JPhysChem}.

Previously, we performed molecular dynamics simulations with a self-developed DFT-adjusted tight-binding method for a system where a $(14,0)$ SWCNT was loaded with carbon pentagons (modelling ferrocene molecules without iron and hydrogen) \cite{laszlo2017molecular}. The calculations were carried out at two temperatures, 2000K and 3000K, and the concentration of the carbon atoms varied between 23 – 59 $atoms/nm$. 
For comparison, the linear concentration of carbon atoms is 29 $atoms/nm$ and 33 $atoms/nm$ for the perfect 6-AGNR and 7-AGNR ribbons, respectively. 
For 33 $atoms/nm$ and 2000K defected ribbon structure was formed whereas in other cases, depending on the actual parameters, chains, inner tubes or even closed cage like molecules were obtained.
{Specifically, at low concentration, chains were obtained at both temperatures; at medium concentration, a nanoribbon at low temperature, and a cage-like structure at high temperature; at high concentration and low temperature, an inner tube; at high concentration and high temperature, a cage-like structure again.}

The aim of our present work was to perform systematic molecular dynamics simulations for the simplest possible situation, namely to investigate what happens when purely carbon atoms are confined in a single-walled carbon nanotube at a suitable concentration and temperature. This can help us better understand the actual experimental observations, where the reactions take place between larger organic molecules inside the nanotube.
The results of \cite{laszlo2017molecular}, together with the new experimental observations by Kuzmany et al mentioned above \cite{kuzmany2021well}, formed the starting point for the present work. However, instead of the time-consuming diagonalization of a Hamiltonian we used interatomic potentials between atoms to calculate the forces acting on the atoms, which is a much faster procedure. 
Calculations were carried out using LAMMPS, which is a versatile code and is one of the most widely used freely available molecular dynamics simulation packages \cite{LAMMPS1995,LAMMPS2022}. LAMMPS has a number of different inter-atomic potentials, the performance of which depends on the purpose for which they are used and the conditions under which they are used.

For simulating carbon-related processes, the ReaxFF (Reactive Force Field) potential is nowadays most commonly used, in which the large number of parameters are fitted to a large training set obtained from DFT calculations \cite{van2001reaxff,chenoweth2008development}. The ReaxFF potential has been successfully applied to a number of systems and reactions, e.g. oxidation and pyrolysis of hydrocarbon fuels \cite{wang2010chemistry}, mechanical properties of graphene \cite{li2018reaxff}, graphitization of amorphous carbon \cite{de2016graphitization}, carbonization of polymers \cite{kowalik2019atomistic,rajabpour2021low}, impact deformation of peapods \cite{de2020carbon}, structural and mechanical properties of graphene, carbon nanotubes and fullerenes \cite{fthenakis2022evaluating}.
Before starting the simulations inside the nanotube, we repeated the procedure described in \cite{de2016graphitization} to see how the method works for 3-dimensional graphitization. Using the ReaxFF potential, we reproduced their published result. The only difference was that in a simple cubic lattice we used a smaller supercell containing only 14 x 14 x 14 carbon atoms instead of 32 x 32 x 32 carbon atoms, otherwise the procedure was exactly the same. 
The method was then adapted to the case where carbon atoms are confined in a nanotube. Unfortunately, the results were {unexpected}: instead of hexagonal structures, we obtained many triangles with carbon chains in between (see the detailes later in the Results and Discussion part). The exact structure depended on the concentration of carbon atoms and the temperature. That is, we have seen from our own experience that there is a significant difference between three-dimensional and one-dimensional simulations, even when using the same interatomic potential. To say the least, there is no room for parallel graphene ribbons inside a nanotube. To the best of our knowledge, the ReaxFF potential has not yet been used to simulate processes inside nanotubes. 
We therefore set out to complement ReaxFF with four additional potentials to systematically investigate how these potentials perform when carbon atoms are confined inside a carbon nanotube. The four additional potentials are: Tersoff \cite{tersoff1988empirical,tersoff1989modeling} and its improved versions, AIREBO (adaptive intermolecular reactive empirical bond order) \cite{AIREBO}, REBO-II (2nd generation reactive empirical bond order) \cite{REBO-II} and LCBOP-I (long range carbon bond order potential) \cite{LCBOP-I}. Each potential is suitable for molecular dynamics studies of similar systems.

{The five potentials used are briefly compared here. The Tersoff potential was the first interatomic potential to be successfully applied to different carbon systems. The Tersoff formula consists of the sum of attractive and repulsive pair potentials. The coefficient of the attractive term, however, depends on the local environment, so that the interaction ends up being many-body in nature. The Tersoff potential does not contain a long-range (van der Waals) interaction. REBO-II is an extension of the Tersoff potential. It describes short-range interactions in empirical bond order terms. The parameter fitting is based on empirical database of equilibrium distances, energies, and stretching force constants of various hydrocarbons. Like the Tersoff potential, REBO-II does not include long-range vdW interactions. AIREBO is another extension of the Tersoff potential, which is also based on the empirical bond order. The main difference compared to REBO is that the AIREBO potential includes the long-range interaction (via the Lennard-Jones potential). The LCBOP potential is similar to AIREBO in that it contains long-range terms in addition to bond order terms. However, while in AIREBO the parametrization of the long-range terms is done independently of the bond order terms, in LCBOP the parameters are fitted together. The philosophy of the ReaxFF potential is different from the others. The functional form of the potential is very broad. Although it includes the usual quantities (bonds, angles, dihedrals, even van der Waals interaction), instead of the chemical considerations used in the other cases, the large number of parameters of ReaxFF are "simply" fitted to a large training set based on DFT.}

\section{Methodology}

All molecular dynamics simulations were performed using the open source software LAMMPS. The conditions were chosen as closely as possible to the experimental conditions, with the significant simplification that instead of organic or organometallic molecules, only single carbon atoms were placed inside the nanotubes, randomly arranged as the starting geometry. The reason for considering only carbon atoms, without hydrogen atoms, was that our main goal was to systematically compare five different interatomic potentials - Tersoff, Rebo, Airebo, Lcbop and ReaxFF - and of these, in case of Tersoff potential one cannot include hydrogen atoms.

The initial geometry was the same in all cases: using PACKMOL software, 99 randomly arranged separate carbon atoms were confined inside a $3 \mathrm{~nm}$ long piece of $(18,0)$ SWCNT. The diameter of the tube was $1.4$ $\mathrm{nm}$. Due to the periodic boundary condition, the geometry was repeated along the nanotube. The density of free carbon atoms of 33 atoms $/ \mathrm{nm}$ was in line with the already mentioned experimental observations \cite{GNR_12}. The nanotube wall was kept rigid, the carbon atoms in the wall did not move at all during the simulation, the wall merely served to lock the carbon atoms inside. Figure 1 shows the initial geometry from the front (left) and from the side (right). To get a better overview of the position of the free carbon atoms, only part of the tube is shown on the right side of Figure 1.

\begin{figure}[ht!]
    {\centering
    \includegraphics[width=\linewidth]{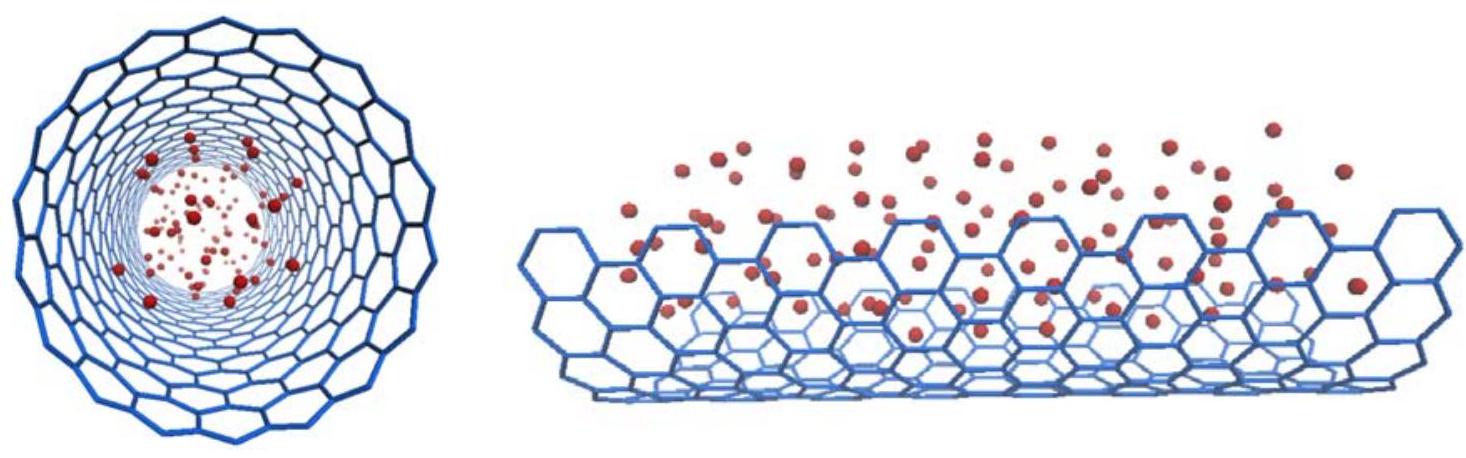}}
    \caption{Initial geometry; 99 carbon atoms randomly arranged inside $(18,0)$ SWCNT with length of $3 \mathrm{~nm}$.}
    \label{fig1}
\end{figure}

After performing many convergence and stability calculations in order to find the optimal numerical parameters, we applied the following procedure in all cases. The time step throughout the modelling was $0.1 \mathrm{fs}$. First, we used a conjugate gradient method for a few steps to minimize the forces on the atoms. In this way we avoided the nonphysical high forces between the randomly arranged carbon atoms. This was followed by equilibration at $300 \mathrm{~K}$ for $10 \mathrm{ps}$ by applying an NPT ensemble, where the environmental temperature was controlled by Nosé-Hoover thermostat. After equilibration, while the SWCNT was kept rigid, the system was heated to raise the temperature from $300 \mathrm{~K}$ to annealing temperatures of $T_{\text {anneal }}=1200 \mathrm{~K}, 3000 \mathrm{~K}$ and $4000 \mathrm{~K}$ in a few ps at a rate of $410 \mathrm{~K} / \mathrm{ps}$. Finally the system was annealed for $100 \mathrm{ps}$ at $\mathrm{T}_{\text {anneal }}$. The heating and annealing processes were carried out using an NVT ensemble, which also adjusts the temperature using a Nosé-Hoover thermostat. 

To quantitatively track the change in structure, we used a self-written Python code that calculates different geometric features from the initial random structure every $100 \mathrm{fs}$ during the simulation. These geometric features are the average number of bonds per carbon atom (a bond is defined when the distance between two carbon atoms is less than $1.7$ \AA ), the distribution of $C-C$ bond lengths, the number of different polygons from triangles to octagons, and the average distance from the best inplane fit of the atom positions (planarity).

\section{Results and Discussion}
\subsection{ReaxFF potential}
The most commonly used potential for simulating carbon-related processes today is ReaxFF \cite{van2001reaxff,chenoweth2008development}. In ReaxFF the large number of parameters are fitted to a large training set obtained from DFT calculations. We will see, however, that the ReaxFF potential performs poorly in describing the formation of hexagonal nanoribbons within carbon nanotubes. 
The final structure using ReaxFF potential after $100 \mathrm{ps}$ annealing at three different temperatures of 1200 $\mathrm{K}, 3000 \mathrm{~K}$ and $4000 \mathrm{~K}$ are shown in Figures 2a, 2b and 2c, respectively.

\begin{figure}[ht!]
     \centering
     \begin{subfigure}{\linewidth}
         \centering
         \includegraphics[width=7cm]{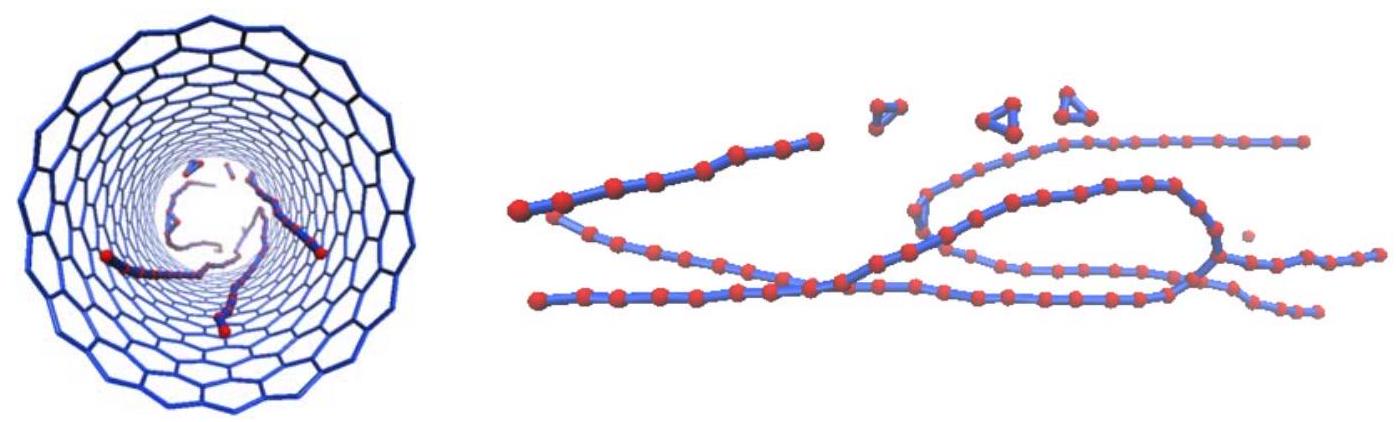}
         \caption{Final result at $1200 \mathrm{~K}$.}
          \label{fig2a}
     \end{subfigure}

     \begin{subfigure}{\linewidth}
         \centering
         \includegraphics[width=7cm]{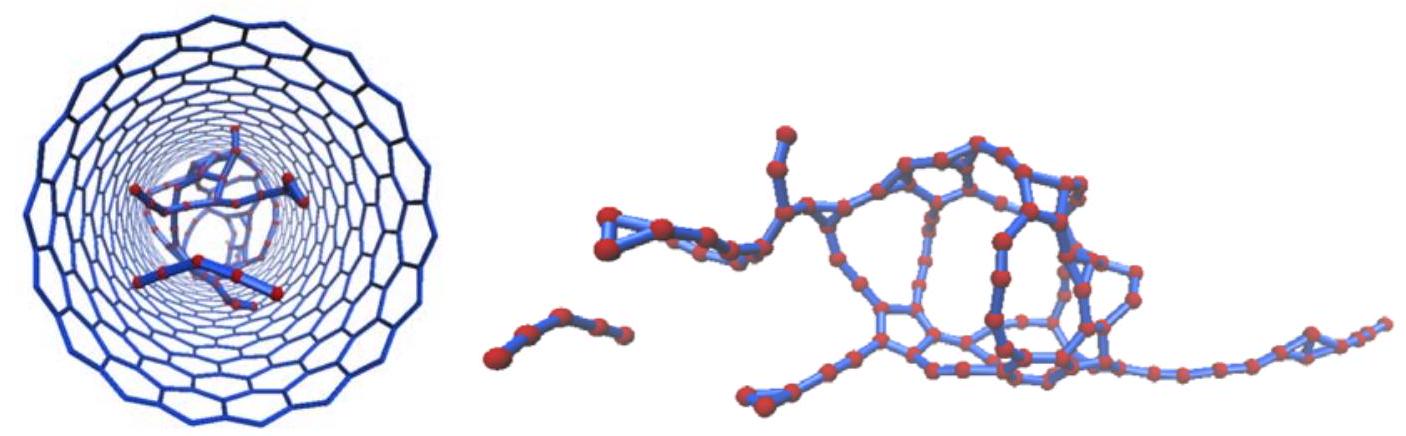}
         \caption{Final result at $3000 \mathrm{~K}$.}
            \label{fig2b}
     \end{subfigure}

     \begin{subfigure}{\linewidth}
         \centering
         \includegraphics[width=7cm]{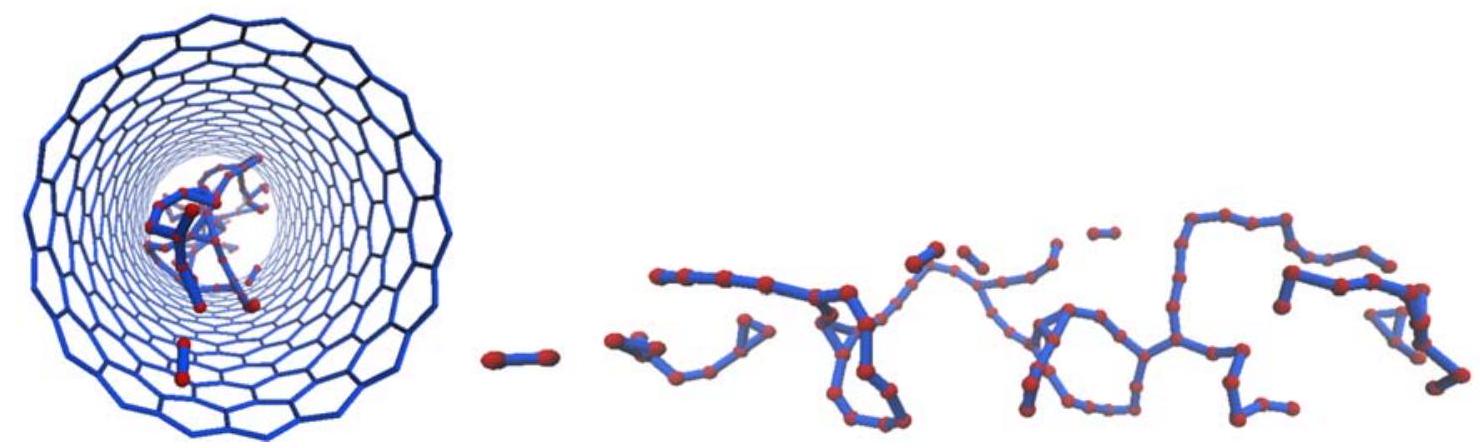}
         \caption{Final result at $4000 \mathrm{~K}$.}
            \label{fig2c}
     \end{subfigure}
     
\caption{ The final structure using ReaxFF potential after 100 ps annealing at three temperatures of $1200 \mathrm{~K}$ (a), $3000\mathrm{~K}$ (b), and $4000 \mathrm{~K}$ (c). At $1200 \mathrm{~K}$ Chains are formed, which are located close to the inner wall of the nanotube. There appeared non-physical triangles, as well. At $3000 \mathrm{~K}$, shorter and longer chains connect several polygons, including non-physical ones such as triangles and rhombuses. At $4000 \mathrm{~K}$, except of triangles, there is no polygon and mostly long chain structures are formed.}
\label{fig2}
\end{figure}

Figure \ref{fig2a} clearly shows that $1200 \mathrm{~K}$ is not an appropriate annealing temperature for the ReaxFF potential, as only chains and non-physical triangles are formed after 100ps. It is noticeable that the chains preferentially "stick" to the inner wall of the nanotube. This is a consequence of the long-range interaction incorporated in ReaxFF. When the annealing temperature is $3000 \mathrm{~K}$, various polygons such as pentagons, hexagons and heptagons are formed, but the non-physical triangles still appear and even rhombuses can be observed, see Figure \ref{fig2b}. At $T_{\text {anneal}}=4000 \mathrm{~K}$, almost no polygons are formed, but chain-like structures with cross-links between the chains, some non-physical triangles and some carbon pairs appear, see Figure \ref{fig2c}.

\begin{figure}[ht!]
    \centering
    \includegraphics[width=\linewidth]{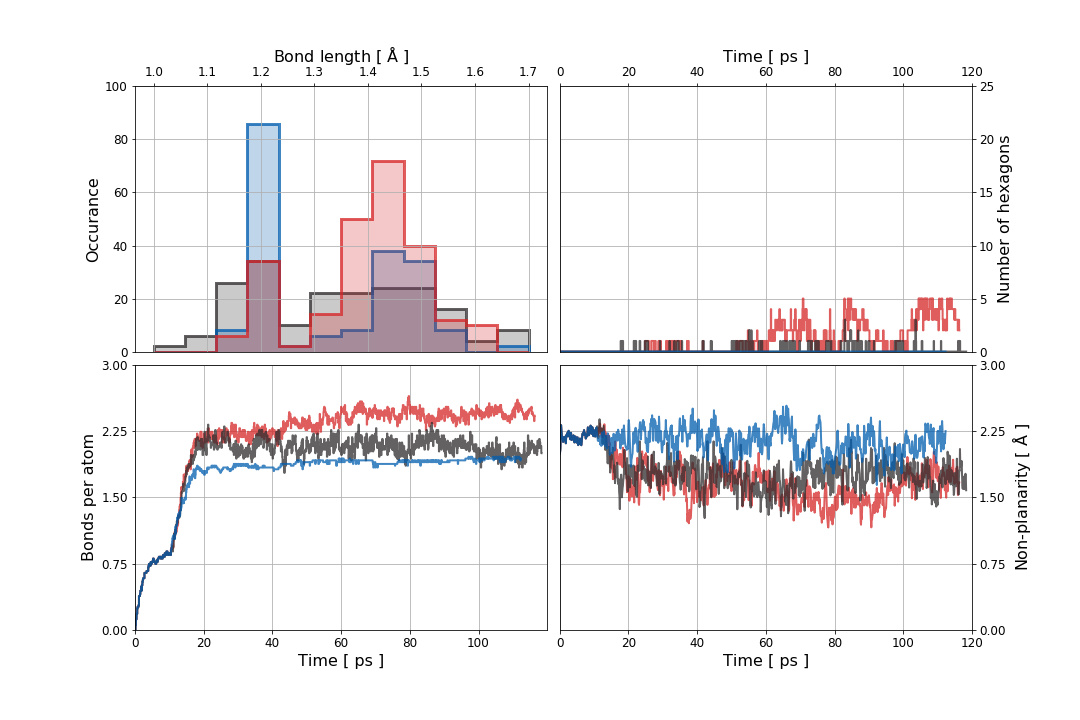}
    \caption{Increase of the average number of bonds per carbon atom during the MD simulation (lower left). The occurrence of $C-C$ bond lengths at the end of the MD simulation (top left). The number of hexagons during the MD simulation (top right). The average distance (in Å) from best planar fit during the MD simulation (lower right). $T_{\text {anneal }}$ is $1200 \mathrm{~K}$ (blue), $3000 \mathrm{~K}$ (red) and $4000 \mathrm{~K}$ (black). All for ReaxFF potential.}
    \label{fig3}
\end{figure}

The change of various geometrical features during the simulation can be followed in Figure \ref{fig3}. The lower left panel of Figure \ref{fig3} shows how the average number of bonds per carbon atom increases and converges during the simulation at the three different temperatures. Remember: a bond is defined when the distance between two carbon atoms is less than $1.7$ \AA. Convergence is achieved after about 60 ps at all three temperatures. The converged value is about (or slightly less than) 2 at $1200 \mathrm{~K}$, and it is about $2.5$ at $3000 \mathrm{~K}$, and it becomes again about 2 at $4000 \mathrm{~K}$. Just for comparison: this value is 2 for infinite linear chain; 3 for infinite graphene and for fullerenes and nanotubes; $2.5$ for coronene, 3-2/N for N-AGNR etc. The converged values in the bottom left of Figure \ref{fig3} correspond nicely to the structures shown in Figure \ref{fig2a}, Figure \ref{fig2b} and Figure \ref{fig2c} . Furthermore, at higher temperatures the fluctuations are larger.

In order to quantify further the qualitative differences seen in Figure \ref{fig2a}, Figure \ref{fig2b} and Figure \ref{fig2c}, additional quantities were calculated. These values in Figure \ref{fig3} refer to the unit cell with a length of $3 \mathrm{~nm}$.

The histograms at the top left panel of Figure \ref{fig3} illustrate the distribution of the $C-C$ distances (bond lengths) at the end of the molecular dynamics simulation for the three different temperatures. The distribution has two maxima at all three temperatures: one at $1.2$ \AA, corresponding to chains, and one at around $1.45$ \AA, corresponding to polygons in the structure. At $1200 \mathrm{~K}$ the contribution of chains and at $3000 \mathrm{~K}$ that of polygons is more significant, in agreement with the structures shown in Figure \ref{fig2a} and Figure \ref{fig2b}. At $4000 \mathrm{~K}$ again more chains and less polygons are formed, in agreement with the structure shown in Figure \ref{fig2c}. Furthermore, there is a general tendency for the distribution to widen with increasing temperature. The number of polygons and especially hexagons is low at any temperature. This is shown in top right panel of Figure \ref{fig3} for the three temperatures. At $1200 \mathrm{~K}$, no hexagons are formed at all, while at $3000 \mathrm{~K}$ there are a few hexagons along with a few pentagons, heptagons and octagons and even with some non-physical triangles and rhombuses. In particular, the number of triangles is high at $3000 \mathrm{~K}$, reaching 10 (not shown in Figure \ref{fig3}). At $4000 \mathrm{~K}$, again, no hexagons are formed, similar to $1200 \mathrm{~K}$. The number of other polygons is also almost zero, except for the triangles, which are almost the same number as at $3000 \mathrm{~K}$.

We are interested in the possibility of creating planar structures, such as carbon nanoribbons. However, carbon atoms are far from being/staying in the same plane at any temperature for ReaxFF potential. The non-planarity at $1200 \mathrm{~K}$ is very high, fluctuating around $2.1$ \AA, which is practically the same as the nonplanarity of the random initial geometry. At $3000 \mathrm{~K}$ and $4000 \mathrm{~K}$ it is slightly lower, but still high, fluctuating around $1.7$ \AA, as shown in the lower right panel of Figure \ref{fig3}.

In summary, the ReaxFF potential performs poorly in describing the formation of hexagonal nanoribbons within SWCNTs. At all three temperatures, carbon atoms are far from forming a planar structure, the tendency to form chains is high, and too many non-physical structures, triangles, are formed.

\subsection{Tersoff potential}
\begin{figure}[ht!]
 \begin{subfigure}{\linewidth}
         \centering
         \includegraphics[width=8cm]{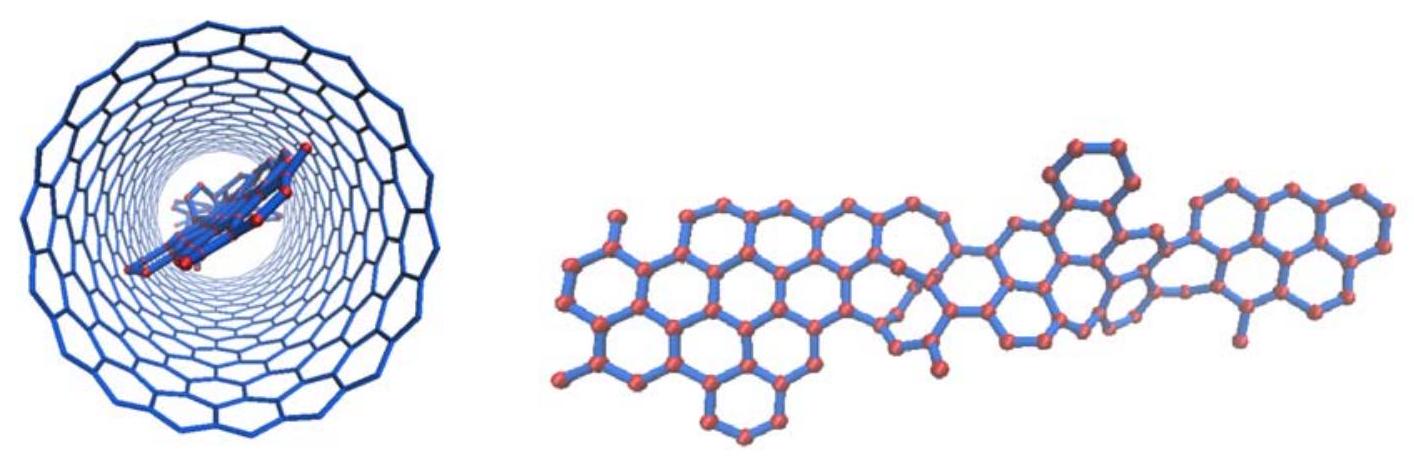}
    \caption{Final Result at $1200 \mathrm{~K}$}
    \label{fig4a}
\end{subfigure}

 \begin{subfigure}{\linewidth}
\centering
         \includegraphics[width=8cm]{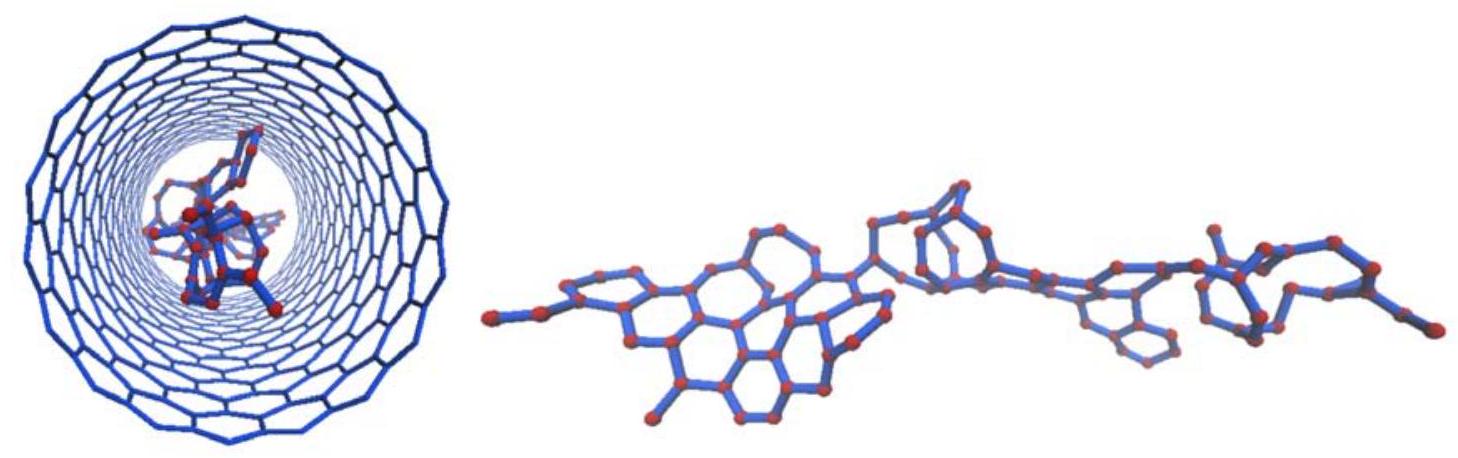}
    \caption{Final Result at $3000 \mathrm{~K}$}
    \label{fig4b}
\end{subfigure}

 \begin{subfigure}{\linewidth}
         \centering
         \includegraphics[width=8cm]{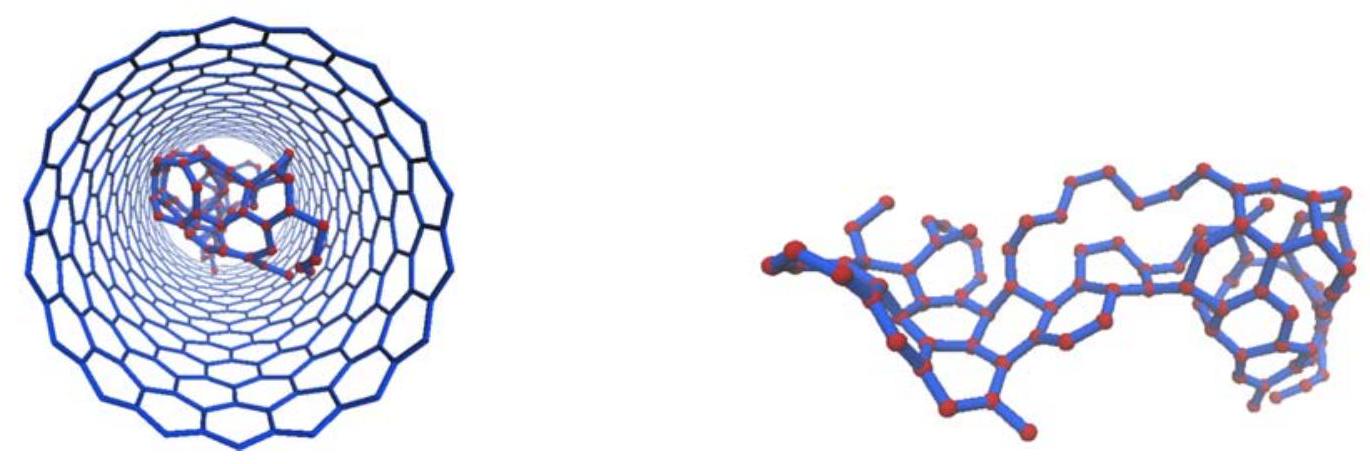}
    \caption{Final Result at $4000 \mathrm{~K}$}
    \label{fig4c}
    \end{subfigure}
  
    \caption{The final structure using Tersoff potential after 100 ps annealing at three temperatures of $1200 \mathrm{~K}$ (a), $3000\mathrm{~K}$ (b), and $4000 \mathrm{~K}$ (c).At $1200 \mathrm{~K}$, 22 hexagons are formed from free carbon atoms inside the tube, and the structure is close to be planar.At $3000 \mathrm{~K}$, there are fewer hexagons and the structure bends.At $4000 \mathrm{~K}$, the non-planarity is increased, the structure bends more and a cage-like geometry appears at the right part.}
    \label{fig4}
\end{figure}

Unlike the ReaxFF potential, the Tersoff potential is more suitable for describing the formation of nanoribbons, as we will see. 
The final structure using Tersoff potential after $100 \mathrm{ps}$ annealing at three different temperatures of $1200\mathrm{K}$, $3000 \mathrm{~K}$ and $4000 \mathrm{~K}$ are shown in Figures \ref{fig4a}, \ref{fig4b} and \ref{fig4c}, respectively.

At $1200 \mathrm{~K}$, a ribbon-like structure with 22 hexagons $(\operatorname{per} 3 \mathrm{~nm}$ ) is formed and the geometry is more or less flat. In contrast, annealing at $3000 \mathrm{~K}$ and $4000 \mathrm{~K}$ results in a reduction in the number of hexagons and the geometry is not flat.

\begin{figure}[ht!]
    \centering
    \includegraphics[width=\linewidth]{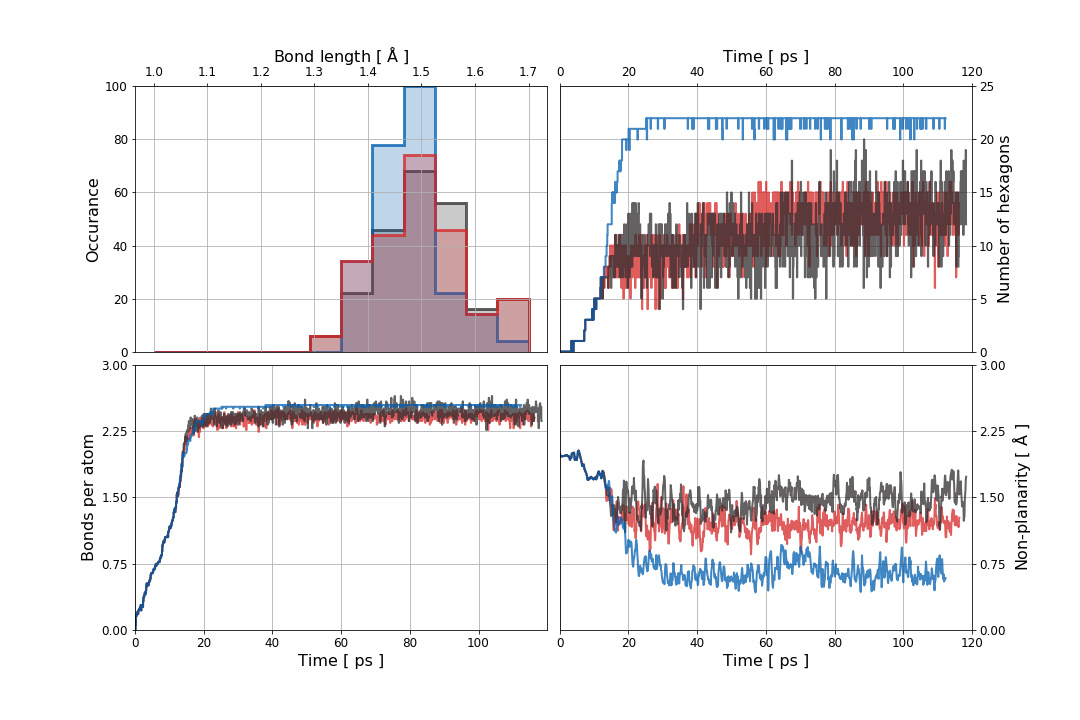}
    \caption{ Increase of the average number of bonds per carbon atom during the MD simulation (lower left). The occurrence of $C-C$ bond lengths at the end of the MD simulation (top left). The number of hexagons during the MD simulation (top right). The average distance (in Å) from best planar fit during the MD simulation (lower right). $T_{\text {anneal }}$ is $1200 \mathrm{~K}$ (blue), $3000 \mathrm{~K}$ (red) and $4000 \mathrm{~K}$ (black). All using Tersoff potential.}
    \label{fig5}
\end{figure}

The change of various geometrical features during the simulation can be followed in Figure \ref{fig5}. The lower left panel of Figure \ref{fig5} shows how the average number of bonds per carbon atom increases and converges during the simulation at the three different temperatures. The convergence is faster than with ReaxFF potential. The converged value is slightly above $2.5$ at $1200 \mathrm{~K}$, whereas it is slightly below $2.5$ at higher temperatures. Furthermore, at higher temperatures the fluctuations are larger.
 
The histograms at the top left panel of Figure \ref{fig5} illustrate the distribution of the $C-C$ distances (bond lengths) at the end of the molecular dynamics simulation for the three different temperatures. At higher temperatures, the distributions are slightly wider. It is worth noting that, looking at the $1200 \mathrm{~K}$ histogram at larger distances (not shown here), after the first maximum of about $1.45$ \AA, the next maxima are at $\sqrt{3} * 1.45$ {\AA} and $2 * 1.45$ \AA, which correspond nicely to the distances of the second and third neighbours in a perfect hexagonal grid.

It is instructive to observe how polygons are formed in the MD simulation. At $1200 \mathrm{~K}$, almost only hexagons are formed, sometimes with two or three heptagons and octagons. At $3000 \mathrm{~K}$ and $4000 \mathrm{~K}$, there are fewer hexagons and more larger polygons. The time dependence of the number of hexagons is shown in top right panel of Figure \ref{fig5} for the three temperatures. At $1200 \mathrm{~K}$, the number of hexagons converges to 22, but at $3000 \mathrm{~K}$ and $4000 \mathrm{~K}$ the number of hexagons fluctuates between 10 and 15.

As shown in Figures \ref{fig4a}, \ref{fig4b} and \ref{fig4c}, the most striking difference between the results of the three temperatures is not only in the number of hexagons but also in the flatness as well. This can be quantified by calculating the average distance of the atoms from the plane fitted to the points using the method of least squares. As shown in the lower right panel of Figure \ref{fig5}, at $1200 \mathrm{~K}$ the non-planarity converges to a small value, with an average deviation from the best in-plane fit of about $0.6$ \AA. At $3000 \mathrm{~K}$ and $4000 \mathrm{~K}$, however, the non-planarity is much larger, fluctuating around $1.25$ {\AA} and $1.5$ \AA, respectively.

In summary, by applying the Tersoff potential at $1200 \mathrm{~K}$, a near-planar hexagonal structure is formed and stabilized a few times in $10 \mathrm{ps}$, although the structure is not defect-free. At annealing temperatures of $3000 \mathrm{~K}$ and $4000 \mathrm{~K}$, the structure is non-planar and consists of far fewer hexagons.

In addition to the Tersoff potential and the ReaxFF potential, three other potentials were also investigated: AIREBO, REBO-II and LCBOP-I.

\subsection{AIREBO potential}

AIREBO is a long-range extension of the Tersoff potential \cite {AIREBO}.
The final structure using AIREBO potential after $100 \mathrm{ps}$ annealing at three different temperatures of $1200 \mathrm{~K}, 3000 \mathrm{~K}$ and $4000 \mathrm{~K}$ are shown in Figures 6a, 6b and 6c , respectively.

\begin{figure}[ht!]
 \begin{subfigure}{\linewidth}
     \centering
    \includegraphics[width=8cm]{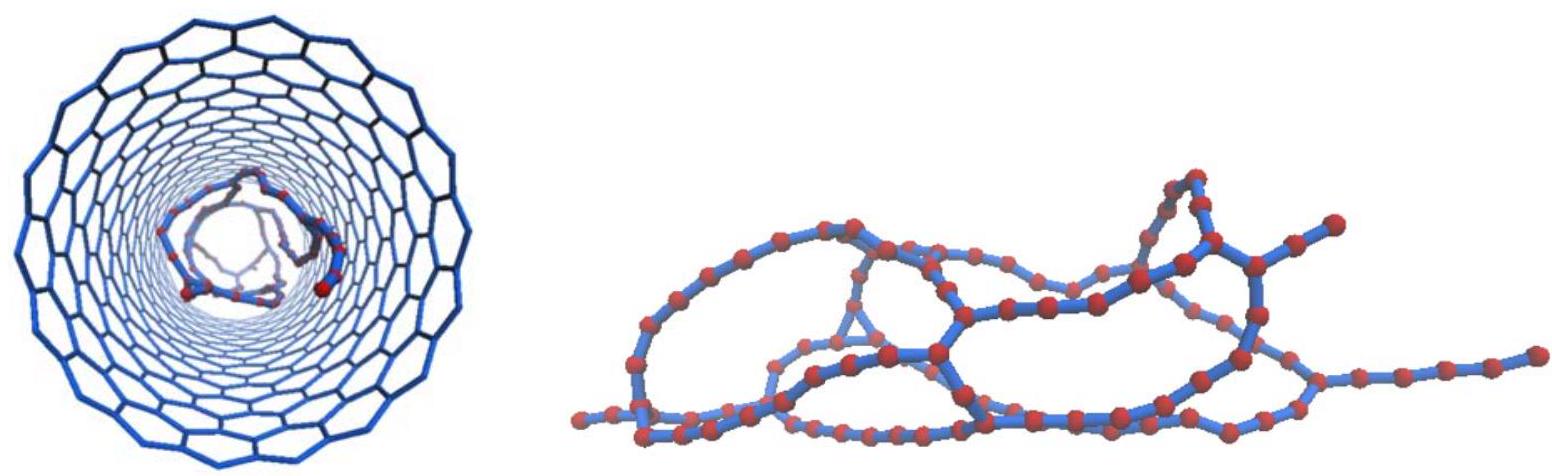}
    \caption{Final Result at $1200 \mathrm{~K}$}
    \label{fig6a}
\end{subfigure}

\begin{subfigure}{\linewidth}
    \centering
    \includegraphics[width=8cm]{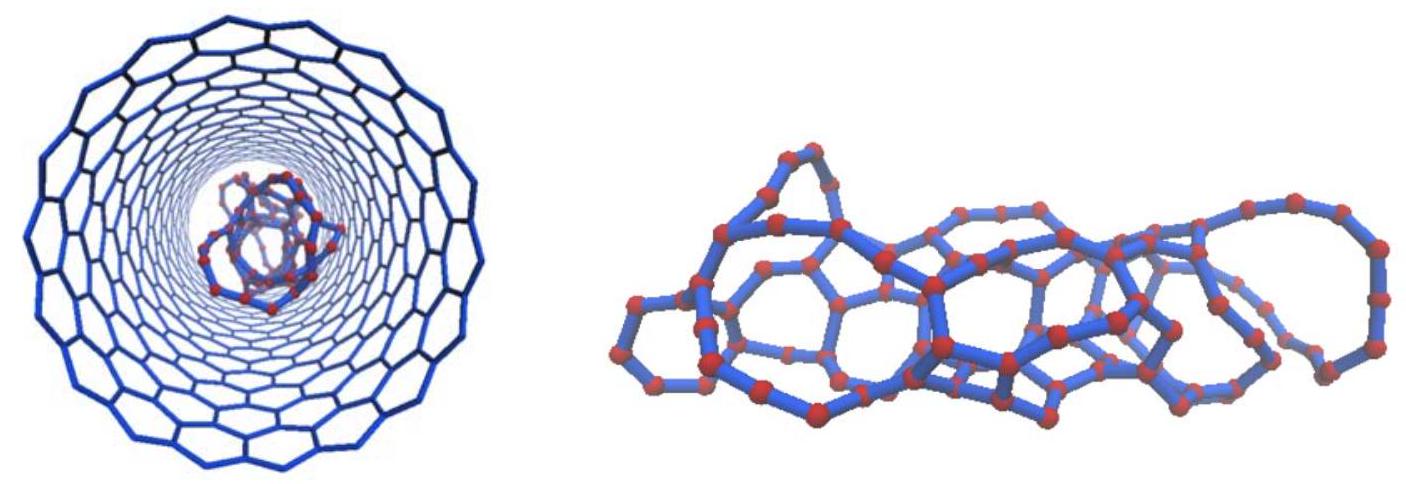}
    \caption{Final Result at $3000 \mathrm{~K}$.}
    \label{fig6b}
\end{subfigure}

\begin{subfigure}{\linewidth}
    \centering
    \includegraphics[width=8cm]{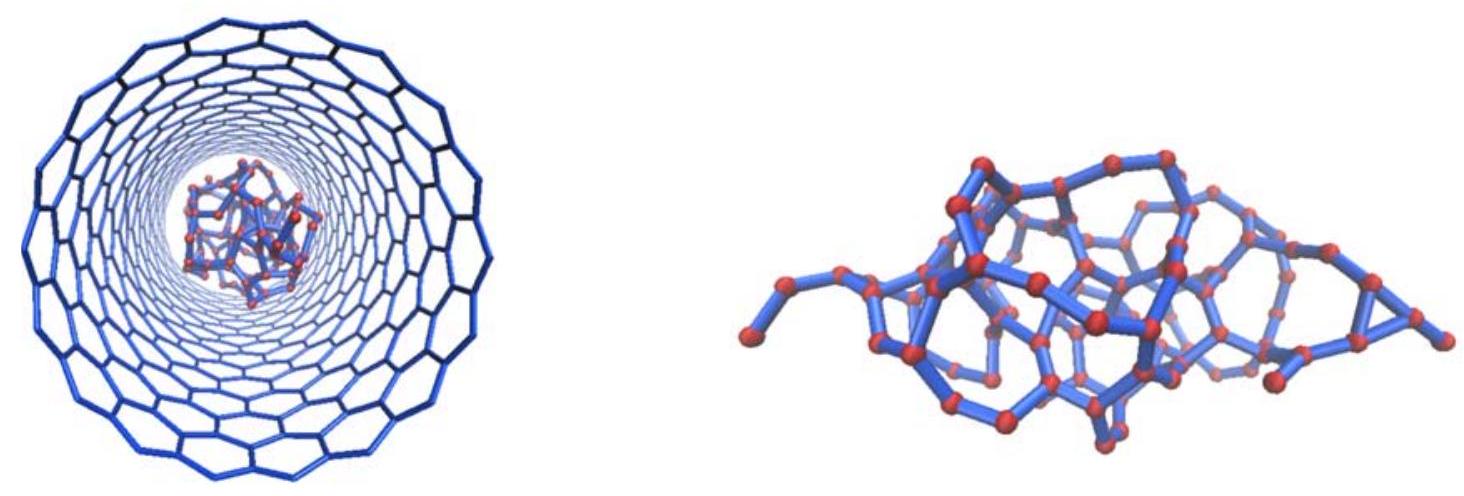}
    \caption{Final Result at $4000 \mathrm{~K}$}
    \label{fig6c}[h]
    \end{subfigure}
    
    \caption{The final structure using AIREBO potential after $100 \mathrm{ps}$ annealing at three temperatures of $1200 \mathrm{~K}$ (a), $3000 \mathrm{~K}$ (b) and $4000 \mathrm{~K}$ (c). At $1200 \mathrm{~K}$ much of the geometry contains chain-like structures, and a non-physical triangle is also formed at this temperature. At $3000 \mathrm{~K}$, the number of polygons is increased, and a tube-like structure is formed.At $4000 \mathrm{~K}$, a more compact structure appeared, the atoms are entangled with each other.}
\end{figure}

Figure \ref{fig6a} clearly shows that a non-planar chain-like structure is formed at $1200 \mathrm{~K}$ and a triangle is also visible. It is striking that the chains are preferably located close to the inner wall of the nanotube. As in ReaxFF, this is due to the presence of long-range interaction in AIREBO. Annealing at $3000 \mathrm{~K}$ results in a larger number of polygons, forming a tube-like structure at this temperature, see Figures \ref{fig6b}. When the temperature is increased to $4000 \mathrm{~K}$, a more compact structure with many defects is formed, see Figures \ref{fig6c}. At higher temperatures, short-term interactions seem to become more important than long-range ones.

The change of various geometrical features during the simulation can be followed in Figure 7. The lower left panel of Figure 7, shows how the average number of bonds per carbon atom increases and converges during the simulation at the three different temperatures. The converged value is about $2.2$ at $1200 \mathrm{~K}$, whereas it is about $2.5$ at higher temperatures. Furthermore, at higher temperatures the fluctuations are larger.

\begin{figure}[ht!]
    \centering
    \includegraphics[width=\linewidth]{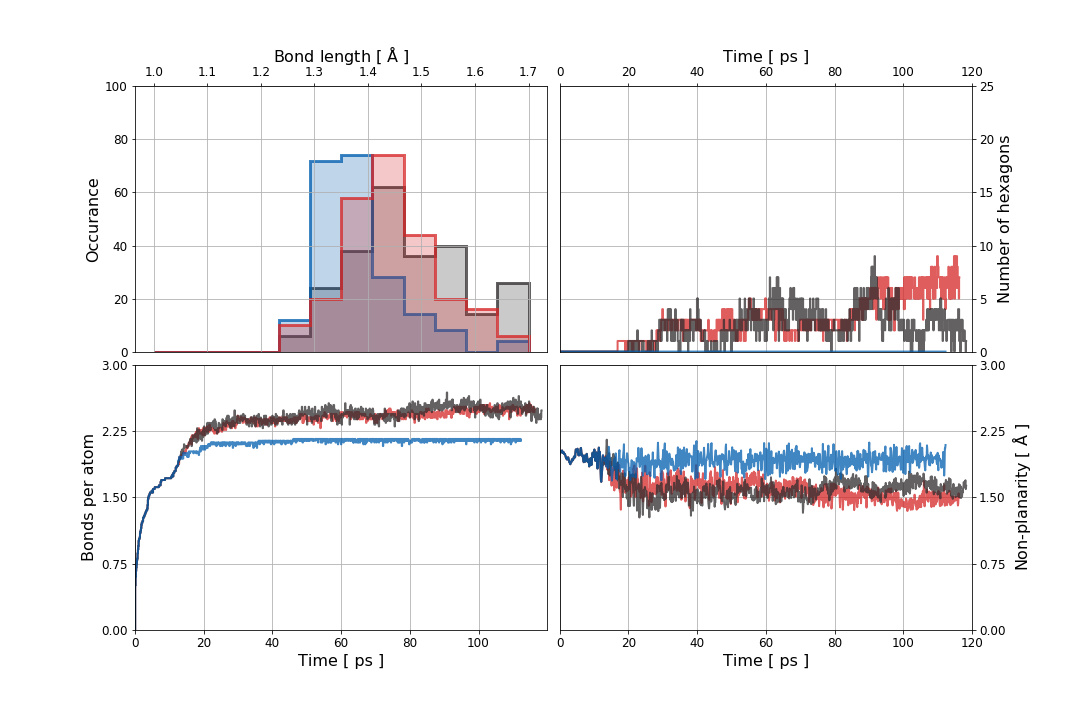}
    \caption{Increase of the average number of bonds per carbon atom during the MD simulation (lower left). The occurrence of $C-C$ bond lengths at the end of the MD simulation (top left). The number of hexagons during the MD simulation (top right). The average distance (in Å) from best planar fit during the MD simulation (lower right). $T_{\text {anneal }}$ is $1200 \mathrm{~K}$ (blue), $3000 \mathrm{~K}$ (red) and $4000 \mathrm{~K}$ (black). All using AIREBO potential.}
    \label{fig7}
\end{figure}

The histograms at the top left panel of Figures \ref{fig7} illustrate the distribution of the $C-C$ distances (bond lengths) at the end of the molecular dynamics simulation for the three different temperatures. The distribution widens as the temperature increases, and the position of the center slides up slightly from $1.35$ {\AA} to $1.45$ \AA.
This is consistent with the fact that the number of chain-like pieces with shorter bond lengths decreases with increasing temperature.

The number of hexagons is low at any temperature. This is shown in top right panel of Figures \ref{fig7} for the three temperatures. At $1200 \mathrm{~K}$, no hexagons are formed at all, while at $3000 \mathrm{~K}$ and $4000 \mathrm{~K}$ there are a few hexagons along with a few pentagons, heptagons and octagons. The structure becomes fullerene like at $4000 \mathrm{~K}$, but with many defects.

Carbon atoms are far from being in the same plane at any temperature. The non-planarity converges to $1.9$ {\AA} at $1200 \mathrm{~K}$, decreases to $1.5$ {\AA} at $3000 \mathrm{~K}$ but is still high, and fluctuates between $1.5$ {\AA} and $1.75$ {\AA} at $4000 \mathrm{~K}$, as shown in the lower right panel of Figure \ref{fig7}. In summary, the use of the AIREBO interatomic potential leads to a non-planar structure full of defects chain-like at lower temperatures and cage-like at higher temperatures and thus performs worse than the Tersoff potential in describing the formation of graphene nanoribbons inside SWCNTs.

\subsection{REBO-II potential}
REBO-II is also an extension of the Tersoff potential, but unlike AIREBO, it does not involve long-range interaction \cite{REBO-II}.
The final structure using REBO-II potential after $100 \mathrm{ps}$ annealing at three different temperatures of $1200 \mathrm{~K}, 3000 \mathrm{~K}$ and $4000 \mathrm{~K}$ are shown in Figures 8a, 8b, and 8c, respectively.

\begin{figure}[ht!]
   \begin{subfigure}{\linewidth}
     \centering
    \includegraphics[width=8cm]{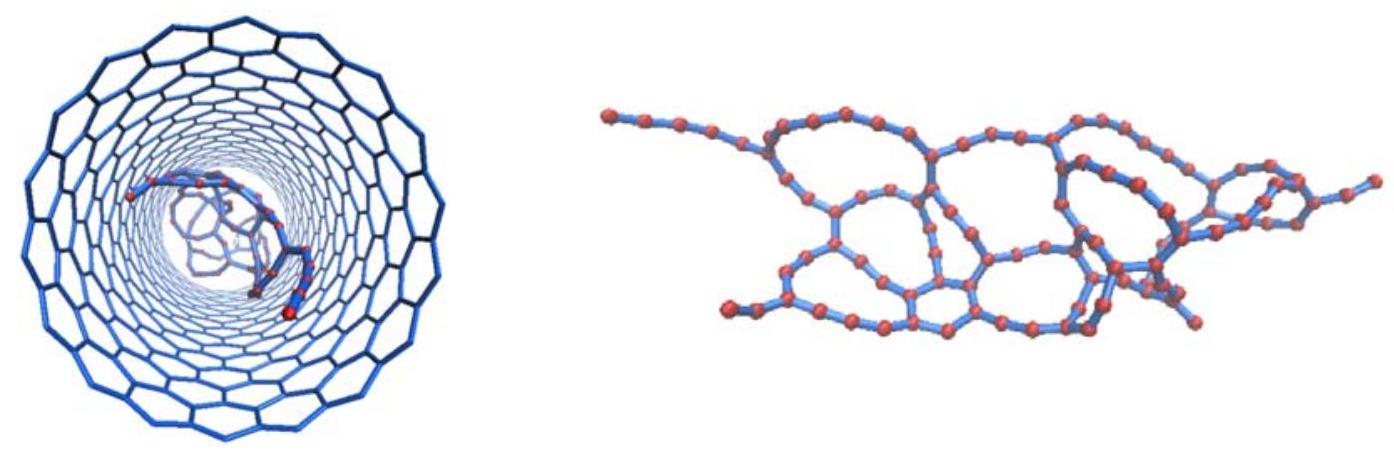}
    \caption{The final result at $1200 \mathrm{~K}$.}
    \label{fig8a}
\end{subfigure}

 \begin{subfigure}{\linewidth}
     \centering
    \includegraphics[width=8cm]{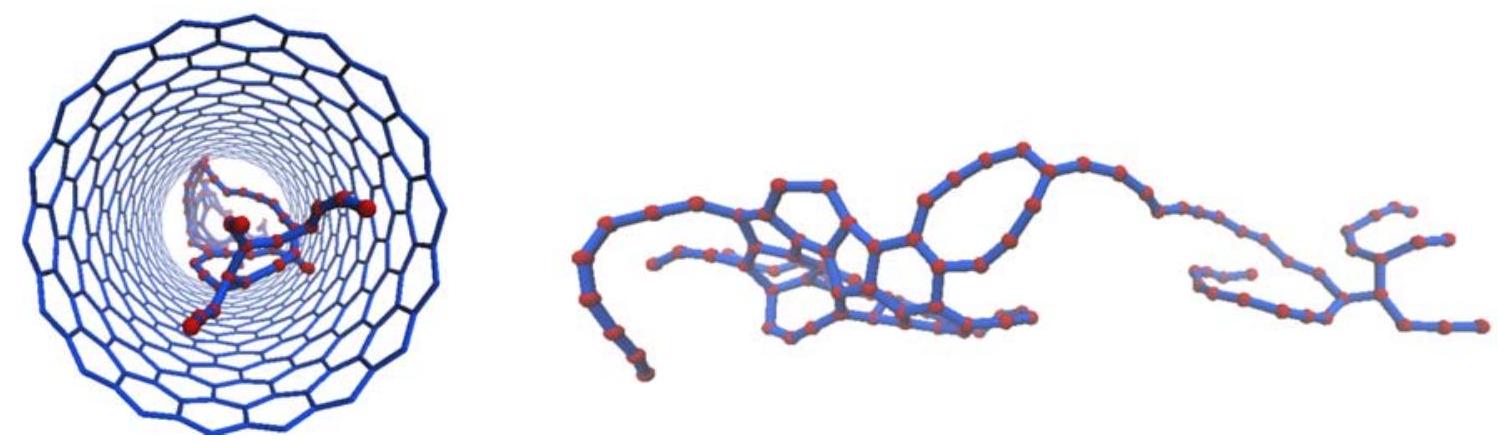}
    \caption{The final result at $3000 \mathrm{~K}$.}
    \label{fig8b}
\end{subfigure}

 \begin{subfigure}{\linewidth}
     \centering
    \includegraphics[width=8cm]{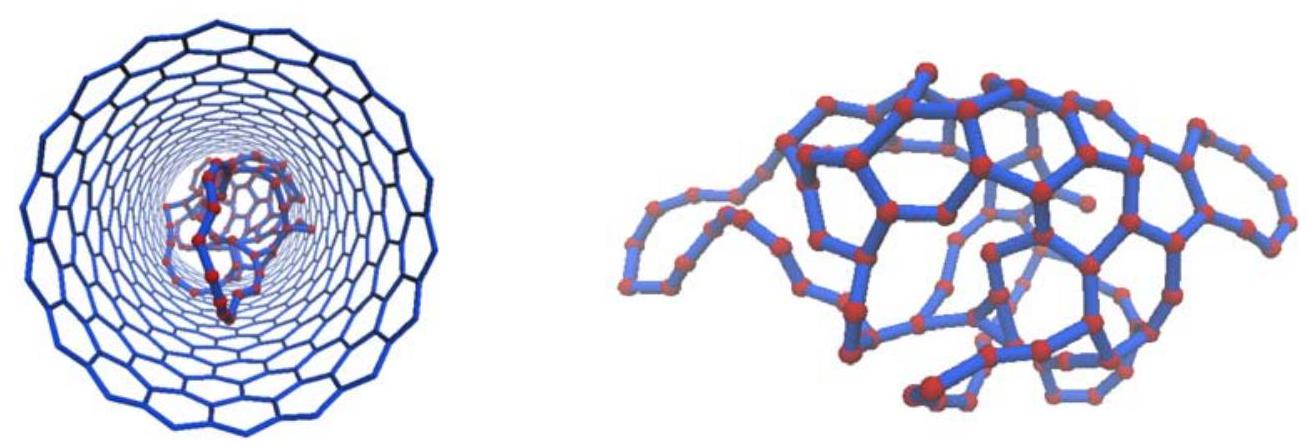}
    \caption{The final result at $4000 \mathrm{~K}$.}
    \label{fig8c}
\end{subfigure}
\caption{The final structure using REBO-II potential after $100 \mathrm{ps}$ annealing at three temperatures of $1200 \mathrm{~K}$ (a), $3000 \mathrm{~K}$ (b), and $4000 \mathrm{~K}$ (c). At $1200 \mathrm{~K}$, only one hexagon is formed from the free carbon atoms in the tube and several chain structures and large polygons appear. At $3000 \mathrm{~K}$ , the structure is mainly chain-like with several polygons. At $4000 \mathrm{~K}$ a cage-like structure appears at the end of the simulations.}
\end{figure}

\begin{figure}[ht!]
    \centering
    \includegraphics[width=\linewidth]{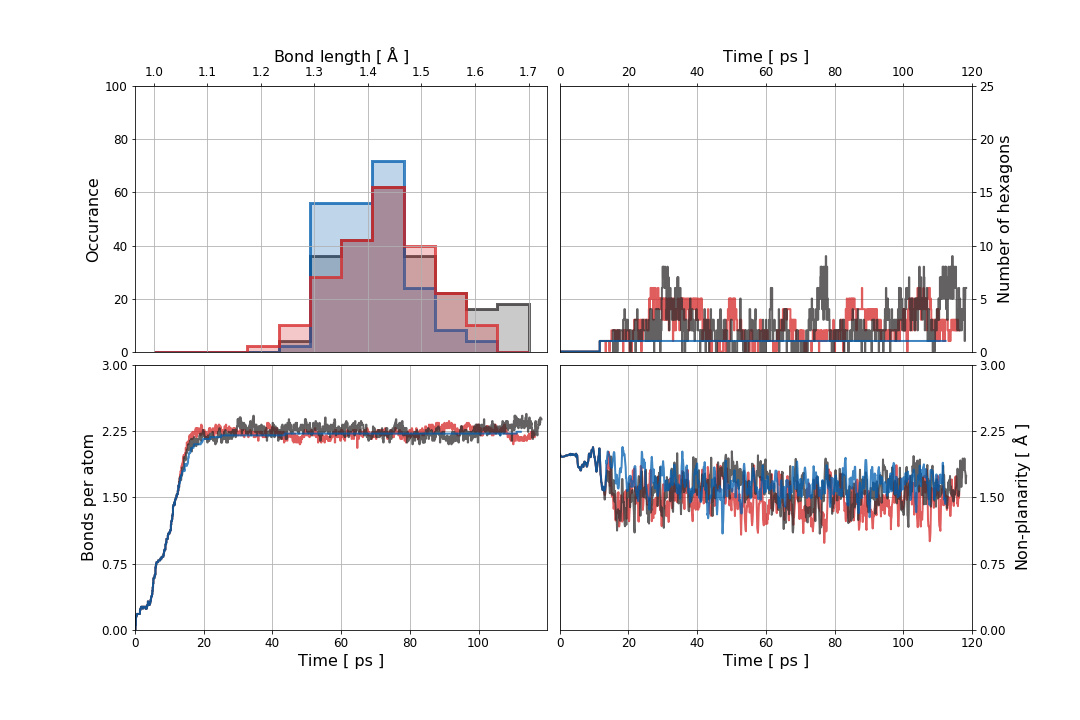}
    \caption{Increase of the average number of bonds per carbon atom during the MD simulation (lower left). The occurrence of $C-C$ bond lengths at the end of the MD simulation (top left). The number of hexagons during the MD simulation (top right). The average distance (in Å) from best planar fit during the MD simulation (lower right). $T_{\text {anneal }}$ is $1200 \mathrm{~K}$ (blue), $3000 \mathrm{~K}$ (red) and $4000 \mathrm{~K}$ (black). All using REBO-II potential.}
    \label{fig9}
\end{figure}

As shown in Figure \ref{fig8a}, at $1200 \mathrm{~K}$, chain-like structures are formed and some larger polygons appear. The chains do not prefer to be close to the inner wall of the nanotubes. This is because, unlike AIREBO, there is no long-range interaction in REBO-II. The structure at $3000 \mathrm{~K}$ is basically similar to that of $1200 \mathrm{~K}$. In contrast, at $4000 \mathrm{~K}$ a cage-like structure is formed.

The change of various geometrical features during the simulation can be followed in Figure \ref{fig9}. The lower left panel of Figure \ref{fig9} shows how the average number of bonds per carbon atom increases and converges during the simulation to about $2.25$ at all three temperatures.

The histograms at the top left panel of Figure 9 illustrate the distribution of the $C-C$ distances at the end of the molecular dynamics simulation for the three different temperatures. The distribution widens as the temperature increases, and the position of the center slides up slightly from $1.4$ {\AA} to $1.5$ \AA.

The number of hexagons is low at any temperature. This is shown in top right panel of Figure 9 for the three temperatures. At $1200 \mathrm{~K}$, only one hexagon is formed, while at $3000 \mathrm{~K}$ and $4000 \mathrm{~K}$ there are a few hexagons along with a few pentagons, heptagons and octagons.

Carbon atoms are far from being in the same plane at any temperature. The non-planarity fluctuates around $1.5$ {\AA} at all three temperatures, as shown in the lower right panel of Figure 9.
Similar to Airebo and Tersoff potentials, at 4000 K, we observe a curved structure. 

In summary, the REBO-II interatomic potential leads to a non-planar structure with a small number of hexagons at all three temperatures, and thus performs worse than the Tersoff potential in describing the formation of graphene nanoribbons inside SWCNTs. 

\subsection{LCBOP potential}

\begin{figure}[ht!]
   \begin{subfigure}{\linewidth}
     \centering
    \includegraphics[width=8cm]{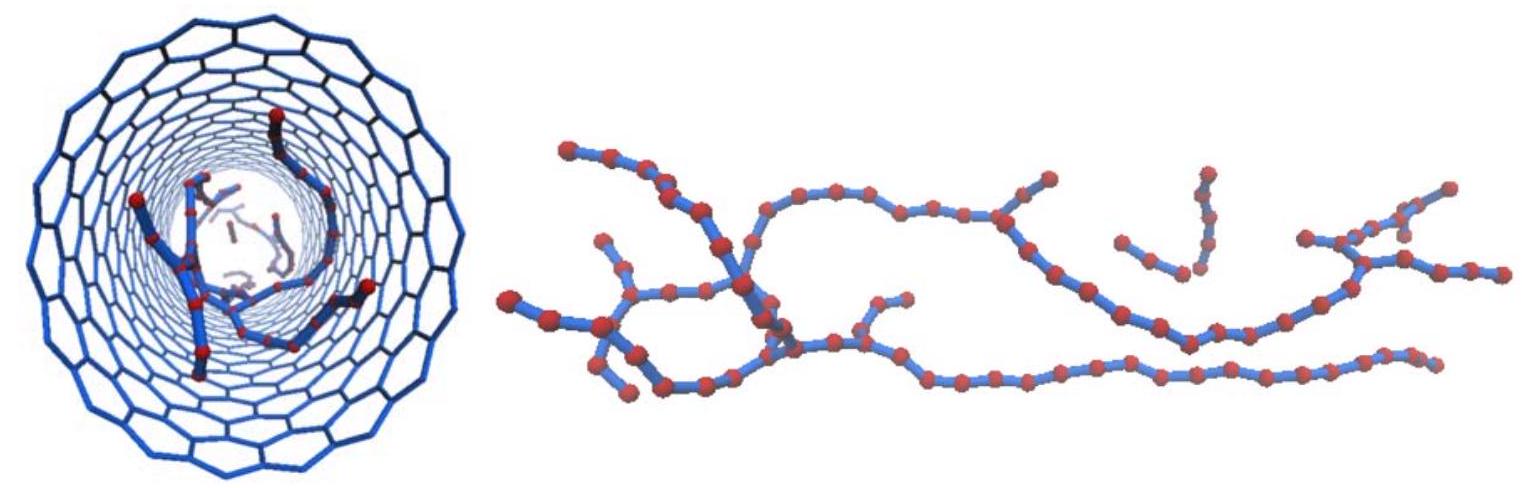}
    \caption{The final result at $1200 \mathrm{~K}$.}
    \label{fig10a}
\end{subfigure}

  \begin{subfigure}{\linewidth}
     \centering
    \includegraphics[width=8cm]{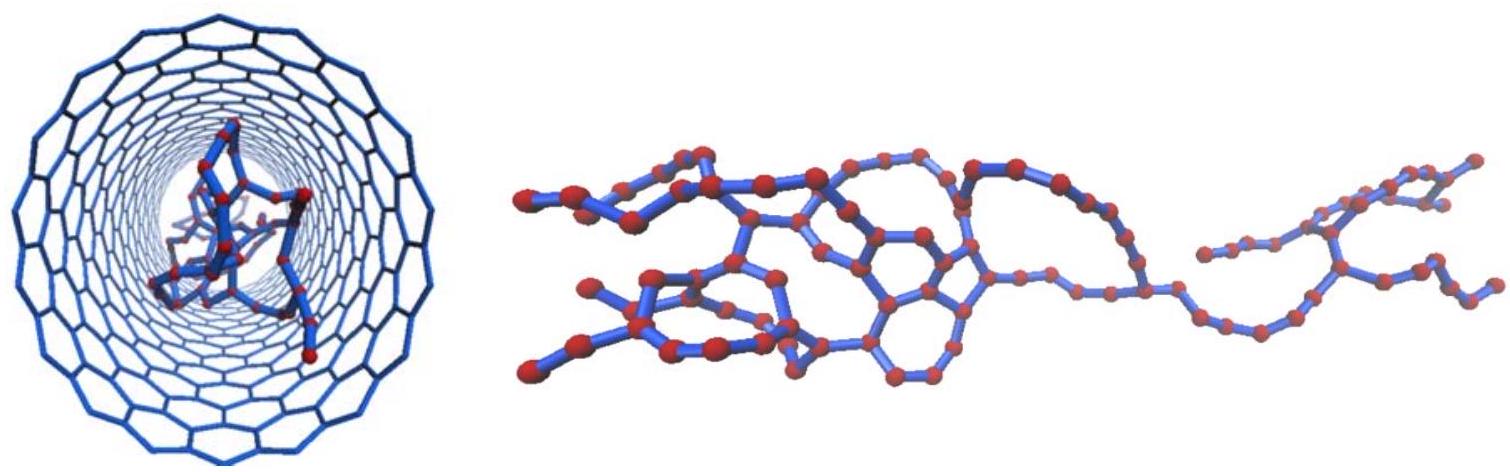}
    \caption{The final result at $3000 \mathrm{~K}$.}
    \label{fig10b}
    \end{subfigure}
    
 \begin{subfigure}{\linewidth}
     \centering
    \includegraphics[width=8cm]{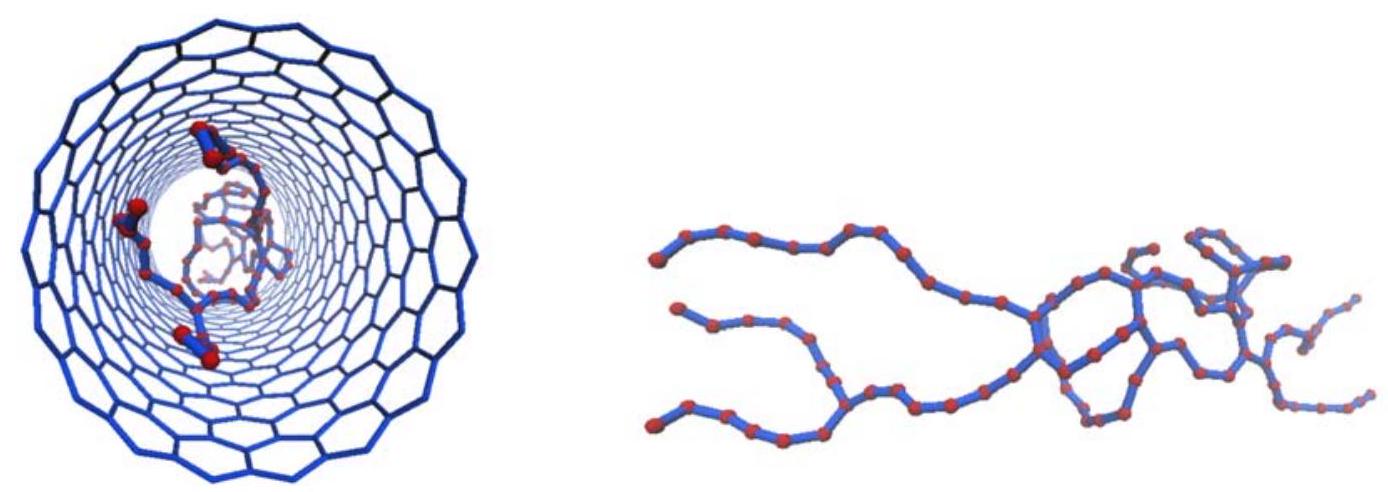}
    \caption{The final result at $4000 \mathrm{~K}$.}
    \label{fig10c}
    \end{subfigure}
    
    \caption{The final structure using LCBOP potential after $100 \mathrm{ps}$ annealing at three temperatures of $1200 \mathrm{~K}$ (a), $3000 \mathrm{~K}$ (b), and $4000 \mathrm{~K}$ (c). At $1200 \mathrm{~K}$, no polygons appear, only chain structures of different lengths are formed. At $3000 \mathrm{~K}$, a few polygons are formed beside chains. At $4000 \mathrm{~K}$, only chains appeared.}
\end{figure}

LCBOP-I is a simultaneous bond-order and long-range extension of REBO \cite{LCBOP-I}.
The final structure using LCBOP potential after 100 ps annealing at three different temperatures of 1200 K, 3000 K and 4000 K are shown in Figures \ref{fig10a}, \ref{fig10b} and \ref{fig10c}, respectively.

As shown in Figure \ref{fig10a}, at $1200 \mathrm{~K}$, a non-planar chain-like structure without any polygon is appeared. The chains are preferably positioned close to the inner wall of the nanotube. This is a consequence of the long-range interaction present in LCBOP. At $3000 \mathrm{~K}$ some polygons are formed while at $4000 \mathrm{~K}$ again only chains can be observed.

\begin{figure}[ht!]
    \centering
    \includegraphics[width=\linewidth]{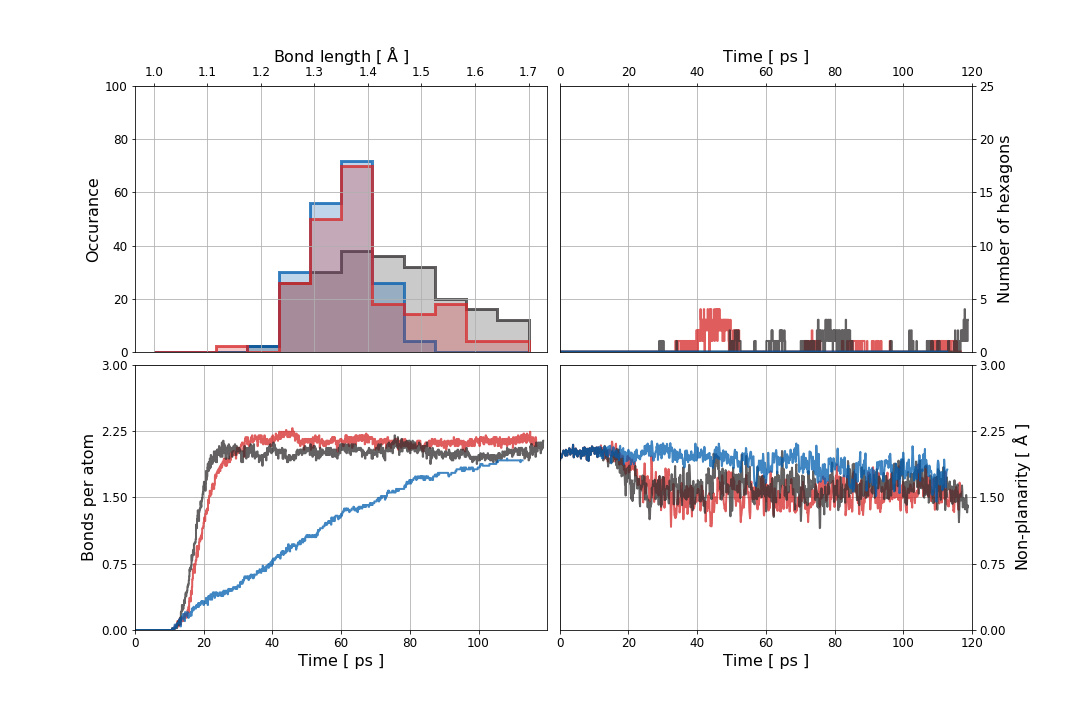}
    \caption{Increase of the average number of bonds per carbon atom during the MD simulation (lower left). The occurrence of $C-C$ bond lengths at the end of the MD simulation (top left). The number of hexagons during the MD simulation (top right). The average distance (in Å) from best planar fit during the MD simulation (lower right). $T_{\text {anneal }}$ is $1200 \mathrm{~K}$ (blue), $3000 \mathrm{~K}$ (red) and $4000 \mathrm{~K}$ (black). All using LCBOP potential.}
    \label{fig11}
\end{figure}

The change of various geometrical features during the simulation can be followed in Figure \ref{fig11}. The lower left panel of Figure \ref{fig11} shows how the average number of bonds per carbon atom increases during the simulation. At $3000 \mathrm{~K}$ and $4000 \mathrm{~K}$ convergence is fast, while at $1200 \mathrm{~K}$ convergence is very slow. The converged value is around 2 at all three temperatures.

The histograms at the top left panel of Figure \ref{fig11} illustrate the distribution of the $C-C$ distances at the end of the molecular dynamics simulation for the three different temperatures. The distribution widens as the temperature increases, and the position of the center slides up slightly from $1.3$ {\AA} to $1.4$ \AA.

The number of polygons and especially hexagons is low at any temperature. This is shown in top right panel of Figure \ref{fig11} for the three temperatures. At $1200 \mathrm{~K}$, no hexagon is formed, while at $3000 \mathrm{~K}$ and 4000 $\mathrm{K}$ there are a few hexagons along with a few pentagons, heptagons and octagons.

Carbon atoms are far from being in the same plane at any temperature. The non-planarity fluctuates around $1.75$ {\AA} at $1200 \mathrm{~K}$ and around $1.5$ {\AA} at $3000 \mathrm{~K}$ and $4000 \mathrm{~K}$, as shown in the lower right panel of Figure \ref{fig11}.

In summary, using the LCBOP potential, a non-planar chain-like structure is formed at $1200 \mathrm{~K}$, while some polygons are formed at an annealing temperature of $3000 \mathrm{~K}$. However, the number of hexagons is small and the atoms are far from being in a plane, and thus LCBOP performs worse than the Tersoff potential in describing the formation of graphene nanoribbons inside SWCNTs.
It is concluded that of the five interatomic potentials investigated, the Tersoff potential is the best for describing the formation of carbon nanoribbon-like structures within SWCNTs. When the Tersoff potential is used, the structure will be most planar and the number of hexagons formed will be the largest.

\subsection{{Discussion of the physics behind the different behaviors of the potentials}}

{Based on the results presented in subsections A-E, it is concluded that of the five potentials investigated, the Tersoff potential is the best for describing the formation of carbon nanoribbon-like structures within SWCNTs. When using the Tersoff potential, the structure will be the most planar and the number of hexagons will be the largest. In this subsection we discuss what is the main physics behind the different behaviors of the different potentials tested in this study during the simulations of carbon nanoribbon formation in a carbon nanotube.

The most important factor is whether a given potential contains long-range interaction, more precisely van der Waals interaction. As we know, this is the case for three of the five potentials we have studied (ReaxFF, AIREBO, LCBOP), the remaining two (Tersoff and REBO-II) do not contain van der Waals interaction. At the concentration we are considering (corresponding to the experimental conditions), the density of atoms inside the tube is 33 atoms per nanometer, while the density of atoms in the wall of the (18,0) nanotube is much (five times) higher (171 atoms per nanometer). The long-range interaction between the carbon atoms inside the tube and the carbon atoms in the tube wall therefore strongly influences the chemical reaction inside the tube. The internal carbon atoms are easily trapped against the nanotube wall, at least at the "lower" 1200 K temperature, as it is seen in Figs. 
{\ref{fig2a}}, {\ref{fig6a}} and {\ref{fig10a}}  
for ReaxFF, AIREBO and LCBOP, respectively. The wall of the nanotube pulls the inner carbon atoms towards itself, almost sweeping them out of the middle of the tube. This prevents the formation of nanoribbons with ReaxFF, AIREBO and LCBOP. At the end of the simulation, the average distance of the trapped carbon atoms from the nanotube wall essentially corresponds to the van der Waals distance (3.4 {\AA} for AIREBO and LCBOP, and 3.3 {\AA} for ReaxFF). As the temperature is raised, the internal carbon atoms can escape from the trap near the wall. At higher temperatures, however, the probability of stable structure formation decreases, as can be clearly seen in all the structural figures.

In the two cases without long-range interactions (Tersoff, REBO-II), the internal carbon atoms are not trapped near the tube wall. However, there is a difference between the two potential functions describing the short-range interactions. Our results show that Tersoff favors bond angles around 120 degrees, while REBO-II does not. This, however, cannot be deduced from the formula describing the interaction simply "at a glance", due to the many-body nature of the potential between atoms. Although the Tersoff formula is a sum of pair potentials, the coefficient of the attractive term depends in a complicated way on the local environment. The fact that the Tersoff potential favors bond angles around 120 degrees is confirmed by our earlier, unpublished observations in addition to the present results. When a carbon nanotube was stretched, we found that using Tersoff potential, the radius of the tube increased(!), such that the angles of the hexagons remained at 120 degrees {\cite{ILaszlo-unpublished}}. The same was no longer true with Brenner potential (REBO), where the radius decreased and the hexagons were distorted when the tube was stretched.}

\subsection{Temperature window for Tersoff potential}

\begin{figure}[ht!]
    \centering
    \includegraphics[width=\linewidth]{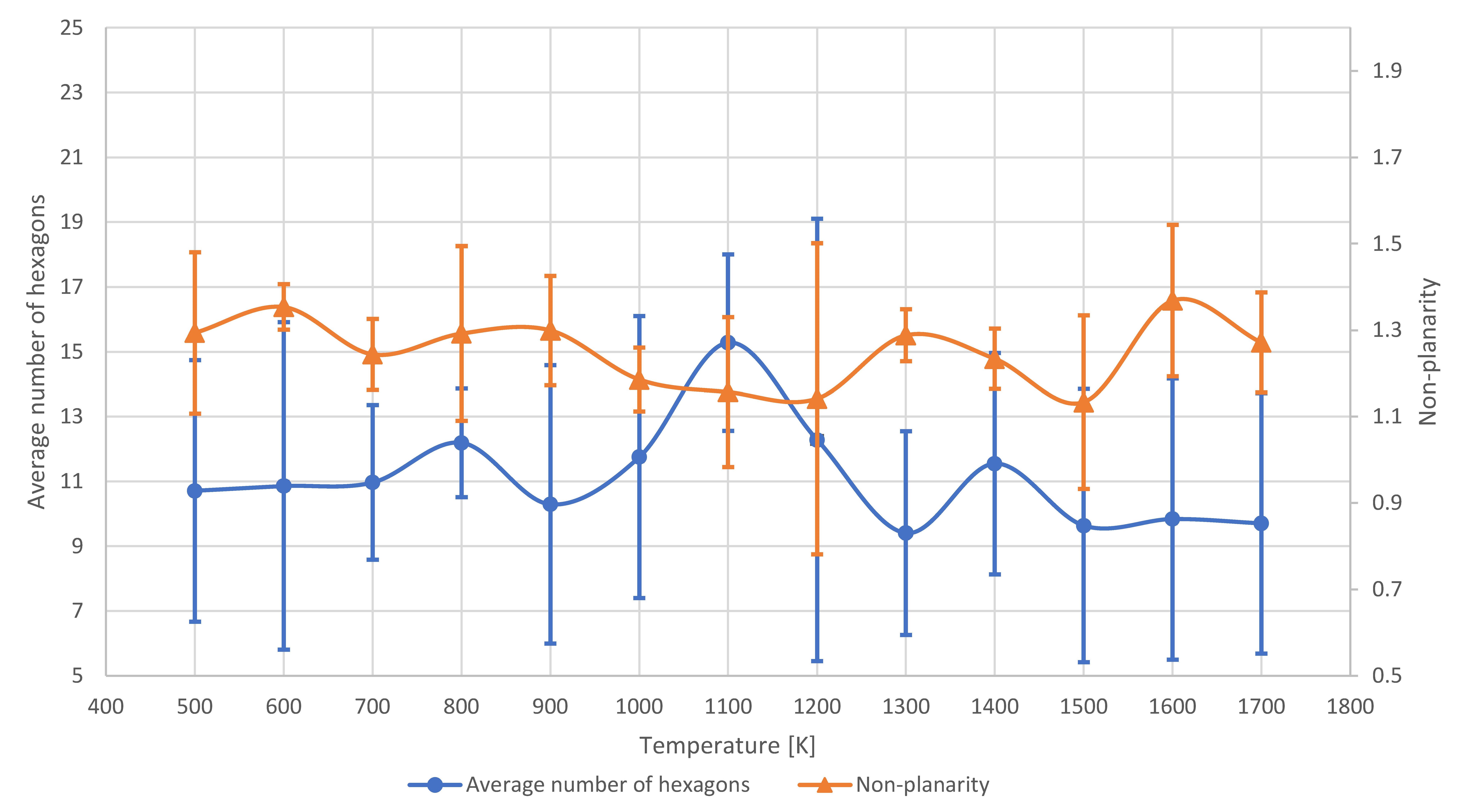}
    \caption{Number of hexagons (circles, blue line, left axis) and non-planarity (triangles, orange line, right axis), averaged over four random initial velocity distribution. The error bars indicate the standard deviations.}
    \label{fig12}
\end{figure}

In agreement with the experiments, our calculations clearly show the important role of the annealing temperature. According to the experiments, the optimal annealing temperature for the formation of carbon nanoribbons from molecules loaded into the SWCNT is around $1100 \mathrm{~K}$. We have investigated in detail the temperature dependence for the best performing Tersoff potential. We carried out the calculations between $500 \mathrm{~K}$ and $1700 \mathrm{~K}$ in steps of $100 \mathrm{~K}$. To reach a more reliable conclusion, we repeated the simulations with four different random seed numbers, giving the carbon atoms four different random initial velocity distribution. The simulation process itself was identical to the one described earlier. Since the convergence in the case of Tersoff potential is already reached after $20 \mathrm{ps}$ (see Figure \ref{fig4a}), the annealing time in the new simulations was set to $40 \mathrm{ps}$ instead of $100 \mathrm{ps}$. In each case, we calculated the evolution of both the number of hexagons and the planarity (as was defined earlier) during the simulation. Figure \ref{fig12} shows the averaged values 
of mentioned parameters for four simulations with different random initial velocity distributions.

As can be seen in the Figure \ref{fig12}, there is a temperature window between about $1000 \mathrm{~K}$ and $1200 \mathrm{~K}$ where the nanoribbons are formed with the least defects. In this interval, both the number of hexagons and the planarity of the system are at a maximum. Surprisingly, this is close to the experimentally observed optimum temperature range \cite{GNR_12,cadena2022molecular}, although this is not actually expected since the simulation was done with carbon atoms only, whereas in the real experiment, annealing occurs for nanotubes filled with organic/organometallic molecules.

\section{Conclusion}

Molecular dynamics simulations were carried out on carbon atoms confined in a carbon nanotube with a diameter of $1.4 \mathrm{~nm}$. By comparing five different interatomic potentials (ReaxFF, Tersoff, AIREBO, REBO-II, LCBOP), we investigated when a nanoribbon-like structure can form from a random initial geometry. From this comparison, we conclude that the Tersoff potential provides by far the best performance in describing the formation of carbon nanoribbons inside SWCNTs. When using the Tersoff potential, the structure will be closest to being planar and the number of hexagons generated will be the largest.
This is {basically} because, as our experience has shown, the Tersoff potential is more prone to hold bond angles around 120° than other potentials, and is therefore more likely to form hexagons {{\cite{ILaszlo-unpublished}}}.

Our experience shows that the other four potentials do not lead to a planar structure. The reasons for this may be complex: the function describing the potential itself, in particular the consideration of long-range interaction, but also whether the simulation is performed in one, two or three dimensions.  The poor performance of ReaxFF may be particularly surprising, since in three dimensions the ReaxFF potential describes graphite formation very well.
However, our results point out that there is a significant difference between three-dimensional simulations and quasi-one-dimensional simulations in confined space, even when using the same interatomic potential. To say the least, there is no room for parallel graphene ribbons inside the nanotube.
Parametrizing the 2D and 1D versions of ReaxFF can be an interesting task to allow it to perform well on systems such as molecules inside SWCNTs.

For the Tersoff potential, we found a temperature window of around $1000 \mathrm{~K}-1200 \mathrm{~K}$, where the nanoribbons formed with the fewest defects, i.e. the highest planarity and the most hexagons. 
This is close to the temperature range that experiments show is most suitable for the formation of carbon nanoribbons inside carbon nanotubes \cite{GNR_12,cadena2022molecular}.

{It should be mentioned that in real experiments, carbon nanoribbons inside SWCNTs are formed from hydrocarbon molecules, whereas here we only investigated the interaction of pure carbon atoms. In terms of ribbon formation, hydrocarbon molecules are preferable to a system containing only carbon atoms in two aspects. First, the molecules studied in the experiments (e.g. coronene, ferrocene, ...)  are inherently planar (or composed of planar parts) which favors the formation of nanoribbons. Furthermore, the hydrogens can passivate the dangling bonds at the edges of the ribbons and thus better stabilize the ribbon-like structures. However, the difficulty of hydrocarbon molecular dynamics is that the initial C-H bonds are not easy to break. Such simulations are therefore much more time consuming. Nevertheless, our results can serve as a good starting point for simulations with molecules used in experiments.}

\begin{acknowledgments}
This research was supported by the Ministry of Culture and Innovation and the National Research, Development and Innovation Office within the Quantum Information National Laboratory of Hungary (Grant No. 2022-2.1.1-NL-2022-00004) and by NKFIH Grant No. K-134437. 
S.E. greatly acknowledges the support from Stipendium Hungaricum No. 23921.
We appreciate the useful discussions with Viktor Zólyomi.
\end{acknowledgments}
\section*{Data Availability}
The data that support the findings of this study are available
from the corresponding authors upon reasonable request.
\section*{References}

\bibliography{manuscript}

\begin{thebibliography}{92}%
\makeatletter
\providecommand \@ifxundefined [1]{%
 \@ifx{#1\undefined}
}%
\providecommand \@ifnum [1]{%
 \ifnum #1\expandafter \@firstoftwo
 \else \expandafter \@secondoftwo
 \fi
}%
\providecommand \@ifx [1]{%
 \ifx #1\expandafter \@firstoftwo
 \else \expandafter \@secondoftwo
 \fi
}%
\providecommand \natexlab [1]{#1}%
\providecommand \enquote  [1]{``#1''}%
\providecommand \bibnamefont  [1]{#1}%
\providecommand \bibfnamefont [1]{#1}%
\providecommand \citenamefont [1]{#1}%
\providecommand \href@noop [0]{\@secondoftwo}%
\providecommand \href [0]{\begingroup \@sanitize@url \@href}%
\providecommand \@href[1]{\@@startlink{#1}\@@href}%
\providecommand \@@href[1]{\endgroup#1\@@endlink}%
\providecommand \@sanitize@url [0]{\catcode `\\12\catcode `\$12\catcode
  `\&12\catcode `\#12\catcode `\^12\catcode `\_12\catcode `\%12\relax}%
\providecommand \@@startlink[1]{}%
\providecommand \@@endlink[0]{}%
\providecommand \url  [0]{\begingroup\@sanitize@url \@url }%
\providecommand \@url [1]{\endgroup\@href {#1}{\urlprefix }}%
\providecommand \urlprefix  [0]{URL }%
\providecommand \Eprint [0]{\href }%
\providecommand \doibase [0]{http://dx.doi.org/}%
\providecommand \selectlanguage [0]{\@gobble}%
\providecommand \bibinfo  [0]{\@secondoftwo}%
\providecommand \bibfield  [0]{\@secondoftwo}%
\providecommand \translation [1]{[#1]}%
\providecommand \BibitemOpen [0]{}%
\providecommand \bibitemStop [0]{}%
\providecommand \bibitemNoStop [0]{.\EOS\space}%
\providecommand \EOS [0]{\spacefactor3000\relax}%
\providecommand \BibitemShut  [1]{\csname bibitem#1\endcsname}%
\let\auto@bib@innerbib\@empty
\bibitem [{\citenamefont {Monthioux}, \citenamefont {Flahaut},\ and\
  \citenamefont {Cleuziou}(2006)}]{ref01}%
  \BibitemOpen
  \bibfield  {author} {\bibinfo {author} {\bibfnamefont {M.}~\bibnamefont
  {Monthioux}}, \bibinfo {author} {\bibfnamefont {E.}~\bibnamefont {Flahaut}},
  \ and\ \bibinfo {author} {\bibfnamefont {J.}~\bibnamefont {Cleuziou}},\
  }\bibfield  {title} {\enquote {\bibinfo {title} {Hybrid carbon nanotubes:
  Strategy, progress, and perspectives},}\ }\href@noop {} {\bibfield  {journal}
  {\bibinfo  {journal} {Journal of Materials Research}\ }\textbf {\bibinfo
  {volume} {21}},\ \bibinfo {pages} {2774--2793} (\bibinfo {year}
  {2006})}\BibitemShut {NoStop}%
\bibitem [{\citenamefont {Khlobystov}(2011)}]{ref02}%
  \BibitemOpen
  \bibfield  {author} {\bibinfo {author} {\bibfnamefont {A.}~\bibnamefont
  {Khlobystov}},\ }\bibfield  {title} {\enquote {\bibinfo {title} {Carbon
  nanotubes: from nano test tube to nano-reactor},}\ }\href@noop {} {\bibfield
  {journal} {\bibinfo  {journal} {ACS Nano}\ }\textbf {\bibinfo {volume} {5}},\
  \bibinfo {pages} {9306--9312} (\bibinfo {year} {2011})}\BibitemShut {NoStop}%
\bibitem [{\citenamefont {Miners}, \citenamefont {Rance},\ and\ \citenamefont
  {Khlobystov}(2016)}]{ref03}%
  \BibitemOpen
  \bibfield  {author} {\bibinfo {author} {\bibfnamefont {S.}~\bibnamefont
  {Miners}}, \bibinfo {author} {\bibfnamefont {G.}~\bibnamefont {Rance}}, \
  and\ \bibinfo {author} {\bibfnamefont {A.}~\bibnamefont {Khlobystov}},\
  }\bibfield  {title} {\enquote {\bibinfo {title} {Chemical reactions confined
  within carbon nanotubes},}\ }\href@noop {} {\bibfield  {journal} {\bibinfo
  {journal} {Chemical Society Reviews}\ }\textbf {\bibinfo {volume} {45}},\
  \bibinfo {pages} {4727--4746} (\bibinfo {year} {2016})}\BibitemShut {NoStop}%
\bibitem [{\citenamefont {Poudel}\ and\ \citenamefont {Li}(2018)}]{ref04}%
  \BibitemOpen
  \bibfield  {author} {\bibinfo {author} {\bibfnamefont {Y.}~\bibnamefont
  {Poudel}}\ and\ \bibinfo {author} {\bibfnamefont {W.}~\bibnamefont {Li}},\
  }\bibfield  {title} {\enquote {\bibinfo {title} {Synthesis, properties, and
  applications of carbon nanotubes filled with foreign materials: A review},}\
  }\href@noop {} {\bibfield  {journal} {\bibinfo  {journal} {Materials Today
  Physics}\ }\textbf {\bibinfo {volume} {7}},\ \bibinfo {pages} {7--34}
  (\bibinfo {year} {2018})}\BibitemShut {NoStop}%
\bibitem [{\citenamefont {Cadena}, \citenamefont {Botka},\ and\ \citenamefont
  {Kamaras}(2021)}]{ref05}%
  \BibitemOpen
  \bibfield  {author} {\bibinfo {author} {\bibfnamefont {A.}~\bibnamefont
  {Cadena}}, \bibinfo {author} {\bibfnamefont {B.}~\bibnamefont {Botka}}, \
  and\ \bibinfo {author} {\bibfnamefont {K.}~\bibnamefont {Kamaras}},\
  }\bibfield  {title} {\enquote {\bibinfo {title} {Organic molecules
  encapsulated in single-walled carbon nanotubes},}\ }\href@noop {} {\bibfield
  {journal} {\bibinfo  {journal} {Oxford Open Materials Science}\ }\textbf
  {\bibinfo {volume} {1}},\ \bibinfo {pages} {itab009} (\bibinfo {year}
  {2021})}\BibitemShut {NoStop}%
\bibitem [{\citenamefont {Ajayan}\ \emph {et~al.}(1993)\citenamefont {Ajayan}
  \emph {et~al.}}]{Inorg_1}%
  \BibitemOpen
  \bibfield  {author} {\bibinfo {author} {\bibfnamefont {P.}~\bibnamefont
  {Ajayan}} \emph {et~al.},\ }\bibfield  {title} {\enquote {\bibinfo {title}
  {Capillarity-induced filling of carbon nanotubes},}\ }\href@noop {}
  {\bibfield  {journal} {\bibinfo  {journal} {Nature}\ }\textbf {\bibinfo
  {volume} {361}},\ \bibinfo {pages} {333--334} (\bibinfo {year}
  {1993})}\BibitemShut {NoStop}%
\bibitem [{\citenamefont {Bao}\ \emph {et~al.}(2002)\citenamefont {Bao},
  \citenamefont {Tie}, \citenamefont {Xu}, \citenamefont {Suo}, \citenamefont
  {Zhou},\ and\ \citenamefont {Hong}}]{Inorg_2}%
  \BibitemOpen
  \bibfield  {author} {\bibinfo {author} {\bibfnamefont {J.}~\bibnamefont
  {Bao}}, \bibinfo {author} {\bibfnamefont {C.}~\bibnamefont {Tie}}, \bibinfo
  {author} {\bibfnamefont {Z.}~\bibnamefont {Xu}}, \bibinfo {author}
  {\bibfnamefont {Z.}~\bibnamefont {Suo}}, \bibinfo {author} {\bibfnamefont
  {Q.}~\bibnamefont {Zhou}}, \ and\ \bibinfo {author} {\bibfnamefont
  {J.}~\bibnamefont {Hong}},\ }\bibfield  {title} {\enquote {\bibinfo {title}
  {A facile method for creating an array of metal-filled carbon nanotubes},}\
  }\href@noop {} {\bibfield  {journal} {\bibinfo  {journal} {Advanced
  Materials}\ }\textbf {\bibinfo {volume} {14}},\ \bibinfo {pages} {1483--1486}
  (\bibinfo {year} {2002})}\BibitemShut {NoStop}%
\bibitem [{\citenamefont {Sloan}\ \emph {et~al.}(2002)\citenamefont {Sloan},
  \citenamefont {Kirkland}, \citenamefont {Hutchison},\ and\ \citenamefont
  {Green}}]{Inorg_3}%
  \BibitemOpen
  \bibfield  {author} {\bibinfo {author} {\bibfnamefont {J.}~\bibnamefont
  {Sloan}}, \bibinfo {author} {\bibfnamefont {A.}~\bibnamefont {Kirkland}},
  \bibinfo {author} {\bibfnamefont {J.}~\bibnamefont {Hutchison}}, \ and\
  \bibinfo {author} {\bibfnamefont {M.}~\bibnamefont {Green}},\ }\bibfield
  {title} {\enquote {\bibinfo {title} {Structural characterization of
  atomically regulated nanocrystals formed within single-walled carbon
  nanotubes using electron microscopy},}\ }\href@noop {} {\bibfield  {journal}
  {\bibinfo  {journal} {Accounts of Chemical Research}\ }\textbf {\bibinfo
  {volume} {35}},\ \bibinfo {pages} {1054--1062} (\bibinfo {year}
  {2002})}\BibitemShut {NoStop}%
\bibitem [{\citenamefont {Guan}\ \emph {et~al.}(2007)\citenamefont {Guan},
  \citenamefont {Suenaga}, \citenamefont {Shi}, \citenamefont {Gu},\ and\
  \citenamefont {Iijima}}]{Inorg_4}%
  \BibitemOpen
  \bibfield  {author} {\bibinfo {author} {\bibfnamefont {L.}~\bibnamefont
  {Guan}}, \bibinfo {author} {\bibfnamefont {K.}~\bibnamefont {Suenaga}},
  \bibinfo {author} {\bibfnamefont {Z.}~\bibnamefont {Shi}}, \bibinfo {author}
  {\bibfnamefont {Z.}~\bibnamefont {Gu}}, \ and\ \bibinfo {author}
  {\bibfnamefont {S.}~\bibnamefont {Iijima}},\ }\bibfield  {title} {\enquote
  {\bibinfo {title} {Polymorphic structures of iodine and their phase
  transition in confined nanospace},}\ }\href@noop {} {\bibfield  {journal}
  {\bibinfo  {journal} {Nano Letters}\ }\textbf {\bibinfo {volume} {7}},\
  \bibinfo {pages} {1532--1535} (\bibinfo {year} {2007})}\BibitemShut {NoStop}%
\bibitem [{\citenamefont {Rodrigues}\ \emph {et~al.}(2008)\citenamefont
  {Rodrigues}, \citenamefont {Saraiva}, \citenamefont {Nascimento},
  \citenamefont {Barros}, \citenamefont {Mendes~Filho}, \citenamefont {Kim},
  \citenamefont {Muramatsu}, \citenamefont {Endo}, \citenamefont {Terrones},
  \citenamefont {Dresselhaus} \emph {et~al.}}]{Inorg_5}%
  \BibitemOpen
  \bibfield  {author} {\bibinfo {author} {\bibfnamefont {O.}~\bibnamefont
  {Rodrigues}}, \bibinfo {author} {\bibfnamefont {G.}~\bibnamefont {Saraiva}},
  \bibinfo {author} {\bibfnamefont {R.}~\bibnamefont {Nascimento}}, \bibinfo
  {author} {\bibfnamefont {E.}~\bibnamefont {Barros}}, \bibinfo {author}
  {\bibfnamefont {J.}~\bibnamefont {Mendes~Filho}}, \bibinfo {author}
  {\bibfnamefont {Y.}~\bibnamefont {Kim}}, \bibinfo {author} {\bibfnamefont
  {H.}~\bibnamefont {Muramatsu}}, \bibinfo {author} {\bibfnamefont
  {M.}~\bibnamefont {Endo}}, \bibinfo {author} {\bibfnamefont {M.}~\bibnamefont
  {Terrones}}, \bibinfo {author} {\bibfnamefont {M.}~\bibnamefont
  {Dresselhaus}},  \emph {et~al.},\ }\bibfield  {title} {\enquote {\bibinfo
  {title} {Synthesis and characterization of selenium- carbon nanocables},}\
  }\href@noop {} {\bibfield  {journal} {\bibinfo  {journal} {Nano Letters}\
  }\textbf {\bibinfo {volume} {8}},\ \bibinfo {pages} {3651--3655} (\bibinfo
  {year} {2008})}\BibitemShut {NoStop}%
\bibitem [{\citenamefont {Kitaura}\ \emph {et~al.}(2009)\citenamefont
  {Kitaura}, \citenamefont {Nakanishi}, \citenamefont {Saito}, \citenamefont
  {Yoshikawa}, \citenamefont {Awaga},\ and\ \citenamefont
  {Shinohara}}]{Inorg_6}%
  \BibitemOpen
  \bibfield  {author} {\bibinfo {author} {\bibfnamefont {R.}~\bibnamefont
  {Kitaura}}, \bibinfo {author} {\bibfnamefont {R.}~\bibnamefont {Nakanishi}},
  \bibinfo {author} {\bibfnamefont {T.}~\bibnamefont {Saito}}, \bibinfo
  {author} {\bibfnamefont {H.}~\bibnamefont {Yoshikawa}}, \bibinfo {author}
  {\bibfnamefont {K.}~\bibnamefont {Awaga}}, \ and\ \bibinfo {author}
  {\bibfnamefont {H.}~\bibnamefont {Shinohara}},\ }\bibfield  {title} {\enquote
  {\bibinfo {title} {High-yield synthesis of ultrathin metal nanowires in
  carbon nanotubes},}\ }\href@noop {} {\bibfield  {journal} {\bibinfo
  {journal} {Angewandte Chemie International Edition}\ }\textbf {\bibinfo
  {volume} {48}},\ \bibinfo {pages} {8298--8302} (\bibinfo {year}
  {2009})}\BibitemShut {NoStop}%
\bibitem [{\citenamefont {Shiozawa}\ \emph {et~al.}(2015)\citenamefont
  {Shiozawa}, \citenamefont {Briones-Leon}, \citenamefont {Domanov},
  \citenamefont {Zechner}, \citenamefont {Sato}, \citenamefont {Suenaga},
  \citenamefont {Saito}, \citenamefont {Eisterer}, \citenamefont {Weschke},
  \citenamefont {Lang} \emph {et~al.}}]{Inorg_7}%
  \BibitemOpen
  \bibfield  {author} {\bibinfo {author} {\bibfnamefont {H.}~\bibnamefont
  {Shiozawa}}, \bibinfo {author} {\bibfnamefont {A.}~\bibnamefont
  {Briones-Leon}}, \bibinfo {author} {\bibfnamefont {O.}~\bibnamefont
  {Domanov}}, \bibinfo {author} {\bibfnamefont {G.}~\bibnamefont {Zechner}},
  \bibinfo {author} {\bibfnamefont {Y.}~\bibnamefont {Sato}}, \bibinfo {author}
  {\bibfnamefont {K.}~\bibnamefont {Suenaga}}, \bibinfo {author} {\bibfnamefont
  {T.}~\bibnamefont {Saito}}, \bibinfo {author} {\bibfnamefont
  {M.}~\bibnamefont {Eisterer}}, \bibinfo {author} {\bibfnamefont
  {E.}~\bibnamefont {Weschke}}, \bibinfo {author} {\bibfnamefont
  {W.}~\bibnamefont {Lang}},  \emph {et~al.},\ }\bibfield  {title} {\enquote
  {\bibinfo {title} {Nickel clusters embedded in carbon nanotubes as high
  performance magnets},}\ }\href@noop {} {\bibfield  {journal} {\bibinfo
  {journal} {Scientific Reports}\ }\textbf {\bibinfo {volume} {5}},\ \bibinfo
  {pages} {1--9} (\bibinfo {year} {2015})}\BibitemShut {NoStop}%
\bibitem [{\citenamefont {Pham}\ \emph {et~al.}(2018)\citenamefont {Pham},
  \citenamefont {Oh}, \citenamefont {Stetz}, \citenamefont {Onishi},
  \citenamefont {Kisielowski}, \citenamefont {Cohen},\ and\ \citenamefont
  {Zettl}}]{Inorg_8}%
  \BibitemOpen
  \bibfield  {author} {\bibinfo {author} {\bibfnamefont {T.}~\bibnamefont
  {Pham}}, \bibinfo {author} {\bibfnamefont {S.}~\bibnamefont {Oh}}, \bibinfo
  {author} {\bibfnamefont {P.}~\bibnamefont {Stetz}}, \bibinfo {author}
  {\bibfnamefont {S.}~\bibnamefont {Onishi}}, \bibinfo {author} {\bibfnamefont
  {C.}~\bibnamefont {Kisielowski}}, \bibinfo {author} {\bibfnamefont
  {M.}~\bibnamefont {Cohen}}, \ and\ \bibinfo {author} {\bibfnamefont
  {A.}~\bibnamefont {Zettl}},\ }\bibfield  {title} {\enquote {\bibinfo {title}
  {Torsional instability in the single-chain limit of a transition metal
  trichalcogenide},}\ }\href@noop {} {\bibfield  {journal} {\bibinfo  {journal}
  {Science}\ }\textbf {\bibinfo {volume} {361}},\ \bibinfo {pages} {263--266}
  (\bibinfo {year} {2018})}\BibitemShut {NoStop}%
\bibitem [{\citenamefont {Smith}, \citenamefont {Monthioux},\ and\
  \citenamefont {Luzzi}(1998)}]{Peapod_1}%
  \BibitemOpen
  \bibfield  {author} {\bibinfo {author} {\bibfnamefont {B.}~\bibnamefont
  {Smith}}, \bibinfo {author} {\bibfnamefont {M.}~\bibnamefont {Monthioux}}, \
  and\ \bibinfo {author} {\bibfnamefont {D.}~\bibnamefont {Luzzi}},\ }\bibfield
   {title} {\enquote {\bibinfo {title} {Encapsulated c60 in carbon
  nanotubes},}\ }\href@noop {} {\bibfield  {journal} {\bibinfo  {journal}
  {Nature}\ }\textbf {\bibinfo {volume} {396}},\ \bibinfo {pages} {323--324}
  (\bibinfo {year} {1998})}\BibitemShut {NoStop}%
\bibitem [{\citenamefont {Smith}\ and\ \citenamefont {Luzzi}(2000)}]{Peapod_2}%
  \BibitemOpen
  \bibfield  {author} {\bibinfo {author} {\bibfnamefont {B.}~\bibnamefont
  {Smith}}\ and\ \bibinfo {author} {\bibfnamefont {D.}~\bibnamefont {Luzzi}},\
  }\bibfield  {title} {\enquote {\bibinfo {title} {Formation mechanism of
  fullerene peapods and coaxial tubes: a path to large scale synthesis},}\
  }\href@noop {} {\bibfield  {journal} {\bibinfo  {journal} {Chemical Physics
  Letters}\ }\textbf {\bibinfo {volume} {321}},\ \bibinfo {pages} {169--174}
  (\bibinfo {year} {2000})}\BibitemShut {NoStop}%
\bibitem [{\citenamefont {Bandow}\ \emph {et~al.}(2001)\citenamefont {Bandow},
  \citenamefont {Takizawa}, \citenamefont {Hirahara}, \citenamefont
  {Yudasaka},\ and\ \citenamefont {Iijima}}]{Peapod_3}%
  \BibitemOpen
  \bibfield  {author} {\bibinfo {author} {\bibfnamefont {S.}~\bibnamefont
  {Bandow}}, \bibinfo {author} {\bibfnamefont {M.}~\bibnamefont {Takizawa}},
  \bibinfo {author} {\bibfnamefont {K.}~\bibnamefont {Hirahara}}, \bibinfo
  {author} {\bibfnamefont {M.}~\bibnamefont {Yudasaka}}, \ and\ \bibinfo
  {author} {\bibfnamefont {S.}~\bibnamefont {Iijima}},\ }\bibfield  {title}
  {\enquote {\bibinfo {title} {Raman scattering study of double-wall carbon
  nanotubes derived from the chains of fullerenes in single-wall carbon
  nanotubes},}\ }\href@noop {} {\bibfield  {journal} {\bibinfo  {journal}
  {Chemical Physics Letters}\ }\textbf {\bibinfo {volume} {337}},\ \bibinfo
  {pages} {48--54} (\bibinfo {year} {2001})}\BibitemShut {NoStop}%
\bibitem [{\citenamefont {Kataura}\ \emph {et~al.}(2002)\citenamefont
  {Kataura}, \citenamefont {Maniwa}, \citenamefont {Abe}, \citenamefont
  {Fujiwara}, \citenamefont {Kodama}, \citenamefont {Kikuchi}, \citenamefont
  {Imahori}, \citenamefont {Misaki}, \citenamefont {Suzuki},\ and\
  \citenamefont {Achiba}}]{Peapod_4}%
  \BibitemOpen
  \bibfield  {author} {\bibinfo {author} {\bibfnamefont {H.}~\bibnamefont
  {Kataura}}, \bibinfo {author} {\bibfnamefont {Y.}~\bibnamefont {Maniwa}},
  \bibinfo {author} {\bibfnamefont {M.}~\bibnamefont {Abe}}, \bibinfo {author}
  {\bibfnamefont {A.}~\bibnamefont {Fujiwara}}, \bibinfo {author}
  {\bibfnamefont {T.}~\bibnamefont {Kodama}}, \bibinfo {author} {\bibfnamefont
  {K.}~\bibnamefont {Kikuchi}}, \bibinfo {author} {\bibfnamefont
  {H.}~\bibnamefont {Imahori}}, \bibinfo {author} {\bibfnamefont
  {Y.}~\bibnamefont {Misaki}}, \bibinfo {author} {\bibfnamefont
  {S.}~\bibnamefont {Suzuki}}, \ and\ \bibinfo {author} {\bibfnamefont
  {Y.}~\bibnamefont {Achiba}},\ }\bibfield  {title} {\enquote {\bibinfo {title}
  {Optical properties of fullerene and non-fullerene peapods},}\ }\href@noop {}
  {\bibfield  {journal} {\bibinfo  {journal} {Applied Physics A}\ }\textbf
  {\bibinfo {volume} {74}},\ \bibinfo {pages} {349--354} (\bibinfo {year}
  {2002})}\BibitemShut {NoStop}%
\bibitem [{\citenamefont {Liu}\ \emph {et~al.}(2002)\citenamefont {Liu},
  \citenamefont {Pichler}, \citenamefont {Knupfer}, \citenamefont {Golden},
  \citenamefont {Fink}, \citenamefont {Kataura}, \citenamefont {Achiba},
  \citenamefont {Hirahara},\ and\ \citenamefont {Iijima}}]{Peapod_5}%
  \BibitemOpen
  \bibfield  {author} {\bibinfo {author} {\bibfnamefont {X.}~\bibnamefont
  {Liu}}, \bibinfo {author} {\bibfnamefont {T.}~\bibnamefont {Pichler}},
  \bibinfo {author} {\bibfnamefont {M.}~\bibnamefont {Knupfer}}, \bibinfo
  {author} {\bibfnamefont {M.}~\bibnamefont {Golden}}, \bibinfo {author}
  {\bibfnamefont {J.}~\bibnamefont {Fink}}, \bibinfo {author} {\bibfnamefont
  {H.}~\bibnamefont {Kataura}}, \bibinfo {author} {\bibfnamefont
  {Y.}~\bibnamefont {Achiba}}, \bibinfo {author} {\bibfnamefont
  {K.}~\bibnamefont {Hirahara}}, \ and\ \bibinfo {author} {\bibfnamefont
  {S.}~\bibnamefont {Iijima}},\ }\bibfield  {title} {\enquote {\bibinfo {title}
  {Filling factors, structural, and electronic properties of c 60 molecules in
  single-wall carbon nanotubes},}\ }\href@noop {} {\bibfield  {journal}
  {\bibinfo  {journal} {Physical Review B}\ }\textbf {\bibinfo {volume} {65}},\
  \bibinfo {pages} {045419} (\bibinfo {year} {2002})}\BibitemShut {NoStop}%
\bibitem [{\citenamefont {Troche}\ \emph {et~al.}(2005)\citenamefont {Troche},
  \citenamefont {Coluci}, \citenamefont {Braga}, \citenamefont {Chinellato},
  \citenamefont {Sato}, \citenamefont {Legoas}, \citenamefont {Rurali},\ and\
  \citenamefont {Galv{\~a}o}}]{Peapod_6}%
  \BibitemOpen
  \bibfield  {author} {\bibinfo {author} {\bibfnamefont {K.}~\bibnamefont
  {Troche}}, \bibinfo {author} {\bibfnamefont {V.}~\bibnamefont {Coluci}},
  \bibinfo {author} {\bibfnamefont {S.}~\bibnamefont {Braga}}, \bibinfo
  {author} {\bibfnamefont {D.}~\bibnamefont {Chinellato}}, \bibinfo {author}
  {\bibfnamefont {F.}~\bibnamefont {Sato}}, \bibinfo {author} {\bibfnamefont
  {S.}~\bibnamefont {Legoas}}, \bibinfo {author} {\bibfnamefont
  {R.}~\bibnamefont {Rurali}}, \ and\ \bibinfo {author} {\bibfnamefont
  {D.}~\bibnamefont {Galv{\~a}o}},\ }\bibfield  {title} {\enquote {\bibinfo
  {title} {Prediction of ordered phases of encapsulated c60, c70, and c78
  inside carbon nanotubes},}\ }\href@noop {} {\bibfield  {journal} {\bibinfo
  {journal} {Nano Letters}\ }\textbf {\bibinfo {volume} {5}},\ \bibinfo {pages}
  {349--355} (\bibinfo {year} {2005})}\BibitemShut {NoStop}%
\bibitem [{\citenamefont {Khlobystov}, \citenamefont {Britz},\ and\
  \citenamefont {Briggs}(2005)}]{Peapod_7}%
  \BibitemOpen
  \bibfield  {author} {\bibinfo {author} {\bibfnamefont {A.}~\bibnamefont
  {Khlobystov}}, \bibinfo {author} {\bibfnamefont {D.}~\bibnamefont {Britz}}, \
  and\ \bibinfo {author} {\bibfnamefont {G.}~\bibnamefont {Briggs}},\
  }\bibfield  {title} {\enquote {\bibinfo {title} {Molecules in carbon
  nanotubes},}\ }\href@noop {} {\bibfield  {journal} {\bibinfo  {journal}
  {Accounts of Chemical Research}\ }\textbf {\bibinfo {volume} {38}},\ \bibinfo
  {pages} {901--909} (\bibinfo {year} {2005})}\BibitemShut {NoStop}%
\bibitem [{\citenamefont {Simon}\ \emph {et~al.}(2006)\citenamefont {Simon},
  \citenamefont {Kuzmany}, \citenamefont {Bernardi}, \citenamefont {Hauke},\
  and\ \citenamefont {Hirsch}}]{Peapod_8}%
  \BibitemOpen
  \bibfield  {author} {\bibinfo {author} {\bibfnamefont {F.}~\bibnamefont
  {Simon}}, \bibinfo {author} {\bibfnamefont {H.}~\bibnamefont {Kuzmany}},
  \bibinfo {author} {\bibfnamefont {J.}~\bibnamefont {Bernardi}}, \bibinfo
  {author} {\bibfnamefont {F.}~\bibnamefont {Hauke}}, \ and\ \bibinfo {author}
  {\bibfnamefont {A.}~\bibnamefont {Hirsch}},\ }\bibfield  {title} {\enquote
  {\bibinfo {title} {Encapsulating c59n azafullerene derivatives inside
  single-wall carbon nanotubes},}\ }\href@noop {} {\bibfield  {journal}
  {\bibinfo  {journal} {Carbon}\ }\textbf {\bibinfo {volume} {44}},\ \bibinfo
  {pages} {1958--1962} (\bibinfo {year} {2006})}\BibitemShut {NoStop}%
\bibitem [{\citenamefont {Kuzmany}\ \emph {et~al.}(2007)\citenamefont
  {Kuzmany}, \citenamefont {Plank}, \citenamefont {Schaman}, \citenamefont
  {Pfeiffer}, \citenamefont {Hasi}, \citenamefont {Simon}, \citenamefont
  {Rotas}, \citenamefont {Pagona},\ and\ \citenamefont
  {Tagmatarchis}}]{Peapod_9}%
  \BibitemOpen
  \bibfield  {author} {\bibinfo {author} {\bibfnamefont {H.}~\bibnamefont
  {Kuzmany}}, \bibinfo {author} {\bibfnamefont {W.}~\bibnamefont {Plank}},
  \bibinfo {author} {\bibfnamefont {C.}~\bibnamefont {Schaman}}, \bibinfo
  {author} {\bibfnamefont {R.}~\bibnamefont {Pfeiffer}}, \bibinfo {author}
  {\bibfnamefont {F.}~\bibnamefont {Hasi}}, \bibinfo {author} {\bibfnamefont
  {F.}~\bibnamefont {Simon}}, \bibinfo {author} {\bibfnamefont
  {G.}~\bibnamefont {Rotas}}, \bibinfo {author} {\bibfnamefont
  {G.}~\bibnamefont {Pagona}}, \ and\ \bibinfo {author} {\bibfnamefont
  {N.}~\bibnamefont {Tagmatarchis}},\ }\bibfield  {title} {\enquote {\bibinfo
  {title} {Raman scattering from nanomaterials encapsulated into single wall
  carbon nanotubes},}\ }\href@noop {} {\bibfield  {journal} {\bibinfo
  {journal} {Journal of Raman Spectroscopy: An International Journal for
  Original Work in all Aspects of Raman Spectroscopy, Including Higher Order
  Processes, and also Brillouin and Rayleigh Scattering}\ }\textbf {\bibinfo
  {volume} {38}},\ \bibinfo {pages} {704--713} (\bibinfo {year}
  {2007})}\BibitemShut {NoStop}%
\bibitem [{\citenamefont {Zou}\ \emph {et~al.}(2007)\citenamefont {Zou},
  \citenamefont {Liu}, \citenamefont {Yao}, \citenamefont {Hou}, \citenamefont
  {Wang}, \citenamefont {Yu}, \citenamefont {Wang}, \citenamefont {Li},
  \citenamefont {Zou}, \citenamefont {Cui} \emph {et~al.}}]{Peapod_10}%
  \BibitemOpen
  \bibfield  {author} {\bibinfo {author} {\bibfnamefont {Y.}~\bibnamefont
  {Zou}}, \bibinfo {author} {\bibfnamefont {B.}~\bibnamefont {Liu}}, \bibinfo
  {author} {\bibfnamefont {M.}~\bibnamefont {Yao}}, \bibinfo {author}
  {\bibfnamefont {Y.}~\bibnamefont {Hou}}, \bibinfo {author} {\bibfnamefont
  {L.}~\bibnamefont {Wang}}, \bibinfo {author} {\bibfnamefont {S.}~\bibnamefont
  {Yu}}, \bibinfo {author} {\bibfnamefont {P.}~\bibnamefont {Wang}}, \bibinfo
  {author} {\bibfnamefont {B.}~\bibnamefont {Li}}, \bibinfo {author}
  {\bibfnamefont {B.}~\bibnamefont {Zou}}, \bibinfo {author} {\bibfnamefont
  {T.}~\bibnamefont {Cui}},  \emph {et~al.},\ }\bibfield  {title} {\enquote
  {\bibinfo {title} {Raman spectroscopy study of carbon nanotube peapods
  excited by near-ir laser under high pressure},}\ }\href@noop {} {\bibfield
  {journal} {\bibinfo  {journal} {Physical Review B}\ }\textbf {\bibinfo
  {volume} {76}},\ \bibinfo {pages} {195417} (\bibinfo {year}
  {2007})}\BibitemShut {NoStop}%
\bibitem [{\citenamefont {Sato}\ \emph {et~al.}(2007)\citenamefont {Sato},
  \citenamefont {Suenaga}, \citenamefont {Okubo}, \citenamefont {Okazaki},\
  and\ \citenamefont {Iijima}}]{Peapod_11}%
  \BibitemOpen
  \bibfield  {author} {\bibinfo {author} {\bibfnamefont {Y.}~\bibnamefont
  {Sato}}, \bibinfo {author} {\bibfnamefont {K.}~\bibnamefont {Suenaga}},
  \bibinfo {author} {\bibfnamefont {S.}~\bibnamefont {Okubo}}, \bibinfo
  {author} {\bibfnamefont {T.}~\bibnamefont {Okazaki}}, \ and\ \bibinfo
  {author} {\bibfnamefont {S.}~\bibnamefont {Iijima}},\ }\bibfield  {title}
  {\enquote {\bibinfo {title} {Structures of $d_{5d}-c_{80}$ and i h-er3n@ c80
  fullerenes and their rotation inside carbon nanotubes demonstrated by
  aberration-corrected electron microscopy},}\ }\href@noop {} {\bibfield
  {journal} {\bibinfo  {journal} {Nano Letters}\ }\textbf {\bibinfo {volume}
  {7}},\ \bibinfo {pages} {3704--3708} (\bibinfo {year} {2007})}\BibitemShut
  {NoStop}%
\bibitem [{\citenamefont {Kitaura}\ \emph {et~al.}(2008)\citenamefont
  {Kitaura}, \citenamefont {Imazu}, \citenamefont {Kobayashi},\ and\
  \citenamefont {Shinohara}}]{Peapod_12}%
  \BibitemOpen
  \bibfield  {author} {\bibinfo {author} {\bibfnamefont {R.}~\bibnamefont
  {Kitaura}}, \bibinfo {author} {\bibfnamefont {N.}~\bibnamefont {Imazu}},
  \bibinfo {author} {\bibfnamefont {K.}~\bibnamefont {Kobayashi}}, \ and\
  \bibinfo {author} {\bibfnamefont {H.}~\bibnamefont {Shinohara}},\ }\bibfield
  {title} {\enquote {\bibinfo {title} {Fabrication of metal nanowires in carbon
  nanotubes via versatile nano-template reaction},}\ }\href@noop {} {\bibfield
  {journal} {\bibinfo  {journal} {Nano Letters}\ }\textbf {\bibinfo {volume}
  {8}},\ \bibinfo {pages} {693--699} (\bibinfo {year} {2008})}\BibitemShut
  {NoStop}%
\bibitem [{\citenamefont {Chamberlain}\ \emph {et~al.}(2008)\citenamefont
  {Chamberlain}, \citenamefont {Pfeiffer}, \citenamefont {Peterlik},
  \citenamefont {Kuzmany}, \citenamefont {Zerbetto}, \citenamefont
  {Melle-Franco}, \citenamefont {Staddon}, \citenamefont {Champness},
  \citenamefont {Briggs},\ and\ \citenamefont {Khlobystov}}]{Peapod_13}%
  \BibitemOpen
  \bibfield  {author} {\bibinfo {author} {\bibfnamefont {T.}~\bibnamefont
  {Chamberlain}}, \bibinfo {author} {\bibfnamefont {R.}~\bibnamefont
  {Pfeiffer}}, \bibinfo {author} {\bibfnamefont {H.}~\bibnamefont {Peterlik}},
  \bibinfo {author} {\bibfnamefont {H.}~\bibnamefont {Kuzmany}}, \bibinfo
  {author} {\bibfnamefont {F.}~\bibnamefont {Zerbetto}}, \bibinfo {author}
  {\bibfnamefont {M.}~\bibnamefont {Melle-Franco}}, \bibinfo {author}
  {\bibfnamefont {L.}~\bibnamefont {Staddon}}, \bibinfo {author} {\bibfnamefont
  {N.}~\bibnamefont {Champness}}, \bibinfo {author} {\bibfnamefont
  {G.}~\bibnamefont {Briggs}}, \ and\ \bibinfo {author} {\bibfnamefont
  {A.}~\bibnamefont {Khlobystov}},\ }\bibfield  {title} {\enquote {\bibinfo
  {title} {Polyarene-functionalized fullerenes in carbon nanotubes: Towards
  controlled geometry of molecular chains},}\ }\href@noop {} {\bibfield
  {journal} {\bibinfo  {journal} {Small}\ }\textbf {\bibinfo {volume} {4}},\
  \bibinfo {pages} {2262--2270} (\bibinfo {year} {2008})}\BibitemShut {NoStop}%
\bibitem [{\citenamefont {Abou-Hamad}\ \emph {et~al.}(2009)\citenamefont
  {Abou-Hamad}, \citenamefont {Kim}, \citenamefont {Talyzin}, \citenamefont
  {Goze-Bac}, \citenamefont {Luzzi}, \citenamefont {Rubio},\ and\ \citenamefont
  {Wagberg}}]{Peapod_14}%
  \BibitemOpen
  \bibfield  {author} {\bibinfo {author} {\bibfnamefont {E.}~\bibnamefont
  {Abou-Hamad}}, \bibinfo {author} {\bibfnamefont {Y.}~\bibnamefont {Kim}},
  \bibinfo {author} {\bibfnamefont {A.}~\bibnamefont {Talyzin}}, \bibinfo
  {author} {\bibfnamefont {C.}~\bibnamefont {Goze-Bac}}, \bibinfo {author}
  {\bibfnamefont {D.}~\bibnamefont {Luzzi}}, \bibinfo {author} {\bibfnamefont
  {A.}~\bibnamefont {Rubio}}, \ and\ \bibinfo {author} {\bibfnamefont
  {T.}~\bibnamefont {Wagberg}},\ }\bibfield  {title} {\enquote {\bibinfo
  {title} {Hydrogenation of c60 in peapods: physical chemistry in nano
  vessels},}\ }\href@noop {} {\bibfield  {journal} {\bibinfo  {journal} {The
  Journal of Physical Chemistry C}\ }\textbf {\bibinfo {volume} {113}},\
  \bibinfo {pages} {8583--8587} (\bibinfo {year} {2009})}\BibitemShut {NoStop}%
\bibitem [{\citenamefont {Zou}\ \emph {et~al.}(2009)\citenamefont {Zou},
  \citenamefont {Liu}, \citenamefont {Wang}, \citenamefont {Liu}, \citenamefont
  {Yu}, \citenamefont {Wang}, \citenamefont {Wang}, \citenamefont {Yao},
  \citenamefont {Li}, \citenamefont {Zou} \emph {et~al.}}]{Peapod_15}%
  \BibitemOpen
  \bibfield  {author} {\bibinfo {author} {\bibfnamefont {Y.}~\bibnamefont
  {Zou}}, \bibinfo {author} {\bibfnamefont {B.}~\bibnamefont {Liu}}, \bibinfo
  {author} {\bibfnamefont {L.}~\bibnamefont {Wang}}, \bibinfo {author}
  {\bibfnamefont {D.}~\bibnamefont {Liu}}, \bibinfo {author} {\bibfnamefont
  {S.}~\bibnamefont {Yu}}, \bibinfo {author} {\bibfnamefont {P.}~\bibnamefont
  {Wang}}, \bibinfo {author} {\bibfnamefont {T.}~\bibnamefont {Wang}}, \bibinfo
  {author} {\bibfnamefont {M.}~\bibnamefont {Yao}}, \bibinfo {author}
  {\bibfnamefont {Q.}~\bibnamefont {Li}}, \bibinfo {author} {\bibfnamefont
  {B.}~\bibnamefont {Zou}},  \emph {et~al.},\ }\bibfield  {title} {\enquote
  {\bibinfo {title} {Rotational dynamics of confined c60 from near-infrared
  raman studies under high pressure},}\ }\href@noop {} {\bibfield  {journal}
  {\bibinfo  {journal} {Proceedings of the National Academy of Sciences}\
  }\textbf {\bibinfo {volume} {106}},\ \bibinfo {pages} {22135--22138}
  (\bibinfo {year} {2009})}\BibitemShut {NoStop}%
\bibitem [{\citenamefont {Maggini}\ \emph {et~al.}(2014)\citenamefont
  {Maggini}, \citenamefont {Fustos}, \citenamefont {Chamberlain}, \citenamefont
  {Cebrian}, \citenamefont {Natali}, \citenamefont {Pietraszkiewicz},
  \citenamefont {Pietraszkiewicz}, \citenamefont {Szekely}, \citenamefont
  {Kamaras}, \citenamefont {De~Cola} \emph {et~al.}}]{Peapod_16}%
  \BibitemOpen
  \bibfield  {author} {\bibinfo {author} {\bibfnamefont {L.}~\bibnamefont
  {Maggini}}, \bibinfo {author} {\bibfnamefont {M.}~\bibnamefont {Fustos}},
  \bibinfo {author} {\bibfnamefont {T.}~\bibnamefont {Chamberlain}}, \bibinfo
  {author} {\bibfnamefont {C.}~\bibnamefont {Cebrian}}, \bibinfo {author}
  {\bibfnamefont {M.}~\bibnamefont {Natali}}, \bibinfo {author} {\bibfnamefont
  {M.}~\bibnamefont {Pietraszkiewicz}}, \bibinfo {author} {\bibfnamefont
  {O.}~\bibnamefont {Pietraszkiewicz}}, \bibinfo {author} {\bibfnamefont
  {E.}~\bibnamefont {Szekely}}, \bibinfo {author} {\bibfnamefont
  {K.}~\bibnamefont {Kamaras}}, \bibinfo {author} {\bibfnamefont
  {L.}~\bibnamefont {De~Cola}},  \emph {et~al.},\ }\bibfield  {title} {\enquote
  {\bibinfo {title} {Fullerene-driven encapsulation of a luminescent eu (iii)
  complex in carbon nanotubes},}\ }\href@noop {} {\bibfield  {journal}
  {\bibinfo  {journal} {Nanoscale}\ }\textbf {\bibinfo {volume} {6}},\ \bibinfo
  {pages} {2887--2894} (\bibinfo {year} {2014})}\BibitemShut {NoStop}%
\bibitem [{\citenamefont {Benjamin}\ \emph {et~al.}(2006)\citenamefont
  {Benjamin}, \citenamefont {Ardavan}, \citenamefont {Briggs}, \citenamefont
  {Britz}, \citenamefont {Gunlycke}, \citenamefont {Jefferson}, \citenamefont
  {Jones}, \citenamefont {Leigh}, \citenamefont {Lovett}, \citenamefont
  {Khlobystov} \emph {et~al.}}]{benjamin2006towards}%
  \BibitemOpen
  \bibfield  {author} {\bibinfo {author} {\bibfnamefont {S.}~\bibnamefont
  {Benjamin}}, \bibinfo {author} {\bibfnamefont {A.}~\bibnamefont {Ardavan}},
  \bibinfo {author} {\bibfnamefont {G.}~\bibnamefont {Briggs}}, \bibinfo
  {author} {\bibfnamefont {D.}~\bibnamefont {Britz}}, \bibinfo {author}
  {\bibfnamefont {D.}~\bibnamefont {Gunlycke}}, \bibinfo {author}
  {\bibfnamefont {J.}~\bibnamefont {Jefferson}}, \bibinfo {author}
  {\bibfnamefont {M.}~\bibnamefont {Jones}}, \bibinfo {author} {\bibfnamefont
  {D.}~\bibnamefont {Leigh}}, \bibinfo {author} {\bibfnamefont
  {B.}~\bibnamefont {Lovett}}, \bibinfo {author} {\bibfnamefont
  {A.}~\bibnamefont {Khlobystov}},  \emph {et~al.},\ }\bibfield  {title}
  {\enquote {\bibinfo {title} {Towards a fullerene-based quantum computer},}\
  }\href@noop {} {\bibfield  {journal} {\bibinfo  {journal} {Journal of
  Physics: Condensed Matter}\ }\textbf {\bibinfo {volume} {18}},\ \bibinfo
  {pages} {S867} (\bibinfo {year} {2006})}\BibitemShut {NoStop}%
\bibitem [{\citenamefont {Takenobu}\ \emph {et~al.}(2003)\citenamefont
  {Takenobu}, \citenamefont {Takano}, \citenamefont {Shiraishi}, \citenamefont
  {Murakami}, \citenamefont {Ata}, \citenamefont {Kataura}, \citenamefont
  {Achiba},\ and\ \citenamefont {Iwasa}}]{Org_1}%
  \BibitemOpen
  \bibfield  {author} {\bibinfo {author} {\bibfnamefont {T.}~\bibnamefont
  {Takenobu}}, \bibinfo {author} {\bibfnamefont {T.}~\bibnamefont {Takano}},
  \bibinfo {author} {\bibfnamefont {M.}~\bibnamefont {Shiraishi}}, \bibinfo
  {author} {\bibfnamefont {Y.}~\bibnamefont {Murakami}}, \bibinfo {author}
  {\bibfnamefont {M.}~\bibnamefont {Ata}}, \bibinfo {author} {\bibfnamefont
  {H.}~\bibnamefont {Kataura}}, \bibinfo {author} {\bibfnamefont
  {Y.}~\bibnamefont {Achiba}}, \ and\ \bibinfo {author} {\bibfnamefont
  {Y.}~\bibnamefont {Iwasa}},\ }\bibfield  {title} {\enquote {\bibinfo {title}
  {Stable and controlled amphoteric doping by encapsulation of organic
  molecules inside carbon nanotubes},}\ }\href@noop {} {\bibfield  {journal}
  {\bibinfo  {journal} {Nature Materials}\ }\textbf {\bibinfo {volume} {2}},\
  \bibinfo {pages} {683--688} (\bibinfo {year} {2003})}\BibitemShut {NoStop}%
\bibitem [{\citenamefont {Steinmetz}\ \emph {et~al.}(2006)\citenamefont
  {Steinmetz}, \citenamefont {Kwon}, \citenamefont {Lee}, \citenamefont
  {Abou-Hamad}, \citenamefont {Almairac}, \citenamefont {Goze-Bac},
  \citenamefont {Kim},\ and\ \citenamefont {Park}}]{Org_2}%
  \BibitemOpen
  \bibfield  {author} {\bibinfo {author} {\bibfnamefont {J.}~\bibnamefont
  {Steinmetz}}, \bibinfo {author} {\bibfnamefont {S.}~\bibnamefont {Kwon}},
  \bibinfo {author} {\bibfnamefont {H.}~\bibnamefont {Lee}}, \bibinfo {author}
  {\bibfnamefont {E.}~\bibnamefont {Abou-Hamad}}, \bibinfo {author}
  {\bibfnamefont {R.}~\bibnamefont {Almairac}}, \bibinfo {author}
  {\bibfnamefont {C.}~\bibnamefont {Goze-Bac}}, \bibinfo {author}
  {\bibfnamefont {H.}~\bibnamefont {Kim}}, \ and\ \bibinfo {author}
  {\bibfnamefont {Y.}~\bibnamefont {Park}},\ }\bibfield  {title} {\enquote
  {\bibinfo {title} {Polymerization of conducting polymers inside carbon
  nanotubes},}\ }\href@noop {} {\bibfield  {journal} {\bibinfo  {journal}
  {Chemical Physics Letters}\ }\textbf {\bibinfo {volume} {431}},\ \bibinfo
  {pages} {139--144} (\bibinfo {year} {2006})}\BibitemShut {NoStop}%
\bibitem [{\citenamefont {Yao}\ \emph {et~al.}(2011)\citenamefont {Yao},
  \citenamefont {Stenmark}, \citenamefont {Abou-Hamad}, \citenamefont {Nitze},
  \citenamefont {Qin}, \citenamefont {Goze-Bac},\ and\ \citenamefont
  {Wagberg}}]{Org_3}%
  \BibitemOpen
  \bibfield  {author} {\bibinfo {author} {\bibfnamefont {M.}~\bibnamefont
  {Yao}}, \bibinfo {author} {\bibfnamefont {P.}~\bibnamefont {Stenmark}},
  \bibinfo {author} {\bibfnamefont {E.}~\bibnamefont {Abou-Hamad}}, \bibinfo
  {author} {\bibfnamefont {F.}~\bibnamefont {Nitze}}, \bibinfo {author}
  {\bibfnamefont {J.}~\bibnamefont {Qin}}, \bibinfo {author} {\bibfnamefont
  {C.}~\bibnamefont {Goze-Bac}}, \ and\ \bibinfo {author} {\bibfnamefont
  {T.}~\bibnamefont {Wagberg}},\ }\bibfield  {title} {\enquote {\bibinfo
  {title} {Confined adamantane molecules assembled to one dimension in carbon
  nanotubes},}\ }\href@noop {} {\bibfield  {journal} {\bibinfo  {journal}
  {Carbon}\ }\textbf {\bibinfo {volume} {49}},\ \bibinfo {pages} {1159--1166}
  (\bibinfo {year} {2011})}\BibitemShut {NoStop}%
\bibitem [{\citenamefont {Okazaki}\ \emph {et~al.}(2011)\citenamefont
  {Okazaki}, \citenamefont {Iizumi}, \citenamefont {Okubo}, \citenamefont
  {Kataura}, \citenamefont {Liu}, \citenamefont {Suenaga}, \citenamefont
  {Tahara}, \citenamefont {Yudasaka}, \citenamefont {Okada},\ and\
  \citenamefont {Iijima}}]{Org_4}%
  \BibitemOpen
  \bibfield  {author} {\bibinfo {author} {\bibfnamefont {T.}~\bibnamefont
  {Okazaki}}, \bibinfo {author} {\bibfnamefont {Y.}~\bibnamefont {Iizumi}},
  \bibinfo {author} {\bibfnamefont {S.}~\bibnamefont {Okubo}}, \bibinfo
  {author} {\bibfnamefont {H.}~\bibnamefont {Kataura}}, \bibinfo {author}
  {\bibfnamefont {Z.}~\bibnamefont {Liu}}, \bibinfo {author} {\bibfnamefont
  {K.}~\bibnamefont {Suenaga}}, \bibinfo {author} {\bibfnamefont
  {Y.}~\bibnamefont {Tahara}}, \bibinfo {author} {\bibfnamefont
  {M.}~\bibnamefont {Yudasaka}}, \bibinfo {author} {\bibfnamefont
  {S.}~\bibnamefont {Okada}}, \ and\ \bibinfo {author} {\bibfnamefont
  {S.}~\bibnamefont {Iijima}},\ }\bibfield  {title} {\enquote {\bibinfo {title}
  {Coaxially stacked coronene columns inside single-walled carbon nanotubes},}\
  }\href@noop {} {\bibfield  {journal} {\bibinfo  {journal} {Angewandte
  Chemie}\ }\textbf {\bibinfo {volume} {123}},\ \bibinfo {pages} {4955--4959}
  (\bibinfo {year} {2011})}\BibitemShut {NoStop}%
\bibitem [{\citenamefont {Botka}\ \emph {et~al.}(2012)\citenamefont {Botka},
  \citenamefont {Fustuo}, \citenamefont {Klupp}, \citenamefont {Kocsis},
  \citenamefont {Szekely}, \citenamefont {Utczas}, \citenamefont {Simandi},
  \citenamefont {Botos}, \citenamefont {Hackl},\ and\ \citenamefont
  {Kamaras}}]{Org_5}%
  \BibitemOpen
  \bibfield  {author} {\bibinfo {author} {\bibfnamefont {B.}~\bibnamefont
  {Botka}}, \bibinfo {author} {\bibfnamefont {M.}~\bibnamefont {Fustuo}},
  \bibinfo {author} {\bibfnamefont {G.}~\bibnamefont {Klupp}}, \bibinfo
  {author} {\bibfnamefont {D.}~\bibnamefont {Kocsis}}, \bibinfo {author}
  {\bibfnamefont {E.}~\bibnamefont {Szekely}}, \bibinfo {author} {\bibfnamefont
  {M.}~\bibnamefont {Utczas}}, \bibinfo {author} {\bibfnamefont
  {B.}~\bibnamefont {Simandi}}, \bibinfo {author} {\bibfnamefont
  {A.}~\bibnamefont {Botos}}, \bibinfo {author} {\bibfnamefont
  {R.}~\bibnamefont {Hackl}}, \ and\ \bibinfo {author} {\bibfnamefont
  {K.}~\bibnamefont {Kamaras}},\ }\bibfield  {title} {\enquote {\bibinfo
  {title} {Low-temperature encapsulation of coronene in carbon nanotubes},}\
  }\href@noop {} {\bibfield  {journal} {\bibinfo  {journal} {Physica Status
  Solidi (b)}\ }\textbf {\bibinfo {volume} {249}},\ \bibinfo {pages}
  {2432--2435} (\bibinfo {year} {2012})}\BibitemShut {NoStop}%
\bibitem [{\citenamefont {Botka}\ \emph {et~al.}(2014)\citenamefont {Botka},
  \citenamefont {Fust{\"o}s}, \citenamefont {T{\'o}h{\'a}ti}, \citenamefont
  {N{\'e}meth}, \citenamefont {Klupp}, \citenamefont {Szekr{\'e}nyes},
  \citenamefont {Kocsis}, \citenamefont {Utcz{\'a}s}, \citenamefont
  {Sz{\'e}kely}, \citenamefont {V{\'a}czi} \emph {et~al.}}]{Org_6}%
  \BibitemOpen
  \bibfield  {author} {\bibinfo {author} {\bibfnamefont {B.}~\bibnamefont
  {Botka}}, \bibinfo {author} {\bibfnamefont {M.}~\bibnamefont {Fust{\"o}s}},
  \bibinfo {author} {\bibfnamefont {H.}~\bibnamefont {T{\'o}h{\'a}ti}},
  \bibinfo {author} {\bibfnamefont {K.}~\bibnamefont {N{\'e}meth}}, \bibinfo
  {author} {\bibfnamefont {G.}~\bibnamefont {Klupp}}, \bibinfo {author}
  {\bibfnamefont {Z.}~\bibnamefont {Szekr{\'e}nyes}}, \bibinfo {author}
  {\bibfnamefont {D.}~\bibnamefont {Kocsis}}, \bibinfo {author} {\bibfnamefont
  {M.}~\bibnamefont {Utcz{\'a}s}}, \bibinfo {author} {\bibfnamefont
  {E.}~\bibnamefont {Sz{\'e}kely}}, \bibinfo {author} {\bibfnamefont
  {T.}~\bibnamefont {V{\'a}czi}},  \emph {et~al.},\ }\bibfield  {title}
  {\enquote {\bibinfo {title} {Interactions and chemical transformations of
  coronene inside and outside carbon nanotubes},}\ }\href@noop {} {\bibfield
  {journal} {\bibinfo  {journal} {Small}\ }\textbf {\bibinfo {volume} {10}},\
  \bibinfo {pages} {1369--1378} (\bibinfo {year} {2014})}\BibitemShut {NoStop}%
\bibitem [{\citenamefont {Li}\ \emph {et~al.}(2005)\citenamefont {Li},
  \citenamefont {Khlobystov}, \citenamefont {Wiltshire}, \citenamefont
  {Briggs},\ and\ \citenamefont {Nicholas}}]{Mc_1}%
  \BibitemOpen
  \bibfield  {author} {\bibinfo {author} {\bibfnamefont {L.}~\bibnamefont
  {Li}}, \bibinfo {author} {\bibfnamefont {A.}~\bibnamefont {Khlobystov}},
  \bibinfo {author} {\bibfnamefont {J.}~\bibnamefont {Wiltshire}}, \bibinfo
  {author} {\bibfnamefont {G.}~\bibnamefont {Briggs}}, \ and\ \bibinfo {author}
  {\bibfnamefont {R.}~\bibnamefont {Nicholas}},\ }\bibfield  {title} {\enquote
  {\bibinfo {title} {Diameter-selective encapsulation of metallocenes in
  single-walled carbon nanotubes},}\ }\href@noop {} {\bibfield  {journal}
  {\bibinfo  {journal} {Nature Materials}\ }\textbf {\bibinfo {volume} {4}},\
  \bibinfo {pages} {481--485} (\bibinfo {year} {2005})}\BibitemShut {NoStop}%
\bibitem [{\citenamefont {Shiozawa}\ \emph {et~al.}(2008)\citenamefont
  {Shiozawa}, \citenamefont {Pichler}, \citenamefont {Gr{\"u}neis},
  \citenamefont {Pfeiffer}, \citenamefont {Kuzmany}, \citenamefont {Liu},
  \citenamefont {Suenaga},\ and\ \citenamefont {Kataura}}]{Mc_2}%
  \BibitemOpen
  \bibfield  {author} {\bibinfo {author} {\bibfnamefont {H.}~\bibnamefont
  {Shiozawa}}, \bibinfo {author} {\bibfnamefont {T.}~\bibnamefont {Pichler}},
  \bibinfo {author} {\bibfnamefont {A.}~\bibnamefont {Gr{\"u}neis}}, \bibinfo
  {author} {\bibfnamefont {R.}~\bibnamefont {Pfeiffer}}, \bibinfo {author}
  {\bibfnamefont {H.}~\bibnamefont {Kuzmany}}, \bibinfo {author} {\bibfnamefont
  {Z.}~\bibnamefont {Liu}}, \bibinfo {author} {\bibfnamefont {K.}~\bibnamefont
  {Suenaga}}, \ and\ \bibinfo {author} {\bibfnamefont {H.}~\bibnamefont
  {Kataura}},\ }\bibfield  {title} {\enquote {\bibinfo {title} {A catalytic
  reaction inside a single-walled carbon nanotube},}\ }\href@noop {} {\bibfield
   {journal} {\bibinfo  {journal} {Advanced Materials}\ }\textbf {\bibinfo
  {volume} {20}},\ \bibinfo {pages} {1443--1449} (\bibinfo {year}
  {2008})}\BibitemShut {NoStop}%
\bibitem [{\citenamefont {Shiozawa}\ \emph {et~al.}(2010)\citenamefont
  {Shiozawa}, \citenamefont {Kramberger}, \citenamefont {Pfeiffer},
  \citenamefont {Kuzmany}, \citenamefont {Pichler}, \citenamefont {Liu},
  \citenamefont {Suenaga}, \citenamefont {Kataura},\ and\ \citenamefont
  {Silva}}]{Mc_3}%
  \BibitemOpen
  \bibfield  {author} {\bibinfo {author} {\bibfnamefont {H.}~\bibnamefont
  {Shiozawa}}, \bibinfo {author} {\bibfnamefont {C.}~\bibnamefont
  {Kramberger}}, \bibinfo {author} {\bibfnamefont {R.}~\bibnamefont
  {Pfeiffer}}, \bibinfo {author} {\bibfnamefont {H.}~\bibnamefont {Kuzmany}},
  \bibinfo {author} {\bibfnamefont {T.}~\bibnamefont {Pichler}}, \bibinfo
  {author} {\bibfnamefont {Z.}~\bibnamefont {Liu}}, \bibinfo {author}
  {\bibfnamefont {K.}~\bibnamefont {Suenaga}}, \bibinfo {author} {\bibfnamefont
  {H.}~\bibnamefont {Kataura}}, \ and\ \bibinfo {author} {\bibfnamefont
  {S.}~\bibnamefont {Silva}},\ }\bibfield  {title} {\enquote {\bibinfo {title}
  {Catalyst and chirality dependent growth of carbon nanotubes determined
  through nano-test tube chemistry},}\ }\href@noop {} {\bibfield  {journal}
  {\bibinfo  {journal} {Advanced Materials}\ }\textbf {\bibinfo {volume}
  {22}},\ \bibinfo {pages} {3685--3689} (\bibinfo {year} {2010})}\BibitemShut
  {NoStop}%
\bibitem [{\citenamefont {Kocsis}\ \emph {et~al.}(2011)\citenamefont {Kocsis},
  \citenamefont {Kapt{\'a}s}, \citenamefont {Botos}, \citenamefont {Pekker},\
  and\ \citenamefont {Kamar{\'a}s}}]{Mc_4}%
  \BibitemOpen
  \bibfield  {author} {\bibinfo {author} {\bibfnamefont {D.}~\bibnamefont
  {Kocsis}}, \bibinfo {author} {\bibfnamefont {D.}~\bibnamefont {Kapt{\'a}s}},
  \bibinfo {author} {\bibfnamefont {A.}~\bibnamefont {Botos}}, \bibinfo
  {author} {\bibfnamefont {A.}~\bibnamefont {Pekker}}, \ and\ \bibinfo {author}
  {\bibfnamefont {K.}~\bibnamefont {Kamar{\'a}s}},\ }\bibfield  {title}
  {\enquote {\bibinfo {title} {Ferrocene encapsulation in carbon nanotubes:
  Various methods of filling and investigation},}\ }\href@noop {} {\bibfield
  {journal} {\bibinfo  {journal} {Physica Status Solidi (b)}\ }\textbf
  {\bibinfo {volume} {248}},\ \bibinfo {pages} {2512--2515} (\bibinfo {year}
  {2011})}\BibitemShut {NoStop}%
\bibitem [{\citenamefont {Liu}\ \emph {et~al.}(2012)\citenamefont {Liu},
  \citenamefont {Kuzmany}, \citenamefont {Ayala}, \citenamefont {Calvaresi},
  \citenamefont {Zerbetto},\ and\ \citenamefont {Pichler}}]{Mc_5}%
  \BibitemOpen
  \bibfield  {author} {\bibinfo {author} {\bibfnamefont {X.}~\bibnamefont
  {Liu}}, \bibinfo {author} {\bibfnamefont {H.}~\bibnamefont {Kuzmany}},
  \bibinfo {author} {\bibfnamefont {P.}~\bibnamefont {Ayala}}, \bibinfo
  {author} {\bibfnamefont {M.}~\bibnamefont {Calvaresi}}, \bibinfo {author}
  {\bibfnamefont {F.}~\bibnamefont {Zerbetto}}, \ and\ \bibinfo {author}
  {\bibfnamefont {T.}~\bibnamefont {Pichler}},\ }\bibfield  {title} {\enquote
  {\bibinfo {title} {Selective enhancement of photoluminescence in filled
  single-walled carbon nanotubes},}\ }\href@noop {} {\bibfield  {journal}
  {\bibinfo  {journal} {Advanced Functional Materials}\ }\textbf {\bibinfo
  {volume} {22}},\ \bibinfo {pages} {3202--3208} (\bibinfo {year}
  {2012})}\BibitemShut {NoStop}%
\bibitem [{\citenamefont {Kharlamova}\ \emph {et~al.}(2015)\citenamefont
  {Kharlamova}, \citenamefont {Sauer}, \citenamefont {Saito}, \citenamefont
  {Sato}, \citenamefont {Suenaga}, \citenamefont {Pichler},\ and\ \citenamefont
  {Shiozawa}}]{Mc_6}%
  \BibitemOpen
  \bibfield  {author} {\bibinfo {author} {\bibfnamefont {M.}~\bibnamefont
  {Kharlamova}}, \bibinfo {author} {\bibfnamefont {M.}~\bibnamefont {Sauer}},
  \bibinfo {author} {\bibfnamefont {T.}~\bibnamefont {Saito}}, \bibinfo
  {author} {\bibfnamefont {Y.}~\bibnamefont {Sato}}, \bibinfo {author}
  {\bibfnamefont {K.}~\bibnamefont {Suenaga}}, \bibinfo {author} {\bibfnamefont
  {T.}~\bibnamefont {Pichler}}, \ and\ \bibinfo {author} {\bibfnamefont
  {H.}~\bibnamefont {Shiozawa}},\ }\bibfield  {title} {\enquote {\bibinfo
  {title} {Doping of single-walled carbon nanotubes controlled via chemical
  transformation of encapsulated nickelocene},}\ }\href@noop {} {\bibfield
  {journal} {\bibinfo  {journal} {Nanoscale}\ }\textbf {\bibinfo {volume}
  {7}},\ \bibinfo {pages} {1383--1391} (\bibinfo {year} {2015})}\BibitemShut
  {NoStop}%
\bibitem [{\citenamefont {Nishide}\ \emph {et~al.}(2006)\citenamefont
  {Nishide}, \citenamefont {Dohi}, \citenamefont {Wakabayashi}, \citenamefont
  {Nishibori}, \citenamefont {Aoyagi}, \citenamefont {Ishida}, \citenamefont
  {Kikuchi}, \citenamefont {Kitaura}, \citenamefont {Sugai}, \citenamefont
  {Sakata} \emph {et~al.}}]{Cchain_1}%
  \BibitemOpen
  \bibfield  {author} {\bibinfo {author} {\bibfnamefont {D.}~\bibnamefont
  {Nishide}}, \bibinfo {author} {\bibfnamefont {H.}~\bibnamefont {Dohi}},
  \bibinfo {author} {\bibfnamefont {T.}~\bibnamefont {Wakabayashi}}, \bibinfo
  {author} {\bibfnamefont {E.}~\bibnamefont {Nishibori}}, \bibinfo {author}
  {\bibfnamefont {S.}~\bibnamefont {Aoyagi}}, \bibinfo {author} {\bibfnamefont
  {M.}~\bibnamefont {Ishida}}, \bibinfo {author} {\bibfnamefont
  {S.}~\bibnamefont {Kikuchi}}, \bibinfo {author} {\bibfnamefont
  {R.}~\bibnamefont {Kitaura}}, \bibinfo {author} {\bibfnamefont
  {T.}~\bibnamefont {Sugai}}, \bibinfo {author} {\bibfnamefont
  {M.}~\bibnamefont {Sakata}},  \emph {et~al.},\ }\bibfield  {title} {\enquote
  {\bibinfo {title} {Single-wall carbon nanotubes encaging linear chain c10h2
  polyyne molecules inside},}\ }\href@noop {} {\bibfield  {journal} {\bibinfo
  {journal} {Chemical Physics Letters}\ }\textbf {\bibinfo {volume} {428}},\
  \bibinfo {pages} {356--360} (\bibinfo {year} {2006})}\BibitemShut {NoStop}%
\bibitem [{\citenamefont {Warner}\ \emph {et~al.}(2010)\citenamefont {Warner},
  \citenamefont {R{\"u}mmeli}, \citenamefont {Bachmatiuk},\ and\ \citenamefont
  {B{\"u}chner}}]{Cchain_2}%
  \BibitemOpen
  \bibfield  {author} {\bibinfo {author} {\bibfnamefont {J.}~\bibnamefont
  {Warner}}, \bibinfo {author} {\bibfnamefont {M.}~\bibnamefont {R{\"u}mmeli}},
  \bibinfo {author} {\bibfnamefont {A.}~\bibnamefont {Bachmatiuk}}, \ and\
  \bibinfo {author} {\bibfnamefont {B.}~\bibnamefont {B{\"u}chner}},\
  }\bibfield  {title} {\enquote {\bibinfo {title} {Structural transformations
  of carbon chains inside nanotubes},}\ }\href@noop {} {\bibfield  {journal}
  {\bibinfo  {journal} {Physical Review B}\ }\textbf {\bibinfo {volume} {81}},\
  \bibinfo {pages} {155419} (\bibinfo {year} {2010})}\BibitemShut {NoStop}%
\bibitem [{\citenamefont {Zhang}\ \emph {et~al.}(2012)\citenamefont {Zhang},
  \citenamefont {Feng}, \citenamefont {Ishiwata}, \citenamefont {Miyata},
  \citenamefont {Kitaura}, \citenamefont {Dahl}, \citenamefont {Carlson},
  \citenamefont {Shinohara},\ and\ \citenamefont {Tom{\'a}nek}}]{Cchain_3}%
  \BibitemOpen
  \bibfield  {author} {\bibinfo {author} {\bibfnamefont {J.}~\bibnamefont
  {Zhang}}, \bibinfo {author} {\bibfnamefont {Y.}~\bibnamefont {Feng}},
  \bibinfo {author} {\bibfnamefont {H.}~\bibnamefont {Ishiwata}}, \bibinfo
  {author} {\bibfnamefont {Y.}~\bibnamefont {Miyata}}, \bibinfo {author}
  {\bibfnamefont {R.}~\bibnamefont {Kitaura}}, \bibinfo {author} {\bibfnamefont
  {J.}~\bibnamefont {Dahl}}, \bibinfo {author} {\bibfnamefont {R.}~\bibnamefont
  {Carlson}}, \bibinfo {author} {\bibfnamefont {H.}~\bibnamefont {Shinohara}},
  \ and\ \bibinfo {author} {\bibfnamefont {D.}~\bibnamefont {Tom{\'a}nek}},\
  }\bibfield  {title} {\enquote {\bibinfo {title} {Synthesis and transformation
  of linear adamantane assemblies inside carbon nanotubes},}\ }\href@noop {}
  {\bibfield  {journal} {\bibinfo  {journal} {ACS Nano}\ }\textbf {\bibinfo
  {volume} {6}},\ \bibinfo {pages} {8674--8683} (\bibinfo {year}
  {2012})}\BibitemShut {NoStop}%
\bibitem [{\citenamefont {Shi}\ \emph {et~al.}(2016)\citenamefont {Shi},
  \citenamefont {Rohringer}, \citenamefont {Suenaga}, \citenamefont {Niimi},
  \citenamefont {Kotakoski}, \citenamefont {Meyer}, \citenamefont {Peterlik},
  \citenamefont {Wanko}, \citenamefont {Cahangirov}, \citenamefont {Rubio}
  \emph {et~al.}}]{Cchain_4}%
  \BibitemOpen
  \bibfield  {author} {\bibinfo {author} {\bibfnamefont {L.}~\bibnamefont
  {Shi}}, \bibinfo {author} {\bibfnamefont {P.}~\bibnamefont {Rohringer}},
  \bibinfo {author} {\bibfnamefont {K.}~\bibnamefont {Suenaga}}, \bibinfo
  {author} {\bibfnamefont {Y.}~\bibnamefont {Niimi}}, \bibinfo {author}
  {\bibfnamefont {J.}~\bibnamefont {Kotakoski}}, \bibinfo {author}
  {\bibfnamefont {J.}~\bibnamefont {Meyer}}, \bibinfo {author} {\bibfnamefont
  {H.}~\bibnamefont {Peterlik}}, \bibinfo {author} {\bibfnamefont
  {M.}~\bibnamefont {Wanko}}, \bibinfo {author} {\bibfnamefont
  {S.}~\bibnamefont {Cahangirov}}, \bibinfo {author} {\bibfnamefont
  {A.}~\bibnamefont {Rubio}},  \emph {et~al.},\ }\bibfield  {title} {\enquote
  {\bibinfo {title} {Confined linear carbon chains as a route to bulk
  carbyne},}\ }\href@noop {} {\bibfield  {journal} {\bibinfo  {journal} {Nature
  Materials}\ }\textbf {\bibinfo {volume} {15}},\ \bibinfo {pages} {634--639}
  (\bibinfo {year} {2016})}\BibitemShut {NoStop}%
\bibitem [{\citenamefont {Shi}\ \emph {et~al.}(2017)\citenamefont {Shi},
  \citenamefont {Rohringer}, \citenamefont {Wanko}, \citenamefont {Rubio},
  \citenamefont {Wa{\ss}erroth}, \citenamefont {Reich}, \citenamefont
  {Cambr{\'e}}, \citenamefont {Wenseleers}, \citenamefont {Ayala},\ and\
  \citenamefont {Pichler}}]{Cchain_5}%
  \BibitemOpen
  \bibfield  {author} {\bibinfo {author} {\bibfnamefont {L.}~\bibnamefont
  {Shi}}, \bibinfo {author} {\bibfnamefont {P.}~\bibnamefont {Rohringer}},
  \bibinfo {author} {\bibfnamefont {M.}~\bibnamefont {Wanko}}, \bibinfo
  {author} {\bibfnamefont {A.}~\bibnamefont {Rubio}}, \bibinfo {author}
  {\bibfnamefont {S.}~\bibnamefont {Wa{\ss}erroth}}, \bibinfo {author}
  {\bibfnamefont {S.}~\bibnamefont {Reich}}, \bibinfo {author} {\bibfnamefont
  {S.}~\bibnamefont {Cambr{\'e}}}, \bibinfo {author} {\bibfnamefont
  {W.}~\bibnamefont {Wenseleers}}, \bibinfo {author} {\bibfnamefont
  {P.}~\bibnamefont {Ayala}}, \ and\ \bibinfo {author} {\bibfnamefont
  {T.}~\bibnamefont {Pichler}},\ }\bibfield  {title} {\enquote {\bibinfo
  {title} {Electronic band gaps of confined linear carbon chains ranging from
  polyyne to carbyne},}\ }\href@noop {} {\bibfield  {journal} {\bibinfo
  {journal} {Physical Review Materials}\ }\textbf {\bibinfo {volume} {1}},\
  \bibinfo {pages} {075601} (\bibinfo {year} {2017})}\BibitemShut {NoStop}%
\bibitem [{\citenamefont {Heeg}\ \emph {et~al.}(2018)\citenamefont {Heeg},
  \citenamefont {Shi}, \citenamefont {Poulikakos}, \citenamefont {Pichler},\
  and\ \citenamefont {Novotny}}]{Cchain_6}%
  \BibitemOpen
  \bibfield  {author} {\bibinfo {author} {\bibfnamefont {S.}~\bibnamefont
  {Heeg}}, \bibinfo {author} {\bibfnamefont {L.}~\bibnamefont {Shi}}, \bibinfo
  {author} {\bibfnamefont {L.}~\bibnamefont {Poulikakos}}, \bibinfo {author}
  {\bibfnamefont {T.}~\bibnamefont {Pichler}}, \ and\ \bibinfo {author}
  {\bibfnamefont {L.}~\bibnamefont {Novotny}},\ }\bibfield  {title} {\enquote
  {\bibinfo {title} {Carbon nanotube chirality determines properties of
  encapsulated linear carbon chain},}\ }\href@noop {} {\bibfield  {journal}
  {\bibinfo  {journal} {Nano Letters}\ }\textbf {\bibinfo {volume} {18}},\
  \bibinfo {pages} {5426--5431} (\bibinfo {year} {2018})}\BibitemShut {NoStop}%
\bibitem [{\citenamefont {Tschannen}\ \emph {et~al.}(2020)\citenamefont
  {Tschannen}, \citenamefont {Gordeev}, \citenamefont {Reich}, \citenamefont
  {Shi}, \citenamefont {Pichler}, \citenamefont {Frimmer}, \citenamefont
  {Novotny},\ and\ \citenamefont {Heeg}}]{Cchain_7}%
  \BibitemOpen
  \bibfield  {author} {\bibinfo {author} {\bibfnamefont {C.}~\bibnamefont
  {Tschannen}}, \bibinfo {author} {\bibfnamefont {G.}~\bibnamefont {Gordeev}},
  \bibinfo {author} {\bibfnamefont {S.}~\bibnamefont {Reich}}, \bibinfo
  {author} {\bibfnamefont {L.}~\bibnamefont {Shi}}, \bibinfo {author}
  {\bibfnamefont {T.}~\bibnamefont {Pichler}}, \bibinfo {author} {\bibfnamefont
  {M.}~\bibnamefont {Frimmer}}, \bibinfo {author} {\bibfnamefont
  {L.}~\bibnamefont {Novotny}}, \ and\ \bibinfo {author} {\bibfnamefont
  {S.}~\bibnamefont {Heeg}},\ }\bibfield  {title} {\enquote {\bibinfo {title}
  {Raman scattering cross section of confined carbyne},}\ }\href@noop {}
  {\bibfield  {journal} {\bibinfo  {journal} {Nano Letters}\ }\textbf {\bibinfo
  {volume} {20}},\ \bibinfo {pages} {6750--6755} (\bibinfo {year}
  {2020})}\BibitemShut {NoStop}%
\bibitem [{\citenamefont {Pfeiffer}\ \emph {et~al.}(2003)\citenamefont
  {Pfeiffer}, \citenamefont {Kuzmany}, \citenamefont {Kramberger},
  \citenamefont {Schaman}, \citenamefont {Pichler}, \citenamefont {Kataura},
  \citenamefont {Achiba}, \citenamefont {K{\"u}rti},\ and\ \citenamefont
  {Z{\'o}lyomi}}]{DW_1}%
  \BibitemOpen
  \bibfield  {author} {\bibinfo {author} {\bibfnamefont {R.}~\bibnamefont
  {Pfeiffer}}, \bibinfo {author} {\bibfnamefont {H.}~\bibnamefont {Kuzmany}},
  \bibinfo {author} {\bibfnamefont {C.}~\bibnamefont {Kramberger}}, \bibinfo
  {author} {\bibfnamefont {C.}~\bibnamefont {Schaman}}, \bibinfo {author}
  {\bibfnamefont {T.}~\bibnamefont {Pichler}}, \bibinfo {author} {\bibfnamefont
  {H.}~\bibnamefont {Kataura}}, \bibinfo {author} {\bibfnamefont
  {Y.}~\bibnamefont {Achiba}}, \bibinfo {author} {\bibfnamefont
  {J.}~\bibnamefont {K{\"u}rti}}, \ and\ \bibinfo {author} {\bibfnamefont
  {V.}~\bibnamefont {Z{\'o}lyomi}},\ }\bibfield  {title} {\enquote {\bibinfo
  {title} {Unusual high degree of unperturbed environment in the interior of
  single-wall carbon nanotubes},}\ }\href@noop {} {\bibfield  {journal}
  {\bibinfo  {journal} {Physical Review Letters}\ }\textbf {\bibinfo {volume}
  {90}},\ \bibinfo {pages} {225501} (\bibinfo {year} {2003})}\BibitemShut
  {NoStop}%
\bibitem [{\citenamefont {Guan}\ \emph {et~al.}(2005)\citenamefont {Guan},
  \citenamefont {Shi}, \citenamefont {Li},\ and\ \citenamefont {Gu}}]{DW_2}%
  \BibitemOpen
  \bibfield  {author} {\bibinfo {author} {\bibfnamefont {L.}~\bibnamefont
  {Guan}}, \bibinfo {author} {\bibfnamefont {Z.}~\bibnamefont {Shi}}, \bibinfo
  {author} {\bibfnamefont {M.}~\bibnamefont {Li}}, \ and\ \bibinfo {author}
  {\bibfnamefont {Z.}~\bibnamefont {Gu}},\ }\bibfield  {title} {\enquote
  {\bibinfo {title} {Ferrocene-filled single-walled carbon nanotubes},}\
  }\href@noop {} {\bibfield  {journal} {\bibinfo  {journal} {Carbon}\ }\textbf
  {\bibinfo {volume} {43}},\ \bibinfo {pages} {2780--2785} (\bibinfo {year}
  {2005})}\BibitemShut {NoStop}%
\bibitem [{\citenamefont {Fujita}, \citenamefont {Bandow},\ and\ \citenamefont
  {Iijima}(2005)}]{DW_3}%
  \BibitemOpen
  \bibfield  {author} {\bibinfo {author} {\bibfnamefont {Y.}~\bibnamefont
  {Fujita}}, \bibinfo {author} {\bibfnamefont {S.}~\bibnamefont {Bandow}}, \
  and\ \bibinfo {author} {\bibfnamefont {S.}~\bibnamefont {Iijima}},\
  }\bibfield  {title} {\enquote {\bibinfo {title} {Formation of small-diameter
  carbon nanotubes from ptcda arranged inside the single-wall carbon
  nanotubes},}\ }\href@noop {} {\bibfield  {journal} {\bibinfo  {journal}
  {Chemical Physics Letters}\ }\textbf {\bibinfo {volume} {413}},\ \bibinfo
  {pages} {410--414} (\bibinfo {year} {2005})}\BibitemShut {NoStop}%
\bibitem [{\citenamefont {Plank}\ \emph {et~al.}(2010)\citenamefont {Plank},
  \citenamefont {Pfeiffer}, \citenamefont {Schaman}, \citenamefont {Kuzmany},
  \citenamefont {Calvaresi}, \citenamefont {Zerbetto},\ and\ \citenamefont
  {Meyer}}]{DW_4}%
  \BibitemOpen
  \bibfield  {author} {\bibinfo {author} {\bibfnamefont {W.}~\bibnamefont
  {Plank}}, \bibinfo {author} {\bibfnamefont {R.}~\bibnamefont {Pfeiffer}},
  \bibinfo {author} {\bibfnamefont {C.}~\bibnamefont {Schaman}}, \bibinfo
  {author} {\bibfnamefont {H.}~\bibnamefont {Kuzmany}}, \bibinfo {author}
  {\bibfnamefont {M.}~\bibnamefont {Calvaresi}}, \bibinfo {author}
  {\bibfnamefont {F.}~\bibnamefont {Zerbetto}}, \ and\ \bibinfo {author}
  {\bibfnamefont {J.}~\bibnamefont {Meyer}},\ }\bibfield  {title} {\enquote
  {\bibinfo {title} {Electronic structure of carbon nanotubes with ultrahigh
  curvature},}\ }\href@noop {} {\bibfield  {journal} {\bibinfo  {journal} {ACS
  Nano}\ }\textbf {\bibinfo {volume} {4}},\ \bibinfo {pages} {4515--4522}
  (\bibinfo {year} {2010})}\BibitemShut {NoStop}%
\bibitem [{\citenamefont {Talyzin}\ \emph {et~al.}(2011)\citenamefont
  {Talyzin}, \citenamefont {Anoshkin}, \citenamefont {Krasheninnikov},
  \citenamefont {Nieminen}, \citenamefont {Nasibulin}, \citenamefont {Jiang},\
  and\ \citenamefont {Kauppinen}}]{GNR_1}%
  \BibitemOpen
  \bibfield  {author} {\bibinfo {author} {\bibfnamefont {A.}~\bibnamefont
  {Talyzin}}, \bibinfo {author} {\bibfnamefont {I.}~\bibnamefont {Anoshkin}},
  \bibinfo {author} {\bibfnamefont {A.}~\bibnamefont {Krasheninnikov}},
  \bibinfo {author} {\bibfnamefont {R.}~\bibnamefont {Nieminen}}, \bibinfo
  {author} {\bibfnamefont {A.}~\bibnamefont {Nasibulin}}, \bibinfo {author}
  {\bibfnamefont {H.}~\bibnamefont {Jiang}}, \ and\ \bibinfo {author}
  {\bibfnamefont {E.}~\bibnamefont {Kauppinen}},\ }\bibfield  {title} {\enquote
  {\bibinfo {title} {Synthesis of graphene nanoribbons encapsulated in
  single-walled carbon nanotubes},}\ }\href@noop {} {\bibfield  {journal}
  {\bibinfo  {journal} {Nano Letters}\ }\textbf {\bibinfo {volume} {11}},\
  \bibinfo {pages} {4352--4356} (\bibinfo {year} {2011})}\BibitemShut {NoStop}%
\bibitem [{\citenamefont {Chamberlain}\ \emph {et~al.}(2012)\citenamefont
  {Chamberlain}, \citenamefont {Biskupek}, \citenamefont {Rance}, \citenamefont
  {Chuvilin}, \citenamefont {Alexander}, \citenamefont {Bichoutskaia},
  \citenamefont {Kaiser},\ and\ \citenamefont {Khlobystov}}]{GNR_2}%
  \BibitemOpen
  \bibfield  {author} {\bibinfo {author} {\bibfnamefont {T.}~\bibnamefont
  {Chamberlain}}, \bibinfo {author} {\bibfnamefont {J.}~\bibnamefont
  {Biskupek}}, \bibinfo {author} {\bibfnamefont {G.}~\bibnamefont {Rance}},
  \bibinfo {author} {\bibfnamefont {A.}~\bibnamefont {Chuvilin}}, \bibinfo
  {author} {\bibfnamefont {T.}~\bibnamefont {Alexander}}, \bibinfo {author}
  {\bibfnamefont {E.}~\bibnamefont {Bichoutskaia}}, \bibinfo {author}
  {\bibfnamefont {U.}~\bibnamefont {Kaiser}}, \ and\ \bibinfo {author}
  {\bibfnamefont {A.}~\bibnamefont {Khlobystov}},\ }\bibfield  {title}
  {\enquote {\bibinfo {title} {Size, structure, and helical twist of graphene
  nanoribbons controlled by confinement in carbon nanotubes},}\ }\href@noop {}
  {\bibfield  {journal} {\bibinfo  {journal} {ACS Nano}\ }\textbf {\bibinfo
  {volume} {6}},\ \bibinfo {pages} {3943--3953} (\bibinfo {year}
  {2012})}\BibitemShut {NoStop}%
\bibitem [{\citenamefont {Fujihara}\ \emph {et~al.}(2012)\citenamefont
  {Fujihara}, \citenamefont {Miyata}, \citenamefont {Kitaura}, \citenamefont
  {Nishimura}, \citenamefont {Camacho}, \citenamefont {Irle}, \citenamefont
  {Iizumi}, \citenamefont {Okazaki},\ and\ \citenamefont {Shinohara}}]{GNR_3}%
  \BibitemOpen
  \bibfield  {author} {\bibinfo {author} {\bibfnamefont {M.}~\bibnamefont
  {Fujihara}}, \bibinfo {author} {\bibfnamefont {Y.}~\bibnamefont {Miyata}},
  \bibinfo {author} {\bibfnamefont {R.}~\bibnamefont {Kitaura}}, \bibinfo
  {author} {\bibfnamefont {Y.}~\bibnamefont {Nishimura}}, \bibinfo {author}
  {\bibfnamefont {C.}~\bibnamefont {Camacho}}, \bibinfo {author} {\bibfnamefont
  {S.}~\bibnamefont {Irle}}, \bibinfo {author} {\bibfnamefont {Y.}~\bibnamefont
  {Iizumi}}, \bibinfo {author} {\bibfnamefont {T.}~\bibnamefont {Okazaki}}, \
  and\ \bibinfo {author} {\bibfnamefont {H.}~\bibnamefont {Shinohara}},\
  }\bibfield  {title} {\enquote {\bibinfo {title} {Dimerization-initiated
  preferential formation of coronene-based graphene nanoribbons in carbon
  nanotubes},}\ }\href@noop {} {\bibfield  {journal} {\bibinfo  {journal} {The
  Journal of Physical Chemistry C}\ }\textbf {\bibinfo {volume} {116}},\
  \bibinfo {pages} {15141--15145} (\bibinfo {year} {2012})}\BibitemShut
  {NoStop}%
\bibitem [{\citenamefont {Fedotov}\ \emph {et~al.}(2013)\citenamefont
  {Fedotov}, \citenamefont {Chernov}, \citenamefont {Talyzin}, \citenamefont
  {Anoshkin}, \citenamefont {Nasibulin}, \citenamefont {Kauppinen},\ and\
  \citenamefont {Obraztsova}}]{GNR_4}%
  \BibitemOpen
  \bibfield  {author} {\bibinfo {author} {\bibfnamefont {P.}~\bibnamefont
  {Fedotov}}, \bibinfo {author} {\bibfnamefont {A.}~\bibnamefont {Chernov}},
  \bibinfo {author} {\bibfnamefont {A.}~\bibnamefont {Talyzin}}, \bibinfo
  {author} {\bibfnamefont {I.}~\bibnamefont {Anoshkin}}, \bibinfo {author}
  {\bibfnamefont {A.}~\bibnamefont {Nasibulin}}, \bibinfo {author}
  {\bibfnamefont {E.}~\bibnamefont {Kauppinen}}, \ and\ \bibinfo {author}
  {\bibfnamefont {E.}~\bibnamefont {Obraztsova}},\ }\bibfield  {title}
  {\enquote {\bibinfo {title} {Optical study of nanotube and coronene
  composites},}\ }\href@noop {} {\bibfield  {journal} {\bibinfo  {journal}
  {Journal of Nanoelectronics and Optoelectronics}\ }\textbf {\bibinfo {volume}
  {8}},\ \bibinfo {pages} {16--22} (\bibinfo {year} {2013})}\BibitemShut
  {NoStop}%
\bibitem [{\citenamefont {Chernov}\ \emph {et~al.}(2013)\citenamefont
  {Chernov}, \citenamefont {Fedotov}, \citenamefont {Talyzin}, \citenamefont
  {Suarez~Lopez}, \citenamefont {Anoshkin}, \citenamefont {Nasibulin},
  \citenamefont {Kauppinen},\ and\ \citenamefont {Obraztsova}}]{GNR_5}%
  \BibitemOpen
  \bibfield  {author} {\bibinfo {author} {\bibfnamefont {A.}~\bibnamefont
  {Chernov}}, \bibinfo {author} {\bibfnamefont {P.}~\bibnamefont {Fedotov}},
  \bibinfo {author} {\bibfnamefont {A.}~\bibnamefont {Talyzin}}, \bibinfo
  {author} {\bibfnamefont {I.}~\bibnamefont {Suarez~Lopez}}, \bibinfo {author}
  {\bibfnamefont {I.}~\bibnamefont {Anoshkin}}, \bibinfo {author}
  {\bibfnamefont {A.}~\bibnamefont {Nasibulin}}, \bibinfo {author}
  {\bibfnamefont {E.}~\bibnamefont {Kauppinen}}, \ and\ \bibinfo {author}
  {\bibfnamefont {E.}~\bibnamefont {Obraztsova}},\ }\bibfield  {title}
  {\enquote {\bibinfo {title} {Optical properties of graphene nanoribbons
  encapsulated in single-walled carbon nanotubes},}\ }\href@noop {} {\bibfield
  {journal} {\bibinfo  {journal} {ACS Nano}\ }\textbf {\bibinfo {volume} {7}},\
  \bibinfo {pages} {6346--6353} (\bibinfo {year} {2013})}\BibitemShut {NoStop}%
\bibitem [{\citenamefont {Lim}\ \emph {et~al.}(2013)\citenamefont {Lim},
  \citenamefont {Miyata}, \citenamefont {Kitaura}, \citenamefont {Nishimura},
  \citenamefont {Nishimoto}, \citenamefont {Irle}, \citenamefont {Warner},
  \citenamefont {Kataura},\ and\ \citenamefont {Shinohara}}]{GNR_7}%
  \BibitemOpen
  \bibfield  {author} {\bibinfo {author} {\bibfnamefont {H.}~\bibnamefont
  {Lim}}, \bibinfo {author} {\bibfnamefont {Y.}~\bibnamefont {Miyata}},
  \bibinfo {author} {\bibfnamefont {R.}~\bibnamefont {Kitaura}}, \bibinfo
  {author} {\bibfnamefont {Y.}~\bibnamefont {Nishimura}}, \bibinfo {author}
  {\bibfnamefont {Y.}~\bibnamefont {Nishimoto}}, \bibinfo {author}
  {\bibfnamefont {S.}~\bibnamefont {Irle}}, \bibinfo {author} {\bibfnamefont
  {J.}~\bibnamefont {Warner}}, \bibinfo {author} {\bibfnamefont
  {H.}~\bibnamefont {Kataura}}, \ and\ \bibinfo {author} {\bibfnamefont
  {H.}~\bibnamefont {Shinohara}},\ }\bibfield  {title} {\enquote {\bibinfo
  {title} {Growth of carbon nanotubes via twisted graphene nanoribbons},}\
  }\href@noop {} {\bibfield  {journal} {\bibinfo  {journal} {Nature
  Communications}\ }\textbf {\bibinfo {volume} {4}},\ \bibinfo {pages} {1--7}
  (\bibinfo {year} {2013})}\BibitemShut {NoStop}%
\bibitem [{\citenamefont {Anoshkin}\ \emph {et~al.}(2014)\citenamefont
  {Anoshkin}, \citenamefont {Talyzin}, \citenamefont {Nasibulin}, \citenamefont
  {Krasheninnikov}, \citenamefont {Jiang}, \citenamefont {Nieminen},\ and\
  \citenamefont {Kauppinen}}]{GNR_8}%
  \BibitemOpen
  \bibfield  {author} {\bibinfo {author} {\bibfnamefont {I.}~\bibnamefont
  {Anoshkin}}, \bibinfo {author} {\bibfnamefont {A.}~\bibnamefont {Talyzin}},
  \bibinfo {author} {\bibfnamefont {A.}~\bibnamefont {Nasibulin}}, \bibinfo
  {author} {\bibfnamefont {A.}~\bibnamefont {Krasheninnikov}}, \bibinfo
  {author} {\bibfnamefont {H.}~\bibnamefont {Jiang}}, \bibinfo {author}
  {\bibfnamefont {R.}~\bibnamefont {Nieminen}}, \ and\ \bibinfo {author}
  {\bibfnamefont {E.}~\bibnamefont {Kauppinen}},\ }\bibfield  {title} {\enquote
  {\bibinfo {title} {Coronene encapsulation in single-walled carbon nanotubes:
  Stacked columns, peapods, and nanoribbons},}\ }\href@noop {} {\bibfield
  {journal} {\bibinfo  {journal} {ChemPhysChem}\ }\textbf {\bibinfo {volume}
  {15}},\ \bibinfo {pages} {1660--1665} (\bibinfo {year} {2014})}\BibitemShut
  {NoStop}%
\bibitem [{\citenamefont {Lim}\ \emph {et~al.}(2015)\citenamefont {Lim},
  \citenamefont {Miyata}, \citenamefont {Fujihara}, \citenamefont {Okada},
  \citenamefont {Liu}, \citenamefont {Sato}, \citenamefont {Omachi},
  \citenamefont {Kitaura}, \citenamefont {Irle}, \citenamefont {Suenaga} \emph
  {et~al.}}]{GNR_9}%
  \BibitemOpen
  \bibfield  {author} {\bibinfo {author} {\bibfnamefont {H.}~\bibnamefont
  {Lim}}, \bibinfo {author} {\bibfnamefont {Y.}~\bibnamefont {Miyata}},
  \bibinfo {author} {\bibfnamefont {M.}~\bibnamefont {Fujihara}}, \bibinfo
  {author} {\bibfnamefont {S.}~\bibnamefont {Okada}}, \bibinfo {author}
  {\bibfnamefont {Z.}~\bibnamefont {Liu}}, \bibinfo {author} {\bibfnamefont
  {K.}~\bibnamefont {Sato}}, \bibinfo {author} {\bibfnamefont {H.}~\bibnamefont
  {Omachi}}, \bibinfo {author} {\bibfnamefont {R.}~\bibnamefont {Kitaura}},
  \bibinfo {author} {\bibfnamefont {S.}~\bibnamefont {Irle}}, \bibinfo {author}
  {\bibfnamefont {K.}~\bibnamefont {Suenaga}},  \emph {et~al.},\ }\bibfield
  {title} {\enquote {\bibinfo {title} {Fabrication and optical probing of
  highly extended, ultrathin graphene nanoribbons in carbon nanotubes},}\
  }\href@noop {} {\bibfield  {journal} {\bibinfo  {journal} {ACS Nano}\
  }\textbf {\bibinfo {volume} {9}},\ \bibinfo {pages} {5034--5040} (\bibinfo
  {year} {2015})}\BibitemShut {NoStop}%
\bibitem [{\citenamefont {Botos}\ \emph {et~al.}(2016)\citenamefont {Botos},
  \citenamefont {Biskupek}, \citenamefont {Chamberlain}, \citenamefont {Rance},
  \citenamefont {Stoppiello}, \citenamefont {Sloan}, \citenamefont {Liu},
  \citenamefont {Suenaga}, \citenamefont {Kaiser},\ and\ \citenamefont
  {Khlobystov}}]{GNR_10}%
  \BibitemOpen
  \bibfield  {author} {\bibinfo {author} {\bibfnamefont {A.}~\bibnamefont
  {Botos}}, \bibinfo {author} {\bibfnamefont {J.}~\bibnamefont {Biskupek}},
  \bibinfo {author} {\bibfnamefont {T.}~\bibnamefont {Chamberlain}}, \bibinfo
  {author} {\bibfnamefont {G.}~\bibnamefont {Rance}}, \bibinfo {author}
  {\bibfnamefont {C.}~\bibnamefont {Stoppiello}}, \bibinfo {author}
  {\bibfnamefont {J.}~\bibnamefont {Sloan}}, \bibinfo {author} {\bibfnamefont
  {Z.}~\bibnamefont {Liu}}, \bibinfo {author} {\bibfnamefont {K.}~\bibnamefont
  {Suenaga}}, \bibinfo {author} {\bibfnamefont {U.}~\bibnamefont {Kaiser}}, \
  and\ \bibinfo {author} {\bibfnamefont {A.}~\bibnamefont {Khlobystov}},\
  }\bibfield  {title} {\enquote {\bibinfo {title} {Carbon nanotubes as
  electrically active nanoreactors for multi-step inorganic synthesis:
  sequential transformations of molecules to nanoclusters and nanoclusters to
  nanoribbons},}\ }\href@noop {} {\bibfield  {journal} {\bibinfo  {journal}
  {Journal of the American Chemical Society}\ }\textbf {\bibinfo {volume}
  {138}},\ \bibinfo {pages} {8175--8183} (\bibinfo {year} {2016})}\BibitemShut
  {NoStop}%
\bibitem [{\citenamefont {Kuzmany}\ \emph {et~al.}(2017)\citenamefont
  {Kuzmany}, \citenamefont {Shi}, \citenamefont {K{\"u}rti}, \citenamefont
  {Koltai}, \citenamefont {Chuvilin}, \citenamefont {Saito},\ and\
  \citenamefont {Pichler}}]{GNR_11}%
  \BibitemOpen
  \bibfield  {author} {\bibinfo {author} {\bibfnamefont {H.}~\bibnamefont
  {Kuzmany}}, \bibinfo {author} {\bibfnamefont {L.}~\bibnamefont {Shi}},
  \bibinfo {author} {\bibfnamefont {J.}~\bibnamefont {K{\"u}rti}}, \bibinfo
  {author} {\bibfnamefont {J.}~\bibnamefont {Koltai}}, \bibinfo {author}
  {\bibfnamefont {A.}~\bibnamefont {Chuvilin}}, \bibinfo {author}
  {\bibfnamefont {T.}~\bibnamefont {Saito}}, \ and\ \bibinfo {author}
  {\bibfnamefont {T.}~\bibnamefont {Pichler}},\ }\bibfield  {title} {\enquote
  {\bibinfo {title} {The growth of new extended carbon nanophases from
  ferrocene inside single-walled carbon nanotubes},}\ }\href@noop {} {\bibfield
   {journal} {\bibinfo  {journal} {Physica Status Solidi (RRL)--Rapid Research
  Letters}\ }\textbf {\bibinfo {volume} {11}},\ \bibinfo {pages} {1700158}
  (\bibinfo {year} {2017})}\BibitemShut {NoStop}%
\bibitem [{\citenamefont {Kuzmany}\ \emph
  {et~al.}(2021{\natexlab{a}})\citenamefont {Kuzmany}, \citenamefont {Shi},
  \citenamefont {Martinati}, \citenamefont {Cambr{\'e}}, \citenamefont
  {Wenseleers}, \citenamefont {K{\"u}rti}, \citenamefont {Koltai},
  \citenamefont {Kukucska}, \citenamefont {Cao}, \citenamefont {Kaiser} \emph
  {et~al.}}]{GNR_12}%
  \BibitemOpen
  \bibfield  {author} {\bibinfo {author} {\bibfnamefont {H.}~\bibnamefont
  {Kuzmany}}, \bibinfo {author} {\bibfnamefont {L.}~\bibnamefont {Shi}},
  \bibinfo {author} {\bibfnamefont {M.}~\bibnamefont {Martinati}}, \bibinfo
  {author} {\bibfnamefont {S.}~\bibnamefont {Cambr{\'e}}}, \bibinfo {author}
  {\bibfnamefont {W.}~\bibnamefont {Wenseleers}}, \bibinfo {author}
  {\bibfnamefont {J.}~\bibnamefont {K{\"u}rti}}, \bibinfo {author}
  {\bibfnamefont {J.}~\bibnamefont {Koltai}}, \bibinfo {author} {\bibfnamefont
  {G.}~\bibnamefont {Kukucska}}, \bibinfo {author} {\bibfnamefont
  {K.}~\bibnamefont {Cao}}, \bibinfo {author} {\bibfnamefont {U.}~\bibnamefont
  {Kaiser}},  \emph {et~al.},\ }\bibfield  {title} {\enquote {\bibinfo {title}
  {Well-defined sub-nanometer graphene ribbons synthesized inside carbon
  nanotubes},}\ }\href@noop {} {\bibfield  {journal} {\bibinfo  {journal}
  {Carbon}\ }\textbf {\bibinfo {volume} {171}},\ \bibinfo {pages} {221--229}
  (\bibinfo {year} {2021}{\natexlab{a}})}\BibitemShut {NoStop}%
\bibitem [{\citenamefont {Martini}\ \emph {et~al.}(2019)\citenamefont
  {Martini}, \citenamefont {Chen}, \citenamefont {Mishra}, \citenamefont
  {Barin}, \citenamefont {Fantuzzi}, \citenamefont {Ruffieux}, \citenamefont
  {Fasel}, \citenamefont {Feng}, \citenamefont {Narita}, \citenamefont
  {Coletti} \emph {et~al.}}]{GNR_13}%
  \BibitemOpen
  \bibfield  {author} {\bibinfo {author} {\bibfnamefont {L.}~\bibnamefont
  {Martini}}, \bibinfo {author} {\bibfnamefont {Z.}~\bibnamefont {Chen}},
  \bibinfo {author} {\bibfnamefont {N.}~\bibnamefont {Mishra}}, \bibinfo
  {author} {\bibfnamefont {G.}~\bibnamefont {Barin}}, \bibinfo {author}
  {\bibfnamefont {P.}~\bibnamefont {Fantuzzi}}, \bibinfo {author}
  {\bibfnamefont {P.}~\bibnamefont {Ruffieux}}, \bibinfo {author}
  {\bibfnamefont {R.}~\bibnamefont {Fasel}}, \bibinfo {author} {\bibfnamefont
  {X.}~\bibnamefont {Feng}}, \bibinfo {author} {\bibfnamefont {A.}~\bibnamefont
  {Narita}}, \bibinfo {author} {\bibfnamefont {C.}~\bibnamefont {Coletti}},
  \emph {et~al.},\ }\bibfield  {title} {\enquote {\bibinfo {title}
  {Structure-dependent electrical properties of graphene nanoribbon devices
  with graphene electrodes},}\ }\href@noop {} {\bibfield  {journal} {\bibinfo
  {journal} {Carbon}\ }\textbf {\bibinfo {volume} {146}},\ \bibinfo {pages}
  {36--43} (\bibinfo {year} {2019})}\BibitemShut {NoStop}%
\bibitem [{\citenamefont {Cadena}\ \emph {et~al.}(2022)\citenamefont {Cadena},
  \citenamefont {Botka}, \citenamefont {Pekker}, \citenamefont {Tschannen},
  \citenamefont {Lombardo}, \citenamefont {Novotny}, \citenamefont
  {Khlobystov},\ and\ \citenamefont {Kamar{\'a}s}}]{cadena2022molecular}%
  \BibitemOpen
  \bibfield  {author} {\bibinfo {author} {\bibfnamefont {A.}~\bibnamefont
  {Cadena}}, \bibinfo {author} {\bibfnamefont {B.}~\bibnamefont {Botka}},
  \bibinfo {author} {\bibfnamefont {{\'A}.}~\bibnamefont {Pekker}}, \bibinfo
  {author} {\bibfnamefont {C.}~\bibnamefont {Tschannen}}, \bibinfo {author}
  {\bibfnamefont {C.}~\bibnamefont {Lombardo}}, \bibinfo {author}
  {\bibfnamefont {L.}~\bibnamefont {Novotny}}, \bibinfo {author} {\bibfnamefont
  {A.}~\bibnamefont {Khlobystov}}, \ and\ \bibinfo {author} {\bibfnamefont
  {K.}~\bibnamefont {Kamar{\'a}s}},\ }\bibfield  {title} {\enquote {\bibinfo
  {title} {Molecular encapsulation from the liquid phase and graphene
  nanoribbon growth in carbon nanotubes},}\ }\href@noop {} {\bibfield
  {journal} {\bibinfo  {journal} {The Journal of Physical Chemistry Letters}\
  }\textbf {\bibinfo {volume} {13}},\ \bibinfo {pages} {9752--9758} (\bibinfo
  {year} {2022})}\BibitemShut {NoStop}%
\bibitem [{\citenamefont {Cadena}\ \emph {et~al.}(2023)\citenamefont {Cadena},
  \citenamefont {Pekker}, \citenamefont {Botka}, \citenamefont {Dodony},
  \citenamefont {Fogarassy}, \citenamefont {P{\'e}cz},\ and\ \citenamefont
  {Kamar{\'a}s}}]{cadena2023encapsulation}%
  \BibitemOpen
  \bibfield  {author} {\bibinfo {author} {\bibfnamefont {A.}~\bibnamefont
  {Cadena}}, \bibinfo {author} {\bibfnamefont {{\'A}.}~\bibnamefont {Pekker}},
  \bibinfo {author} {\bibfnamefont {B.}~\bibnamefont {Botka}}, \bibinfo
  {author} {\bibfnamefont {E.}~\bibnamefont {Dodony}}, \bibinfo {author}
  {\bibfnamefont {Z.}~\bibnamefont {Fogarassy}}, \bibinfo {author}
  {\bibfnamefont {B.}~\bibnamefont {P{\'e}cz}}, \ and\ \bibinfo {author}
  {\bibfnamefont {K.}~\bibnamefont {Kamar{\'a}s}},\ }\bibfield  {title}
  {\enquote {\bibinfo {title} {Encapsulation of the graphene nanoribbon
  precursor 1, 2, 4-trichlorobenzene in boron nitride nanotubes at room
  temperature},}\ }\href@noop {} {\bibfield  {journal} {\bibinfo  {journal}
  {Physica Status Solidi (RRL)--Rapid Research Letters}\ }\textbf {\bibinfo
  {volume} {17}},\ \bibinfo {pages} {2200284} (\bibinfo {year}
  {2023})}\BibitemShut {NoStop}%
\bibitem [{\citenamefont {Han}\ \emph {et~al.}(2004)\citenamefont {Han},
  \citenamefont {Yoon}, \citenamefont {Berber}, \citenamefont {Park},
  \citenamefont {Osawa}, \citenamefont {Ihm},\ and\ \citenamefont
  {Tom{\'a}nek}}]{Tomanek2004_PRB}%
  \BibitemOpen
  \bibfield  {author} {\bibinfo {author} {\bibfnamefont {S.}~\bibnamefont
  {Han}}, \bibinfo {author} {\bibfnamefont {M.}~\bibnamefont {Yoon}}, \bibinfo
  {author} {\bibfnamefont {S.}~\bibnamefont {Berber}}, \bibinfo {author}
  {\bibfnamefont {N.}~\bibnamefont {Park}}, \bibinfo {author} {\bibfnamefont
  {E.}~\bibnamefont {Osawa}}, \bibinfo {author} {\bibfnamefont
  {J.}~\bibnamefont {Ihm}}, \ and\ \bibinfo {author} {\bibfnamefont
  {D.}~\bibnamefont {Tom{\'a}nek}},\ }\bibfield  {title} {\enquote {\bibinfo
  {title} {Microscopic mechanism of fullerene fusion},}\ }\href@noop {}
  {\bibfield  {journal} {\bibinfo  {journal} {Physical Review B}\ }\textbf
  {\bibinfo {volume} {70}},\ \bibinfo {pages} {113402} (\bibinfo {year}
  {2004})}\BibitemShut {NoStop}%
\bibitem [{\citenamefont {Nishio}\ \emph {et~al.}(2008)\citenamefont {Nishio},
  \citenamefont {Ozaki}, \citenamefont {Morishita},\ and\ \citenamefont
  {Mikami}}]{nishio2008formation}%
  \BibitemOpen
  \bibfield  {author} {\bibinfo {author} {\bibfnamefont {K.}~\bibnamefont
  {Nishio}}, \bibinfo {author} {\bibfnamefont {T.}~\bibnamefont {Ozaki}},
  \bibinfo {author} {\bibfnamefont {T.}~\bibnamefont {Morishita}}, \ and\
  \bibinfo {author} {\bibfnamefont {M.}~\bibnamefont {Mikami}},\ }\bibfield
  {title} {\enquote {\bibinfo {title} {Formation of silicon-fullerene-linked
  nanowires inside carbon nanotubes: A molecular-dynamics and first-principles
  study},}\ }\href@noop {} {\bibfield  {journal} {\bibinfo  {journal} {Physical
  Review B}\ }\textbf {\bibinfo {volume} {77}},\ \bibinfo {pages} {201401}
  (\bibinfo {year} {2008})}\BibitemShut {NoStop}%
\bibitem [{\citenamefont {Tersoff}(1988)}]{tersoff1988empirical}%
  \BibitemOpen
  \bibfield  {author} {\bibinfo {author} {\bibfnamefont {J.}~\bibnamefont
  {Tersoff}},\ }\bibfield  {title} {\enquote {\bibinfo {title} {Empirical
  interatomic potential for carbon, with applications to amorphous carbon},}\
  }\href@noop {} {\bibfield  {journal} {\bibinfo  {journal} {Physical Review
  Letters}\ }\textbf {\bibinfo {volume} {61}},\ \bibinfo {pages} {2879}
  (\bibinfo {year} {1988})}\BibitemShut {NoStop}%
\bibitem [{\citenamefont {Tersoff}(1989)}]{tersoff1989modeling}%
  \BibitemOpen
  \bibfield  {author} {\bibinfo {author} {\bibfnamefont {J.}~\bibnamefont
  {Tersoff}},\ }\bibfield  {title} {\enquote {\bibinfo {title} {Modeling
  solid-state chemistry: Interatomic potentials for multicomponent systems},}\
  }\href@noop {} {\bibfield  {journal} {\bibinfo  {journal} {Physical Review
  B}\ }\textbf {\bibinfo {volume} {39}},\ \bibinfo {pages} {5566} (\bibinfo
  {year} {1989})}\BibitemShut {NoStop}%
\bibitem [{\citenamefont {Z{\'o}lyomi}\ \emph {et~al.}(2010)\citenamefont
  {Z{\'o}lyomi}, \citenamefont {Koltai}, \citenamefont {Visontai},
  \citenamefont {Oroszl{\'a}ny}, \citenamefont {Ruszny{\'a}k}, \citenamefont
  {L{\'a}szl{\'o}},\ and\ \citenamefont
  {K{\"u}rti}}]{zolyomi2010characteristics}%
  \BibitemOpen
  \bibfield  {author} {\bibinfo {author} {\bibfnamefont {V.}~\bibnamefont
  {Z{\'o}lyomi}}, \bibinfo {author} {\bibfnamefont {J.}~\bibnamefont {Koltai}},
  \bibinfo {author} {\bibfnamefont {D.}~\bibnamefont {Visontai}}, \bibinfo
  {author} {\bibfnamefont {L.}~\bibnamefont {Oroszl{\'a}ny}}, \bibinfo {author}
  {\bibfnamefont {{\'A}.}~\bibnamefont {Ruszny{\'a}k}}, \bibinfo {author}
  {\bibfnamefont {I.}~\bibnamefont {L{\'a}szl{\'o}}}, \ and\ \bibinfo {author}
  {\bibfnamefont {J.}~\bibnamefont {K{\"u}rti}},\ }\bibfield  {title} {\enquote
  {\bibinfo {title} {Characteristics of bamboo defects in peapod-grown
  double-walled carbon nanotubes},}\ }\href@noop {} {\bibfield  {journal}
  {\bibinfo  {journal} {Physical Review B}\ }\textbf {\bibinfo {volume} {82}},\
  \bibinfo {pages} {195423} (\bibinfo {year} {2010})}\BibitemShut {NoStop}%
\bibitem [{\citenamefont {Smith}, \citenamefont {Iyer},\ and\ \citenamefont
  {Corry}(2014)}]{smith2014confined}%
  \BibitemOpen
  \bibfield  {author} {\bibinfo {author} {\bibfnamefont {N.}~\bibnamefont
  {Smith}}, \bibinfo {author} {\bibfnamefont {K.}~\bibnamefont {Iyer}}, \ and\
  \bibinfo {author} {\bibfnamefont {B.}~\bibnamefont {Corry}},\ }\bibfield
  {title} {\enquote {\bibinfo {title} {The confined space inside carbon
  nanotubes can dictate the stereo-and regioselectivity of diels--alder
  reactions},}\ }\href@noop {} {\bibfield  {journal} {\bibinfo  {journal}
  {Physical Chemistry Chemical Physics}\ }\textbf {\bibinfo {volume} {16}},\
  \bibinfo {pages} {6986--6989} (\bibinfo {year} {2014})}\BibitemShut {NoStop}%
\bibitem [{\citenamefont {Calvaresi}\ and\ \citenamefont
  {Zerbetto}(2014)}]{Calvaresi2014_JMaterChemA}%
  \BibitemOpen
  \bibfield  {author} {\bibinfo {author} {\bibfnamefont {M.}~\bibnamefont
  {Calvaresi}}\ and\ \bibinfo {author} {\bibfnamefont {F.}~\bibnamefont
  {Zerbetto}},\ }\bibfield  {title} {\enquote {\bibinfo {title} {Atomistic
  molecular dynamics simulations reveal insights into adsorption, packing, and
  fluxes of molecules with carbon nanotubes},}\ }\href@noop {} {\bibfield
  {journal} {\bibinfo  {journal} {Journal of Materials Chemistry A}\ }\textbf
  {\bibinfo {volume} {2}},\ \bibinfo {pages} {12123--12135} (\bibinfo {year}
  {2014})}\BibitemShut {NoStop}%
\bibitem [{\citenamefont {Marforio}\ \emph {et~al.}(2017)\citenamefont
  {Marforio}, \citenamefont {Bottoni}, \citenamefont {Giacinto}, \citenamefont
  {Zerbetto},\ and\ \citenamefont {Calvaresi}}]{Marforio2017_JPhysChem}%
  \BibitemOpen
  \bibfield  {author} {\bibinfo {author} {\bibfnamefont {T.}~\bibnamefont
  {Marforio}}, \bibinfo {author} {\bibfnamefont {A.}~\bibnamefont {Bottoni}},
  \bibinfo {author} {\bibfnamefont {P.}~\bibnamefont {Giacinto}}, \bibinfo
  {author} {\bibfnamefont {F.}~\bibnamefont {Zerbetto}}, \ and\ \bibinfo
  {author} {\bibfnamefont {M.}~\bibnamefont {Calvaresi}},\ }\bibfield  {title}
  {\enquote {\bibinfo {title} {Aromatic bromination of n-phenylacetamide inside
  cnts. are cnts real nanoreactors controlling regioselectivity and kinetics? a
  qm/mm investigation},}\ }\href@noop {} {\bibfield  {journal} {\bibinfo
  {journal} {The Journal of Physical Chemistry C}\ }\textbf {\bibinfo {volume}
  {121}},\ \bibinfo {pages} {27674--27682} (\bibinfo {year}
  {2017})}\BibitemShut {NoStop}%
\bibitem [{\citenamefont {L{\'a}szl{\'o}}\ \emph {et~al.}(2017)\citenamefont
  {L{\'a}szl{\'o}}, \citenamefont {Gyimesi}, \citenamefont {Koltai},\ and\
  \citenamefont {K{\"u}rti}}]{laszlo2017molecular}%
  \BibitemOpen
  \bibfield  {author} {\bibinfo {author} {\bibfnamefont {I.}~\bibnamefont
  {L{\'a}szl{\'o}}}, \bibinfo {author} {\bibfnamefont {B.}~\bibnamefont
  {Gyimesi}}, \bibinfo {author} {\bibfnamefont {J.}~\bibnamefont {Koltai}}, \
  and\ \bibinfo {author} {\bibfnamefont {J.}~\bibnamefont {K{\"u}rti}},\
  }\bibfield  {title} {\enquote {\bibinfo {title} {Molecular dynamics
  simulation of carbon structures inside small diameter carbon nanotubes},}\
  }\href@noop {} {\bibfield  {journal} {\bibinfo  {journal} {Physica Status
  Solidi (b)}\ }\textbf {\bibinfo {volume} {254}},\ \bibinfo {pages} {1700206}
  (\bibinfo {year} {2017})}\BibitemShut {NoStop}%
\bibitem [{\citenamefont {Kuzmany}\ \emph
  {et~al.}(2021{\natexlab{b}})\citenamefont {Kuzmany}, \citenamefont {Shi},
  \citenamefont {Martinati}, \citenamefont {Cambr{\'e}}, \citenamefont
  {Wenseleers}, \citenamefont {K{\"u}rti}, \citenamefont {Koltai},
  \citenamefont {Kukucska}, \citenamefont {Cao}, \citenamefont {Kaiser} \emph
  {et~al.}}]{kuzmany2021well}%
  \BibitemOpen
  \bibfield  {author} {\bibinfo {author} {\bibfnamefont {H.}~\bibnamefont
  {Kuzmany}}, \bibinfo {author} {\bibfnamefont {L.}~\bibnamefont {Shi}},
  \bibinfo {author} {\bibfnamefont {M.}~\bibnamefont {Martinati}}, \bibinfo
  {author} {\bibfnamefont {S.}~\bibnamefont {Cambr{\'e}}}, \bibinfo {author}
  {\bibfnamefont {W.}~\bibnamefont {Wenseleers}}, \bibinfo {author}
  {\bibfnamefont {J.}~\bibnamefont {K{\"u}rti}}, \bibinfo {author}
  {\bibfnamefont {J.}~\bibnamefont {Koltai}}, \bibinfo {author} {\bibfnamefont
  {G.}~\bibnamefont {Kukucska}}, \bibinfo {author} {\bibfnamefont
  {K.}~\bibnamefont {Cao}}, \bibinfo {author} {\bibfnamefont {U.}~\bibnamefont
  {Kaiser}},  \emph {et~al.},\ }\bibfield  {title} {\enquote {\bibinfo {title}
  {Well-defined sub-nanometer graphene ribbons synthesized inside carbon
  nanotubes},}\ }\href@noop {} {\bibfield  {journal} {\bibinfo  {journal}
  {Carbon}\ }\textbf {\bibinfo {volume} {171}},\ \bibinfo {pages} {221--229}
  (\bibinfo {year} {2021}{\natexlab{b}})}\BibitemShut {NoStop}%
\bibitem [{\citenamefont {Plimpton}(1995)}]{LAMMPS1995}%
  \BibitemOpen
  \bibfield  {author} {\bibinfo {author} {\bibfnamefont {S.}~\bibnamefont
  {Plimpton}},\ }\bibfield  {title} {\enquote {\bibinfo {title} {Fast parallel
  algorithms for short-range molecular dynamics},}\ }\href@noop {} {\bibfield
  {journal} {\bibinfo  {journal} {Journal of Computational Physics}\ }\textbf
  {\bibinfo {volume} {117}},\ \bibinfo {pages} {1--19} (\bibinfo {year}
  {1995})}\BibitemShut {NoStop}%
\bibitem [{\citenamefont {Thompson}\ \emph {et~al.}(2022)\citenamefont
  {Thompson}, \citenamefont {Aktulga}, \citenamefont {Berger}, \citenamefont
  {Bolintineanu}, \citenamefont {Brown}, \citenamefont {Crozier}, \citenamefont
  {in't Veld}, \citenamefont {Kohlmeyer}, \citenamefont {Moore}, \citenamefont
  {Nguyen} \emph {et~al.}}]{LAMMPS2022}%
  \BibitemOpen
  \bibfield  {author} {\bibinfo {author} {\bibfnamefont {A.}~\bibnamefont
  {Thompson}}, \bibinfo {author} {\bibfnamefont {H.}~\bibnamefont {Aktulga}},
  \bibinfo {author} {\bibfnamefont {R.}~\bibnamefont {Berger}}, \bibinfo
  {author} {\bibfnamefont {D.}~\bibnamefont {Bolintineanu}}, \bibinfo {author}
  {\bibfnamefont {W.}~\bibnamefont {Brown}}, \bibinfo {author} {\bibfnamefont
  {P.}~\bibnamefont {Crozier}}, \bibinfo {author} {\bibfnamefont
  {P.}~\bibnamefont {in't Veld}}, \bibinfo {author} {\bibfnamefont
  {A.}~\bibnamefont {Kohlmeyer}}, \bibinfo {author} {\bibfnamefont
  {S.}~\bibnamefont {Moore}}, \bibinfo {author} {\bibfnamefont
  {T.}~\bibnamefont {Nguyen}},  \emph {et~al.},\ }\bibfield  {title} {\enquote
  {\bibinfo {title} {Lammps-a flexible simulation tool for particle-based
  materials modeling at the atomic, meso, and continuum scales},}\ }\href@noop
  {} {\bibfield  {journal} {\bibinfo  {journal} {Computer Physics
  Communications}\ }\textbf {\bibinfo {volume} {271}},\ \bibinfo {pages}
  {108171} (\bibinfo {year} {2022})}\BibitemShut {NoStop}%
\bibitem [{\citenamefont {Van~Duin}\ \emph {et~al.}(2001)\citenamefont
  {Van~Duin}, \citenamefont {Dasgupta}, \citenamefont {Lorant},\ and\
  \citenamefont {Goddard}}]{van2001reaxff}%
  \BibitemOpen
  \bibfield  {author} {\bibinfo {author} {\bibfnamefont {A.}~\bibnamefont
  {Van~Duin}}, \bibinfo {author} {\bibfnamefont {S.}~\bibnamefont {Dasgupta}},
  \bibinfo {author} {\bibfnamefont {F.}~\bibnamefont {Lorant}}, \ and\ \bibinfo
  {author} {\bibfnamefont {W.}~\bibnamefont {Goddard}},\ }\bibfield  {title}
  {\enquote {\bibinfo {title} {Reaxff: a reactive force field for
  hydrocarbons},}\ }\href@noop {} {\bibfield  {journal} {\bibinfo  {journal}
  {The Journal of Physical Chemistry A}\ }\textbf {\bibinfo {volume} {105}},\
  \bibinfo {pages} {9396--9409} (\bibinfo {year} {2001})}\BibitemShut {NoStop}%
\bibitem [{\citenamefont {Chenoweth}\ \emph {et~al.}(2008)\citenamefont
  {Chenoweth}, \citenamefont {Van~Duin}, \citenamefont {Persson}, \citenamefont
  {Cheng}, \citenamefont {Oxgaard},\ and\ \citenamefont
  {Goddard~Iii}}]{chenoweth2008development}%
  \BibitemOpen
  \bibfield  {author} {\bibinfo {author} {\bibfnamefont {K.}~\bibnamefont
  {Chenoweth}}, \bibinfo {author} {\bibfnamefont {A.}~\bibnamefont {Van~Duin}},
  \bibinfo {author} {\bibfnamefont {P.}~\bibnamefont {Persson}}, \bibinfo
  {author} {\bibfnamefont {M.}~\bibnamefont {Cheng}}, \bibinfo {author}
  {\bibfnamefont {J.}~\bibnamefont {Oxgaard}}, \ and\ \bibinfo {author}
  {\bibfnamefont {W.}~\bibnamefont {Goddard~Iii}},\ }\bibfield  {title}
  {\enquote {\bibinfo {title} {Development and application of a reaxff reactive
  force field for oxidative dehydrogenation on vanadium oxide catalysts},}\
  }\href@noop {} {\bibfield  {journal} {\bibinfo  {journal} {The Journal of
  Physical Chemistry C}\ }\textbf {\bibinfo {volume} {112}},\ \bibinfo {pages}
  {14645--14654} (\bibinfo {year} {2008})}\BibitemShut {NoStop}%
\bibitem [{\citenamefont {Wang}, \citenamefont {Shi},\ and\ \citenamefont
  {Gu}(2010)}]{wang2010chemistry}%
  \BibitemOpen
  \bibfield  {author} {\bibinfo {author} {\bibfnamefont {Z.}~\bibnamefont
  {Wang}}, \bibinfo {author} {\bibfnamefont {Z.}~\bibnamefont {Shi}}, \ and\
  \bibinfo {author} {\bibfnamefont {Z.}~\bibnamefont {Gu}},\ }\bibfield
  {title} {\enquote {\bibinfo {title} {Chemistry in the nanospace of carbon
  nanotubes},}\ }\href@noop {} {\bibfield  {journal} {\bibinfo  {journal}
  {Chemistry--An Asian Journal}\ }\textbf {\bibinfo {volume} {5}},\ \bibinfo
  {pages} {1030--1038} (\bibinfo {year} {2010})}\BibitemShut {NoStop}%
\bibitem [{\citenamefont {Li}\ \emph {et~al.}(2018)\citenamefont {Li},
  \citenamefont {Zhang}, \citenamefont {Li}, \citenamefont {Zhang},
  \citenamefont {Bouhadja}, \citenamefont {Liu}, \citenamefont {Skelton},\ and\
  \citenamefont {Barati}}]{li2018reaxff}%
  \BibitemOpen
  \bibfield  {author} {\bibinfo {author} {\bibfnamefont {K.}~\bibnamefont
  {Li}}, \bibinfo {author} {\bibfnamefont {H.}~\bibnamefont {Zhang}}, \bibinfo
  {author} {\bibfnamefont {G.}~\bibnamefont {Li}}, \bibinfo {author}
  {\bibfnamefont {J.}~\bibnamefont {Zhang}}, \bibinfo {author} {\bibfnamefont
  {M.}~\bibnamefont {Bouhadja}}, \bibinfo {author} {\bibfnamefont
  {Z.}~\bibnamefont {Liu}}, \bibinfo {author} {\bibfnamefont {A.}~\bibnamefont
  {Skelton}}, \ and\ \bibinfo {author} {\bibfnamefont {M.}~\bibnamefont
  {Barati}},\ }\bibfield  {title} {\enquote {\bibinfo {title} {Reaxff molecular
  dynamics simulation for the graphitization of amorphous carbon: a parametric
  study},}\ }\href@noop {} {\bibfield  {journal} {\bibinfo  {journal} {Journal
  of Chemical Theory and Computation}\ }\textbf {\bibinfo {volume} {14}},\
  \bibinfo {pages} {2322--2331} (\bibinfo {year} {2018})}\BibitemShut {NoStop}%
\bibitem [{\citenamefont {de~Tomas}, \citenamefont {Suarez-Martinez},\ and\
  \citenamefont {Marks}(2016)}]{de2016graphitization}%
  \BibitemOpen
  \bibfield  {author} {\bibinfo {author} {\bibfnamefont {C.}~\bibnamefont
  {de~Tomas}}, \bibinfo {author} {\bibfnamefont {I.}~\bibnamefont
  {Suarez-Martinez}}, \ and\ \bibinfo {author} {\bibfnamefont {N.}~\bibnamefont
  {Marks}},\ }\bibfield  {title} {\enquote {\bibinfo {title} {Graphitization of
  amorphous carbons: A comparative study of interatomic potentials},}\
  }\href@noop {} {\bibfield  {journal} {\bibinfo  {journal} {Carbon}\ }\textbf
  {\bibinfo {volume} {109}},\ \bibinfo {pages} {681--693} (\bibinfo {year}
  {2016})}\BibitemShut {NoStop}%
\bibitem [{\citenamefont {Kowalik}\ \emph {et~al.}(2019)\citenamefont
  {Kowalik}, \citenamefont {Ashraf}, \citenamefont {Damirchi}, \citenamefont
  {Akbarian}, \citenamefont {Rajabpour},\ and\ \citenamefont
  {Van~Duin}}]{kowalik2019atomistic}%
  \BibitemOpen
  \bibfield  {author} {\bibinfo {author} {\bibfnamefont {M.}~\bibnamefont
  {Kowalik}}, \bibinfo {author} {\bibfnamefont {C.}~\bibnamefont {Ashraf}},
  \bibinfo {author} {\bibfnamefont {B.}~\bibnamefont {Damirchi}}, \bibinfo
  {author} {\bibfnamefont {D.}~\bibnamefont {Akbarian}}, \bibinfo {author}
  {\bibfnamefont {S.}~\bibnamefont {Rajabpour}}, \ and\ \bibinfo {author}
  {\bibfnamefont {A.}~\bibnamefont {Van~Duin}},\ }\bibfield  {title} {\enquote
  {\bibinfo {title} {Atomistic scale analysis of the carbonization process for
  c/h/o/n-based polymers with the reaxff reactive force field},}\ }\href@noop
  {} {\bibfield  {journal} {\bibinfo  {journal} {The Journal of Physical
  Chemistry B}\ }\textbf {\bibinfo {volume} {123}},\ \bibinfo {pages}
  {5357--5367} (\bibinfo {year} {2019})}\BibitemShut {NoStop}%
\bibitem [{\citenamefont {Rajabpour}\ \emph {et~al.}(2021)\citenamefont
  {Rajabpour}, \citenamefont {Mao}, \citenamefont {Gao}, \citenamefont
  {Talkhoncheh}, \citenamefont {Zhu}, \citenamefont {Schwab}, \citenamefont
  {Kowalik}, \citenamefont {Li},\ and\ \citenamefont {van
  Duin}}]{rajabpour2021low}%
  \BibitemOpen
  \bibfield  {author} {\bibinfo {author} {\bibfnamefont {S.}~\bibnamefont
  {Rajabpour}}, \bibinfo {author} {\bibfnamefont {Q.}~\bibnamefont {Mao}},
  \bibinfo {author} {\bibfnamefont {Z.}~\bibnamefont {Gao}}, \bibinfo {author}
  {\bibfnamefont {M.}~\bibnamefont {Talkhoncheh}}, \bibinfo {author}
  {\bibfnamefont {J.}~\bibnamefont {Zhu}}, \bibinfo {author} {\bibfnamefont
  {Y.}~\bibnamefont {Schwab}}, \bibinfo {author} {\bibfnamefont
  {M.}~\bibnamefont {Kowalik}}, \bibinfo {author} {\bibfnamefont
  {X.}~\bibnamefont {Li}}, \ and\ \bibinfo {author} {\bibfnamefont
  {A.}~\bibnamefont {van Duin}},\ }\bibfield  {title} {\enquote {\bibinfo
  {title} {Low-temperature carbonization of polyacrylonitrile/graphene carbon
  fibers: A combined reaxff molecular dynamics and experimental study},}\
  }\href@noop {} {\bibfield  {journal} {\bibinfo  {journal} {Carbon}\ }\textbf
  {\bibinfo {volume} {174}},\ \bibinfo {pages} {345--356} (\bibinfo {year}
  {2021})}\BibitemShut {NoStop}%
\bibitem [{\citenamefont {De~Sousa}\ \emph {et~al.}(2020)\citenamefont
  {De~Sousa}, \citenamefont {Woellner}, \citenamefont {Machado}, \citenamefont
  {Autreto},\ and\ \citenamefont {Galvao}}]{de2020carbon}%
  \BibitemOpen
  \bibfield  {author} {\bibinfo {author} {\bibfnamefont {J.}~\bibnamefont
  {De~Sousa}}, \bibinfo {author} {\bibfnamefont {C.}~\bibnamefont {Woellner}},
  \bibinfo {author} {\bibfnamefont {L.}~\bibnamefont {Machado}}, \bibinfo
  {author} {\bibfnamefont {P.}~\bibnamefont {Autreto}}, \ and\ \bibinfo
  {author} {\bibfnamefont {D.}~\bibnamefont {Galvao}},\ }\bibfield  {title}
  {\enquote {\bibinfo {title} {Carbon nanotube peapods under high-strain rate
  conditions: A molecular dynamics investigation},}\ }\href@noop {} {\bibfield
  {journal} {\bibinfo  {journal} {MRS Advances}\ }\textbf {\bibinfo {volume}
  {5}},\ \bibinfo {pages} {1723--1730} (\bibinfo {year} {2020})}\BibitemShut
  {NoStop}%
\bibitem [{\citenamefont {Fthenakis}\ \emph {et~al.}(2022)\citenamefont
  {Fthenakis}, \citenamefont {Petsalakis}, \citenamefont {Tozzini},\ and\
  \citenamefont {Lathiotakis}}]{fthenakis2022evaluating}%
  \BibitemOpen
  \bibfield  {author} {\bibinfo {author} {\bibfnamefont {Z.}~\bibnamefont
  {Fthenakis}}, \bibinfo {author} {\bibfnamefont {I.}~\bibnamefont
  {Petsalakis}}, \bibinfo {author} {\bibfnamefont {V.}~\bibnamefont {Tozzini}},
  \ and\ \bibinfo {author} {\bibfnamefont {N.}~\bibnamefont {Lathiotakis}},\
  }\bibfield  {title} {\enquote {\bibinfo {title} {Evaluating the performance
  of reaxff potentials for sp2 carbon systems (graphene, carbon nanotubes,
  fullerenes) and a new reaxff potential},}\ }\href@noop {} {\bibfield
  {journal} {\bibinfo  {journal} {Frontiers in Chemistry}\ }\textbf {\bibinfo
  {volume} {10}} (\bibinfo {year} {2022})}\BibitemShut {NoStop}%
\bibitem [{\citenamefont {Stuart}, \citenamefont {Tutein},\ and\ \citenamefont
  {Harrison}(2000)}]{AIREBO}%
  \BibitemOpen
  \bibfield  {author} {\bibinfo {author} {\bibfnamefont {S.}~\bibnamefont
  {Stuart}}, \bibinfo {author} {\bibfnamefont {A.}~\bibnamefont {Tutein}}, \
  and\ \bibinfo {author} {\bibfnamefont {J.}~\bibnamefont {Harrison}},\
  }\bibfield  {title} {\enquote {\bibinfo {title} {A reactive potential for
  hydrocarbons with intermolecular interactions},}\ }\href@noop {} {\bibfield
  {journal} {\bibinfo  {journal} {The Journal of Chemical Physics}\ }\textbf
  {\bibinfo {volume} {112}},\ \bibinfo {pages} {6472--6486} (\bibinfo {year}
  {2000})}\BibitemShut {NoStop}%
\bibitem [{\citenamefont {Brenner}\ \emph {et~al.}(2002)\citenamefont
  {Brenner}, \citenamefont {Shenderova}, \citenamefont {Harrison},
  \citenamefont {Stuart}, \citenamefont {Ni},\ and\ \citenamefont
  {Sinnott}}]{REBO-II}%
  \BibitemOpen
  \bibfield  {author} {\bibinfo {author} {\bibfnamefont {D.}~\bibnamefont
  {Brenner}}, \bibinfo {author} {\bibfnamefont {O.}~\bibnamefont {Shenderova}},
  \bibinfo {author} {\bibfnamefont {J.}~\bibnamefont {Harrison}}, \bibinfo
  {author} {\bibfnamefont {S.}~\bibnamefont {Stuart}}, \bibinfo {author}
  {\bibfnamefont {B.}~\bibnamefont {Ni}}, \ and\ \bibinfo {author}
  {\bibfnamefont {S.}~\bibnamefont {Sinnott}},\ }\bibfield  {title} {\enquote
  {\bibinfo {title} {A second-generation reactive empirical bond order (rebo)
  potential energy expression for hydrocarbons},}\ }\href@noop {} {\bibfield
  {journal} {\bibinfo  {journal} {Journal of Physics: Condensed Matter}\
  }\textbf {\bibinfo {volume} {14}},\ \bibinfo {pages} {783} (\bibinfo {year}
  {2002})}\BibitemShut {NoStop}%
\bibitem [{\citenamefont {Los}\ and\ \citenamefont {Fasolino}(2003)}]{LCBOP-I}%
  \BibitemOpen
  \bibfield  {author} {\bibinfo {author} {\bibfnamefont {J.}~\bibnamefont
  {Los}}\ and\ \bibinfo {author} {\bibfnamefont {A.}~\bibnamefont {Fasolino}},\
  }\bibfield  {title} {\enquote {\bibinfo {title} {Intrinsic long-range
  bond-order potential for carbon: Performance in monte carlo simulations of
  graphitization},}\ }\href@noop {} {\bibfield  {journal} {\bibinfo  {journal}
  {Physical Review B}\ }\textbf {\bibinfo {volume} {68}},\ \bibinfo {pages}
  {024107} (\bibinfo {year} {2003})}\BibitemShut {NoStop}%
\bibitem [{\citenamefont {L{\'a}szl{\'o}}()}]{ILaszlo-unpublished}%
  \BibitemOpen
  \bibfield  {author} {\bibinfo {author} {\bibfnamefont {I.}~\bibnamefont
  {L{\'a}szl{\'o}}},\ }\href@noop {} {\enquote {\bibinfo {title}
  {unpublished},}\ }\BibitemShut {NoStop}%
\end{thebibliography}%

\end{document}